# Long and short range multi-locus QTL interactions in a complex trait of yeast

Evgeny M Mirkes[1], Thomas Walsh[2], Edward J Louis[2] and Alexander N Gorban[1]

[1]Centre for Mathematical Modelling and

[2]Centre for Genetic Architecture of Complex Traits

University of Leicester

Leicester LE1 7RH, UK

**Abstract**

We analyse interactions of Quantitative Trait Loci (QTL) in heat selected yeast by comparing them to an unselected pool of random individuals. Here we re-examine data on individual F12 progeny selected for heat tolerance, which have been genotyped at 25 locations identified by sequencing a selected pool [Parts, L., Cubillos, F. A., Warringer, J., Jain, K., Salinas, F., Bumpstead, S. J., Molin, M., Zia, A., Simpson, J. T., Quail, M. A., Moses, A., Louis, E. J., Durbin, R., & Liti, G. (2011). *Genome research*, *21*(7), 1131-1138]. 960 individuals were genotyped at these locations and multi-locus genotype frequencies were compared to 172 sequenced individuals from the original unselected pool (a control group). Various non-random associations were found across the genome, both within chromosomes and between chromosomes. Some of the non-random associations are likely due to retention of linkage disequilibrium in the F12 population, however many, including the inter-chromosomal interactions, must be due to genetic interactions in heat tolerance. One region of particular interest involves 3 linked loci on chromosome IV where the central variant responsible for heat tolerance is antagonistic, coming from the heat sensitive parent and the flanking ones are from the more heat tolerant parent. The 3-locus haplotypes in the selected individuals represent a highly biased sample of the population haplotypes with rare double recombinants in high frequency. These were missed in the original analysis and would never be seen without the multigenerational approach. We show that a statistical analysis of entropy and information gain in genotypes of a selected population can reveal further interactions than previously seen. Importantly this must be done in comparison to the unselected population's genotypes to account for inherent biases in the original population.

## Introduction

The determination of the underlying genetic causes of particular phenotypes has progressed greatly in recent years with studies in yeast being at the forefront of the application and development of new techniques [1]. Recent advances in quantitative genetic analysis in yeast have resulted in an unprecedented dissection of the genetic architecture of complex traits. The sequencing of a pool of selected progeny of a hybrid cross has provided high resolution mapping of QTLs for a number of traits [2-5]. Adding multiple generations via an advanced intercross line approach increases the resolution and sensitivity [6, 7]. In this paper, we analyse data on individual F12 progeny selected for heat tolerance, which have been genotyped at 25 locations identified by sequencing a selected pool. Determining epistatic interactions among QTLs requires the knowledge of genotypes in selected individuals and has been successfully used on progeny of F1 hybrids [8, 9]. Basic quantitative trait analysis of the progeny of F1 crosses, combined with the knowledge of gene function gained through decades of analysis, allows for the determination of causal genetic variation within large

QTL regions [8, 10]. Some genetic interactions between these can also be detected but only for very strong non-random associations [8]. Backcrosses help resolve QTL regions and reveal linked sets of causal variants in some cases [11, 12, 13]. Improvements in resolution have been made by pooling selected segregants of F1 hybrids [2, 14], and by using multi-generational hybrid populations [6, 7]. The issue of interactions/associations of genetic variants is still problematic, however, some progress has been made by sequencing large numbers of individual F1 progeny [9].

Heat tolerance has been one phenotype studied extensively, first with crosses involving a lab strain [10, 11, 13], then in 6 pairwise crosses between 4 different populations [8], then in a pairwise multi-generational cross [6] followed by a 4-way multigenerational cross incorporating the 4 populations originally studied [7]. In the first study large regions were identified that explained some of the phenotypic variation in heat tolerance with an analysis of candidate genes revealing some responsible genetic variation [10]. Further backcrosses revealed a linked set of QTLs within the regions, some of the variation providing heat tolerance coming from the more heat sensitive parent [11, 13]. In the 6 pairwise cross study, a total of 11 QTLs were identified for heat tolerance with low resolution, though none of the crosses had more than 4 QTLs segregating. Deleting one or the other allele in the hybrid, and measuring the heat tolerance in the resulting hemizygote, confirmed candidate gene involvement. The pairwise 12 generation cross, between the North American (NA) and West African (WA) populations, resulted in a very high resolution determination of 21 to 22 QTLs as opposed to the 4 large QTL regions determined from the F1 cross of the same parents [6, 8]. Here 8 of the variants providing heat tolerance came from the more heat sensitive parent (WA). The 4-way, 12-generation study increased the number of QTLs responsible for heat tolerance identified to 34 [7]. In the pairwise 12-generation cross, 960 individual heat resistant segregants were genotyped at the 22 QTLs and a few other segregating markers. No 2-locus inter-chromosomal interactions were found under the hypothesis of independence before selection and correction for multiple comparisons when analysing 19 of the segregating sites in the selected pool [6]. This is despite strong evidence of epistasis among the QTLs responsible for heat tolerance and the finding of negative epistatis between two loci in hemizygous double allele deletions. This analysis didn't detect interactions of linked QTLs nor could detect those of multiple gene interactions.

The problem of the "apparent lack of allele fixation and strong interchromosomal interactions after 12 d under selection" was discussed and several hypotheses were proposed [6]. In our work, we apply more sensitive technics for analysis of multiple testing results.

For analysis of associations in the heat selected population, which contains 896 multilocus genotypes after removing those with missing data, we apply a bootstrap based analysis of ordered Relative Information Gain (RIG, [15]) (developed specially for this study) and a more sophisticated version of false discovery rate control procedure [16]. These methods allow the identification of more than 20 pairs of QTLs with significantly non-random dependences. The requirement of significant differences of RIG between the unselected and heat selected pools measured by fraction of RIG and relative entropy reduces the set of significant pairs to 18. Weak correlations can be significant: statistically significant association does not necessarily mean strong correlation. We consider associations with RIG greater than 0.01 as moderate. There are four moderately correlated significantly dependent pairs: chr15-0172xxx & chr15-1032xxx, chr07-0859xxx & chr15-0172xxx, chr01-0119xxx & chr13-0910xxx and chr10-0420xxx & chr15-0172xxx. These four pairs and three constant loci (only one of two alleles present in the heat selected pool) can be considered as associated with heat tolerance.

**Materials and Methods**

We used the genotype data for the 960 heat selected individuals from [6] (provided in a supplemental spreadsheet – Table S1). Firstly 64 individuals with missing data were identified and removed. There was no bias in those with missing data. Indeed, we consider two samples: the original sample and sample with complete genotypes. To check the hypothesis that complete sample is not biased we test the hypothesis of coincidence of the distributions of NA and WA for each attribute in both samples. We apply $\chi^2$ tests [17] to check this hypothesis. For this test *p*-value is the probability to observe by chance the same or greater deviation in two samples if both samples are equally distributed. The minimal *p*-value for all attributes is 64% (p-values are provided in a supplemental spreadsheet – Table S2). As a result there is no evidence to reject the hypothesis of coincidence. It can be interpreted that missed values are missed completely at random (MCAR). It means that removing of incomplete records does not bias sample distributions.

In order to remove the possibility that there may have been bias in the original population before selection, leading to this apparent association due to the selection for heat tolerance, we needed to have the genotypes of individuals from the original pool prior to heat selection. Fortunately, 172 individuals from this study were sequenced as part of the 4-way study [7] to compare sizes of LD blocks (associations due to linkage). We scored each at the 25 loci for the allele to create a control data set for comparison (provided in a supplemental spreadsheet – Table S3). As a control group in our study we used these 172 genome sequences of unselected individuals from the pairwise cross [7], to determine the genotypes at the QTLs and other segregating loci used in the selected individuals. Testing the hypothesis that this complete sample is not biased shows that missed values (present in 7 genomes) in this unselected pool can be interpreted as missed completely at random (MCAR) (p-values are provided in a supplemental spreadsheet – Table S4).

To find associations between loci we use several approaches. Relative Information Gain (RIG) [15] is widely used in data mining to measure dependence. RIG is not symmetric. The greater value of RIG means the stronger the correlation and it is zero for independent attributes. RIG of the locus $X$ with respect the locus $Y$ is defined as:

$$\mathrm{RIG}(X|Y) = \frac{Entropy(X) - Entropy(X|Y)}{Entropy(X)},$$

where $Entropy(X)$ is the entropy of the allele distribution for the locus $X$:

$$Entropy(X) = -\mu \ln \mu - (1-\mu)\ln(1-\mu),$$

where $\mu$ is the fraction of genotypes with the NA allele in the locus $X$ among all genotypes, $Entropy(X|Y)$ is the relative entropy:

$$Entropy(X|Y) = \nu\, Entropy(X|y = \text{NA}) + (1 - \nu)\, Entropy(X|y = \text{WA}),$$

where $\nu$ is the fraction of genotypes with the NA allele in the locus $Y$ among all genotypes, $Entropy(X|y = \text{NA})$ and $Entropy(X|y = \text{WA})$ are the specific conditional entropies:

$$Entropy(X|y = \text{NA}) = -\mu_{y=\text{NA}} \ln \mu_{y=\text{NA}} - \left(1 - \mu_{y=\text{NA}}\right) \ln\left(1 - \mu_{y=\text{NA}}\right),$$

$$Entropy(X|y = \text{WA}) = -\mu_{y=\text{WA}} \ln \mu_{y=\text{WA}} - \left(1 - \mu_{y=\text{WA}}\right) \ln\left(1 - \mu_{y=\text{WA}}\right),$$

where $\mu_{y=\text{NA}}$ is the fraction of genotypes with NA allele in the locus $X$ among all genotypes with NA allele in the locus $Y$ and $\mu_{y=\text{WA}}$ is the fraction of genotypes with NA allele in the locus $X$ among all genotypes with WA allele in the locus *Y.*

One of the most widely used measures of correlation is Pearson's correlation coefficient (PCC):

$$\rho(X, Y) = \frac{p_{xy} - p_x p_y}{\sqrt{p_x(1 - p_x)p_y(1 - p_y)}},$$

where $p_{xy}$ is the fraction of genotypes with the NA alleles in the loci $X$ and $Y$ simultaneously among all genotypes, $p_x, p_y$ are the marginal frequencies of the NA alleles in the loci $X$ and $Y$ correspondingly. PCC is also used as a Linkage disequilibrium measure [18].

In signal recognition the Hamming correlation coefficient [19] is an alternative to the PCC. The normalized Hamming's correlation coefficient (NHCC) is the number of coincident symbols minus the number of different symbols divided by the length of sequences:

$$h = \frac{coincident - different}{n}.$$

For independent binary random variables, the probability of observing by chance the same or greater deviation in contingency table is equal to the probability of observing by chance the same or greater correlation (PCC, NHCC or RIG) for uncorrelated variables. Therefore, usage of PCC, NHCC or RIG is equivalent. Since RIG is evaluates correlation of categorical features, we use this measure in our study.

The test of independence [17] provides us a direct technique for independence analysis. To check the hypothesis of independence we apply the two-sided Fisher's exact test. It is closely related to the $D(\Delta)$ measure of linkage disequilibrium [20].

Revealing significant associations between different loci is a typical multiple testing problem. There are several techniques of accounting for multiple testing. The simplest one is the Bonferroni correction. Unfortunately the Bonferroni correction is very conservative [16, 21, 22]. The widely used BH step-up procedure [23] is less conservative than the Bonferroni correction, but is less powerful and more conservative [16, 22] than the $q$-value technique suggested by Storey and Tibshirani [16]. To define the significance of dependency we apply the calculation of $q$-values, which characterizes the False Discovery Rate (FDR) in the version suggested by Storey and Tibshirani [16].

Also we apply a *Bootstrap Test for ordered RIG* (BToRIG) to define the significance of each RIG. We have $L$ loci and $N$ observations for each locus. Calculate frequencies of NA for all loci. RIG is not symmetric and we calculate $m = L(L-1)$ values of RIG. Let us select a large number $B \gg m$ (in our study we have $m = 240$ and use $B = 10{,}100$). Let us sort the RIGs in descending order: $r_1 \geq r_2 \geq \cdots \geq r_m$. Our test calculates $p_i$-values which is the probability to observe by chance the same or greater value of RIG for $i$th maximal value of RIG if all loci are independently distributed with frequencies defined by original sample. We consider the $i$th RIG as significant with significance level $\alpha$ if $p_j \leq \alpha, \forall j \leq i$. The algorithm of this test is:

1. Calculate the frequencies $f_i$ for all loci, $i = 1, \dots, L$.
2. Calculate two RIGs $r_i$ for each pair of loci.
3. Sort RIGs in descending order $r_1 \geq r_2 \geq \cdots \geq r_m$.
4. Select large number $B \gg m$ and perform bootstrap procedure:
   4.1. Generate artificial loci with frequencies $f_i$ (number of NA can slightly fluctuate).
   4.2. Calculate two RIGs $R_{i,k}$ for each pair of loci, where $k = 1, \dots, B$ is the number of generation.
   4.3. Sort RIGs in order $R_{1,k} \geq R_{2,k} \geq \cdots \geq R_{m,k}$.
5. Calculate $p$-value for $i$th correlation
   5.1. $K_i = \sum_{k=1}^{B} h(r_i - R_{i,k})$, where $h(x)$ is the Heaviside step function $h(x) = 1$ if $x > 0$ and $h(x) = 0$ if $a \leq 0$.
   5.2. $p_i = K_i / B$.
6. Correlation $r_i$ is significant with significance level $\alpha$ if $r_j \leq \alpha, j \leq i$.

This approach can be also applied for ordered PCC and for any other measure of correlation.

To check the identity of distributions of NA and WA in the same locus in heat selected and unselected pools we consider two random variables for each locus: the first variable is whether it is in the heat selected or unselected pool and the second variable is which allele, NA or WA. Independence of these two variables means that selection has no effect for this locus. We apply Fisher's exact test and the $\chi^2$ test.

'Statistically significant association' does not necessarily mean 'large correlation'. The multiple testing procedures return lists of statistically significant associations between loci in the heat selected population. But the correlations may be quite small. To select a reasonable level of correlation we retain significant links with RIG>$\varepsilon$ for some threshold $\varepsilon$>0 only (for example, $\varepsilon$=0.01).
A second selection is necessary to compare associations in the heat selected populations with the unselected population (control group). For this purpose, we consider the allele distributions in pairs of associated (after selection) loci and calculate the relative entropy [24] with respect to this distribution in the unselected population. A value of zero for relative entropy means that the association is the same in both the selected and unselected groups. Associations with small relative entropy with respect to the unselected population should be additionally tested as they may be caused not by the heat tolerance but be a property of the unselected population. Another measure for change of association after selection gives the RIG ratio: ($\mathrm{RIG}_{selected}/\mathrm{RIG}_{unselected}$).

**Results**

Several methods of association analysis, i.e. lack of independence, were applied to these data. Statistics of PCC, NHCC, and RIG were analysed for each pool separately. The false discovery rate control procedure [16] was also implemented and applied. Both inter and intra chromosomal associations were found. Comparison of associations in the two pools allows identification of real connections associated with heat tolerance.

There are two pools: unselected and heat selected. The first reasonable question is ‘Are these pools significantly different in allele distribution inside loci?’ Tests of identity of distributions of NA and WA in selected and unselected pools reveals six loci with identical distributions. Four loci are identified with $p$-value greater than 0.1 and two loci with $p$-value between 0.01 and 0.1 (see Figure 1 and Table S5 in supplementary material). In this test the $p$-value is the probability of observing by chance the same or greater dependence in contingency table if two random variables are independent.

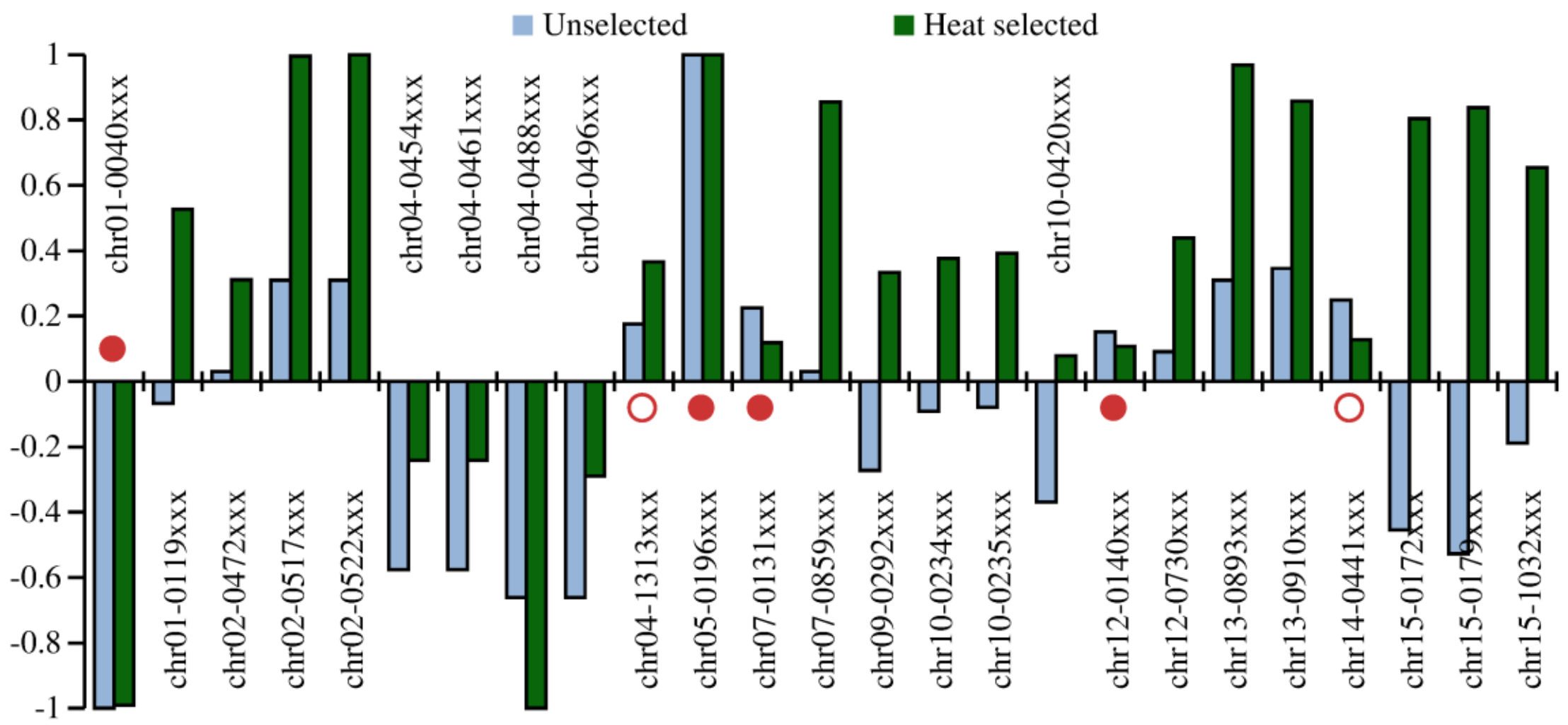

Figure 1. The diagrams of $(N_{\text{NA}} - N_{\text{WA}})/(N_{\text{NA}} + N_{\text{WA}})$ for unselected and heat selected samples. Loci in which the hypothesis of independence of allele

distributions of the heat selected or unselected pools cannot be rejected are marked by solid circle if $p$-value is greater than 0.1 and by circle if $p$-value is between 0.01 and 0.1.

Figure 1 shows the significant differences between heat selected and unselected pools for most of loci. As might be expected the fraction of NA is increased after selection in most of the loci as this is the heat tolerant parent. A perhaps unexpected elimination of NA is observed for locus chr04-0488xxx, however such antagonisitic alleles are known and this one has been discussed in the previous analysis [6]. For eight loci we can see the changing of sign, antagonisim, of differences $N_{\mathrm{NA}} - N_{\mathrm{WA}}$. For three loci of chromosome IV an increase of variability is observed.

Figure 1 shows that two of the markers (chr01-0040xxx and chr05-0196xxx) in the unselected pool and three of the markers (chr02-0522xxx, chr04-0488xxx and chr05-0196xxx) in the heat selected pool had alleles from only one of the parents, i.e. they were fixed. Furthermore there are two almost constant markers in the heat selected pool: chr01-0040xxx contains 99.6% WA and chr02-0517xxx contains 99.8% NA. These two markers can be interpreted as constant loci because numbers of observed NA and WA correspondingly are too small. Three markers chr02-0522xxx, chr04-0488xxx and chr02-0517xxx can be interpreted as exactly associated with heat resistance. We exclude these loci from the further analysis of associations.

We call two loci linked if these loci are adjacent and fraction of mixed genes (NA-WA and WA-NA) is low – i.e. there is linkage disequilibrium. The linked loci (both for the unselected and heat selected pools) have the same colour in Figure 2. For each group of linked loci, one locus with the most balanced (nearest to 0.5) frequencies of NA and WA is retained for further analysis.

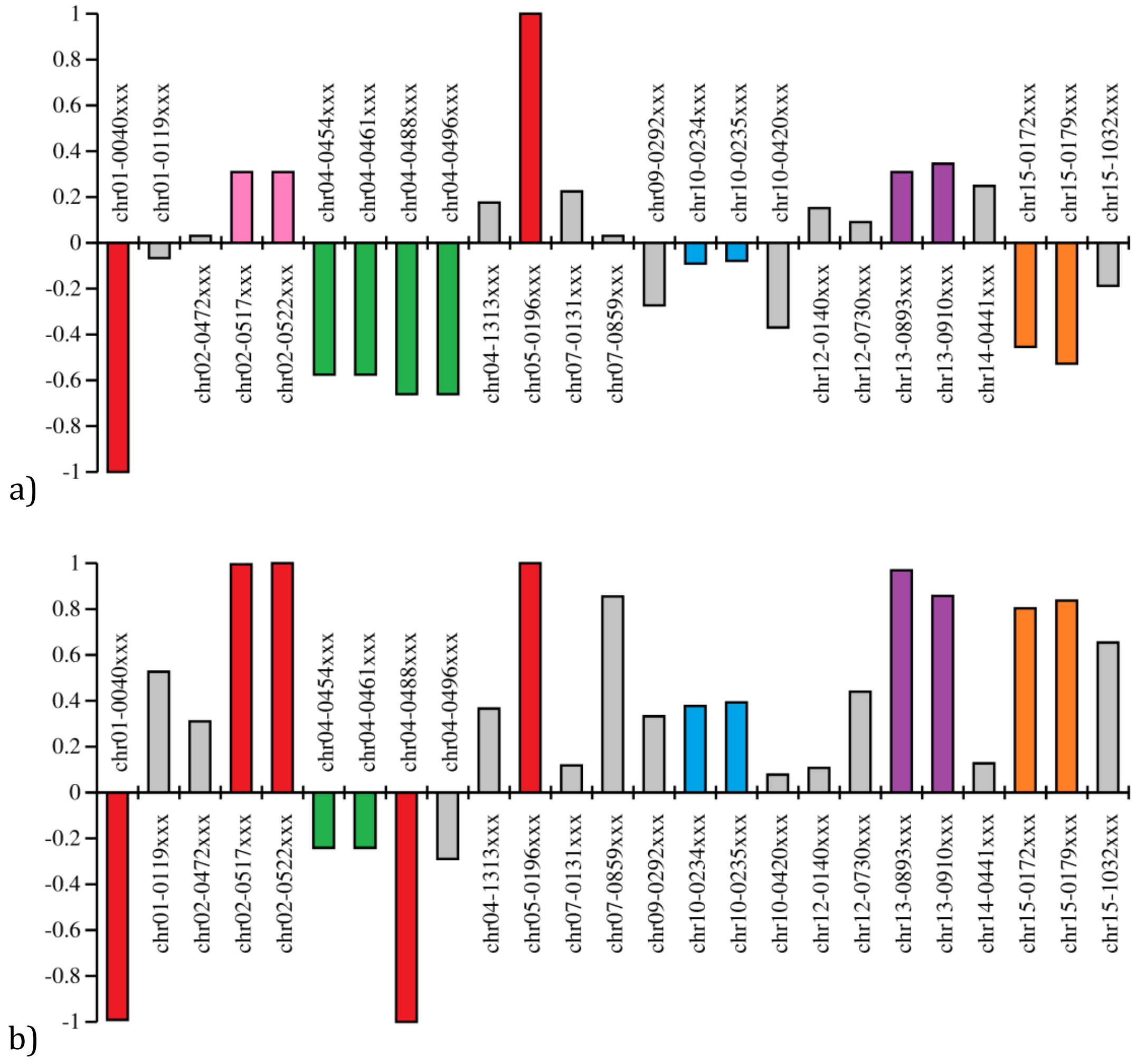

Figure 2. Distribution of $(N_{\mathrm{NA}} - N_{\mathrm{WA}})/(N_{\mathrm{NA}} + N_{\mathrm{WA}})$ versus QTLs for (a) unselected and (b) heat selected pools. Red corresponds to one parent allele (constant loci) and almost constant loci (the fraction of one of the gene is greater than 99%), magenta, green, blue, violet and brown correspond to different groups of linked loci and grey colour corresponds to all other loci.

To find correlated loci in the heat selected pool we apply a bootstrap test for PCC, NHCC, and RIG.

The bootstrap test of significance of PCC defines two pairs (chr04-0461xxx & chr04-0496xxx, and chr10-0234xxx & chr12-0730xxx) with significance level *p*=0.05. This result means that for binary random variables PCC is not an appropriate measure of correlation. The bootstrap test of significance of NHCC defines two pairs (chr04-0461xxx & chr04-0496xxx, and chr07-0859xxx & chr13-0910xxx) with significance level *p*=0.05. This result means that NHCC is not an appropriate measure of correlation for this genomic study. However, the sum of NHCC for all pairs of loci, except constant and linked loci in the unselected pool is equal to 9.72 and for the heat selected pool 19.73. This indicates a significant increase of correlation (measured by the sum of NHCC). This effect (growth of correlations under stress) is well known [25, 26].

For BToRIG, the number of significant connections with respect to *p*-value is depicted in Figure 3a. We consider as significant connections with a *p*-value which is not greater than 0.005 (29 pairs in Figure 4a). This set includes three pairs found in the previous analysis [6].

a)
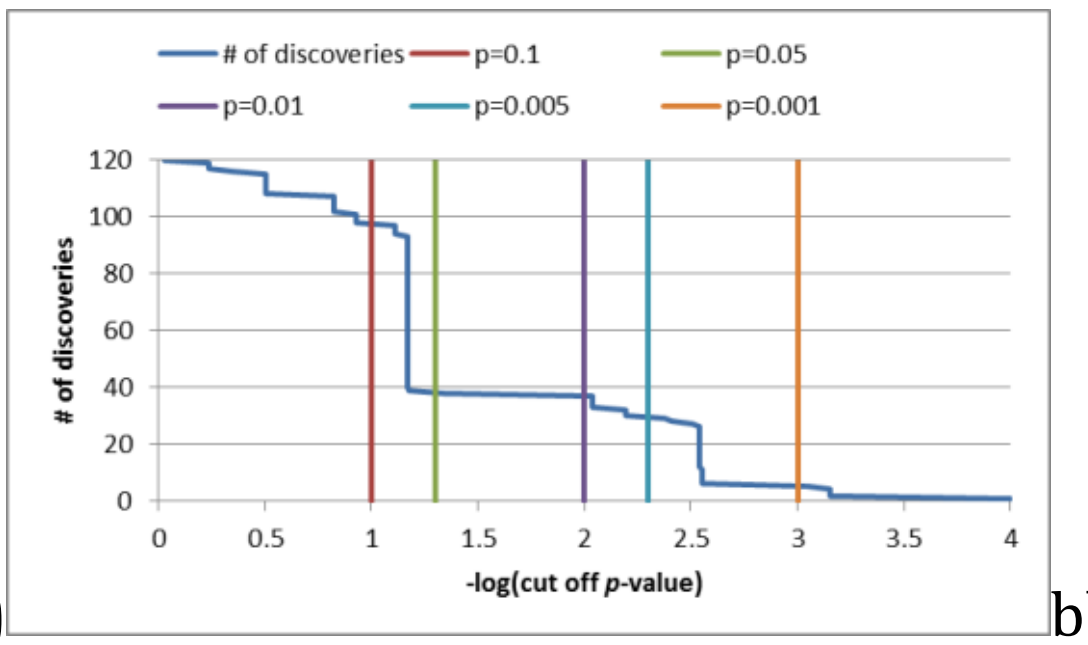

b)
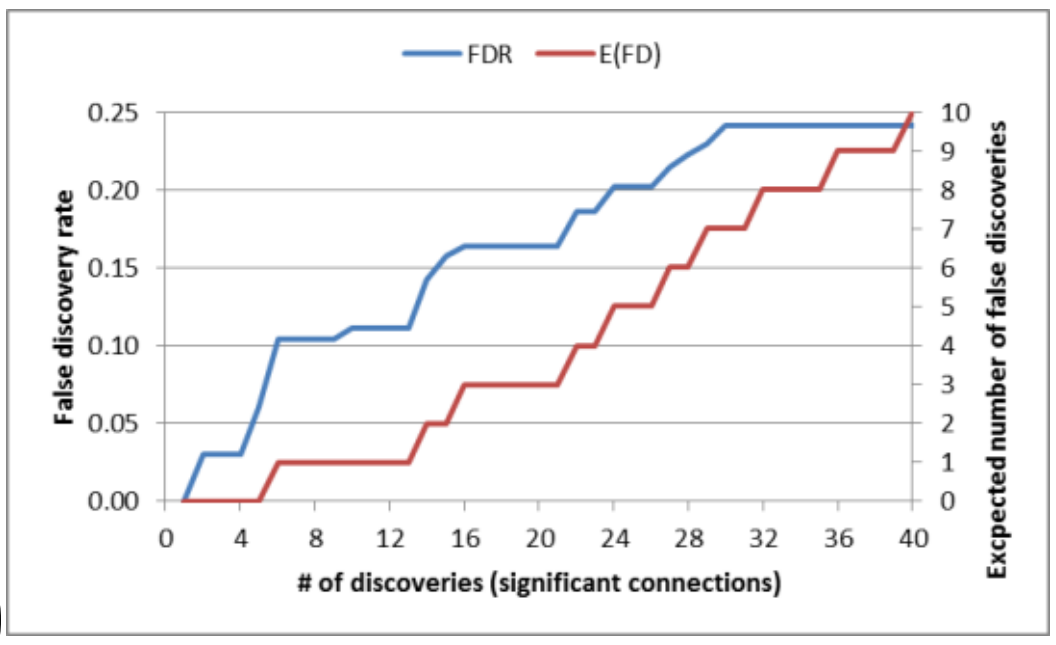

Figure 3. The number of significant correlations for BToRIG (a) and FDR and estimated number of false discoveries for FDR approach.

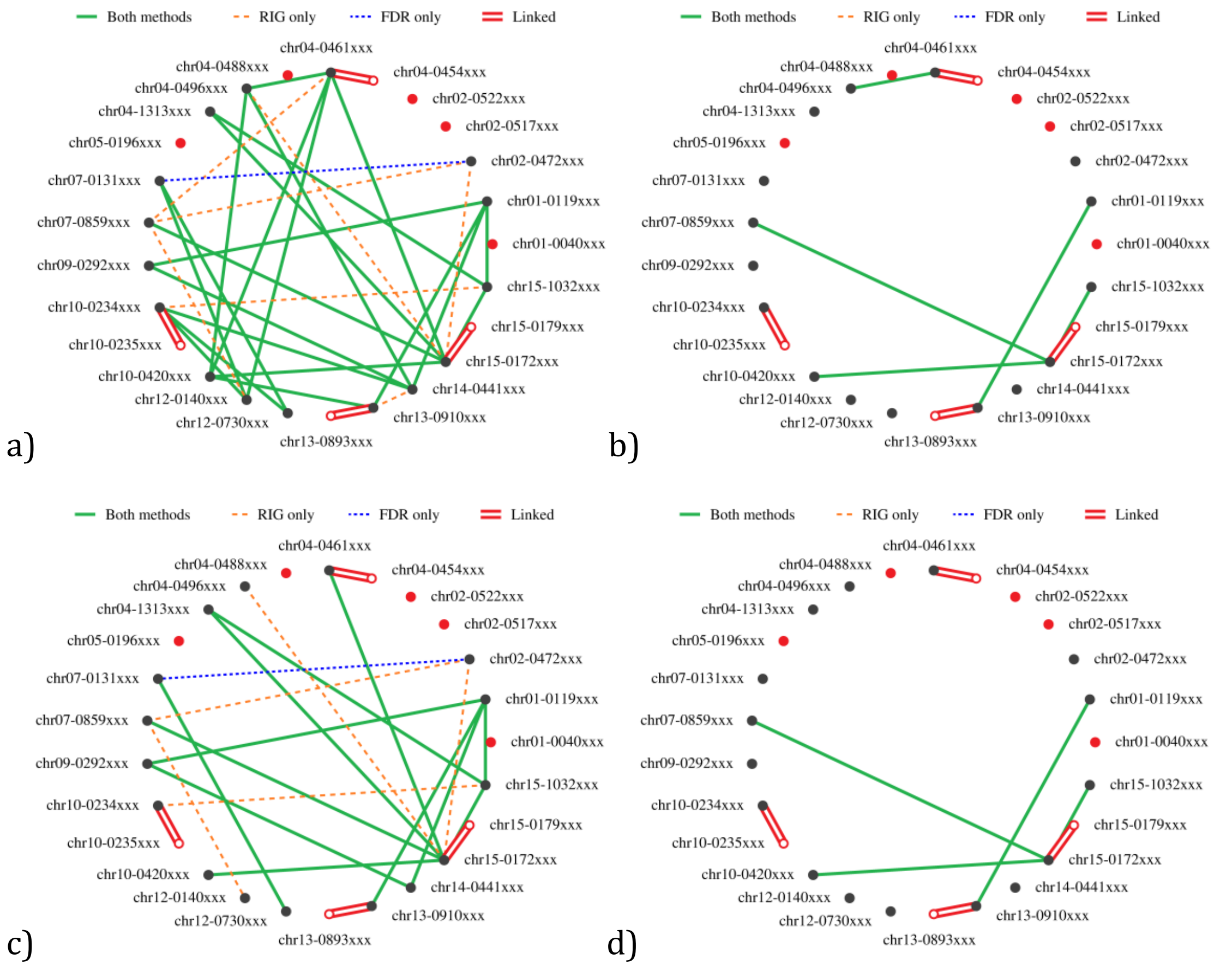

Figure 4. Sets of significantly dependent loci for heat selected pools: a) all connections selected as significant, b) strong and moderate connections with RIG $\geq 0.01$, c) significant connections with $RIG_{selected}/RIG_{unselected} \geq 2$ or relative entropy (selected with respect to unselected group) is greater than 0.5 and d) significant connections with RIG $\geq 0.01$, and one of the following conditions: $RIG_{selected}/RIG_{unselected} \geq 2$ or relative entropy is greater than 0.5. Red solid circle depict the constant loci. Red circle with white centre depicts the loci excluded because there are in linkage disequilibrium with other loci (doubled red line). Solid green lines connect loci defined as significantly dependent by DFR and BToRIG. Brown dashed lines connect loci defined as significantly correlated by the Bootstrap test only. Blue dotted line connects loci defined as significant by DFR only.

We also apply the False Discovery Rate (FDR) approach to identify significantly dependent loci. Graphs of FDR and expected number of false discovery with respect to number of significant connections are depicted in Figure 3b. For FDR we consider as significantly dependent the first $k$ connections with $q_k < 0.2$, where $q_k$ is the $q$-value of $k$th test. It means that expected number of false discovery is not greater than $kq_k \leq 0.2k$ for $k$ significant connections. Number of significantly dependent connections is 23. All these connections also were detected as significantly correlated by BToRIG except pair chr02-0472xxx & chr07-0131xxx. This set includes three pairs found previously [6].

The heat selected pool contains 896 genomes. This is large enough a sample so that a weak correlation can be identified as significant. To exclude significantly correlated pairs with really weak correlation we remove all pairs with RIG which is less than 0.01. All pairs with strong and moderate correlation are depicted in Figure 4b.

To identify correlations associated with heat tolerance it is necessary to compare RIG for the same pair of loci in unselected and heat selected pools. We apply two approaches for this purpose: estimation of RIG ratio ($\mathrm{RIG}_{selected}/\mathrm{RIG}_{unselected}$) and the relative entropy. We consider change of RIG as significant if RIG ratio is greater than 2 or relative entropy is greater than 0.5. Significantly correlated associations with significant changes of RIG are depicted in Figure 4c.

Finally we unite two requirements: a pair of loci can be associated with heat tolerance if the correlation of this pair is significant, the RIG of this pair has significant changes and the RIG is not less than 0.01. The final set of pairs of loci associated with heat tolerance is depicted in Figure 4d. The expected number of false discoveries for this set of pairs is less than one.

Chromosome IV needs additional discussion. Table 1 shows that alleles in the first four loci in IV chromosome (chr04-0454xxx, chr04-0461xxx, chr04-0488xxx and chr04-0496xxx) are not distributed independently with probability 1/2 of NA and WA. For such an equidistribution, the probability of each combination of alleles in four loci is 1/16. The probability of observing by chance the same or greater than in Table 1 deviation in contingency table from this independent equidistribution is less than $10^{-300}$. Alleles in the first four loci are also not distributed independently with probabilities calculated by samples (these probabilities are different for unselected and heat selected pools, the probability of observing by chance the same or greater deviation in contingency table is less than $10^{-300}$). In the unselected pool these four loci are in linkage disequilibrium because the fraction of genotypes with the same marker in all loci is 91% (NA-NA-NA-NA 15% and WA-WA-WA-WA 76%). If we remove the constant locus chr04-0488xxx from the heat selected pool then we also can consider the other three loci as in linkage disequilibrium because the fraction of genotypes with the same marker in all loci is 89% (NA-NA-NA 31% and WA-WA-WA 58%). However, the locus chr04-0488xxx is located between loci chr04-0461xxx and chr04-0496xxx. Therefore, the strong correlation between loci chr04-0461xxx and chr04-0496xxx cannot be explained by lack of crossovers as can be done for the unselected pool.

Table 1. Frequencies of markers in four loci of IV chromosome.

| Chromosome IV locus | | | | Observed | | | | Independent with marginal probabilities | | | |
|---|---|---|---|---|---|---|---|---|---|---|---|
| 0454 xxx | 0461 xxx | 0488 xxx | 0496 xxx | Unselected | | Heat selected | | Unselected | | Heat selected | |
| | | | | # | % | # | % | # | % | # | % |
| NA | NA | NA | NA | 24 | 15 | 0 | 0 | 0.21 | 0 | 0.00 | 0 |
| NA | NA | NA | WA | 1 | 1 | 0 | 0 | 1.05 | 1 | 0.00 | 0 |
| NA | NA | WA | NA | 0 | 0 | 279 | 31 | 1.05 | 1 | 45.79 | 5 |
| NA | NA | WA | WA | 9 | 5 | 53 | 6 | 5.12 | 3 | 83.23 | 9 |
| NA | WA | NA | NA | 0 | 0 | 0 | 0 | 0.79 | 0 | 0.00 | 0 |
| NA | WA | NA | WA | 0 | 0 | 0 | 0 | 3.89 | 2 | 0.00 | 0 |
| NA | WA | WA | NA | 0 | 0 | 3 | 0 | 3.89 | 2 | 74.88 | 8 |
| NA | WA | WA | WA | 1 | 1 | 5 | 1 | 19.01 | 12 | 136.10 | 15 |
| WA | NA | NA | NA | 1 | 1 | 0 | 0 | 0.79 | 0 | 0.00 | 0 |
| WA | NA | NA | WA | 0 | 0 | 0 | 0 | 3.89 | 2 | 0.00 | 0 |
| WA | NA | WA | NA | 0 | 0 | 4 | 0 | 3.89 | 2 | 74.88 | 8 |
| WA | NA | WA | WA | 0 | 0 | 4 | 0 | 19.01 | 12 | 136.10 | 15 |
| WA | WA | NA | NA | 2 | 1 | 0 | 0 | 2.95 | 2 | 0.00 | 0 |
| WA | WA | NA | WA | 0 | 0 | 0 | 0 | 14.43 | 9 | 0.00 | 0 |
| WA | WA | WA | NA | 1 | 1 | 32 | 4 | 14.43 | 9 | 122.45 | 14 |
| WA | WA | WA | WA | 126 | 76 | 516 | 58 | 70.61 | 43 | 222.57 | 25 |

Statistical comparisons of the heat selected population to the control population now reveal associations due to the selection for heat tolerance above any pre-existing biases in the genotypes due to the process of generation of the test population (linkage disequilibrium and inadvertent selection for some variants during the 12 rounds of meiosis and random mating). There remain several 2-way up to an 8-way association between unlinked markers (see Figure 4). Importantly the linked associations where the NA-WA-NA haplotype on chromosome IV, extremely rare in the unselected population (none were seen in the 172 unselected individuals and have a predicted frequency of 0.08% if 12 generations of crossing result in independent genetic intervals), was found in 30% of the heat selected individuals.

## Discussion/Conclusions

Thirty significant associations between QTLs are detected by two statistical approaches, FDR (False Discovery Rate) and BToRIG (Bootstrap Test for ordered RIG). 22 of them are detected by both approaches simultaneously (see Figure 4). These include both inter-chromosomal, unlinked, interactions as seen in Figure 4as well as interactions among linked QTLs as seen on chromosome IV, Figure 4.

Previous analysis of these data [6] found three pairs of significant connections: chr15-0172xxx & chr15-1032xxx, chr10-0234xxx & chr12-0730xxx and chr07-0131xxx & chr12-0140xxx. The remarkable difference in findings is caused by the more conservative BH step up procedure of multiple testing in [6] (it is well known that this procedure has relatively low power). All three of these pairs are found by both applied technique: FDR and BToRIG (see Figure 4a). However two of them, chr10-0234xxx & chr12-0730xxx and chr07-0131xxx & chr12-0140xxx, have relatively small value of RIG and small changing of RIG (in comparison of the heat selected and unselected pools). It means that these associations are significant but are not considered here as sufficiently strongly correlated. Furthermore, both these associations have approximately the same correlations in the heat selected and unselected pools. As a result the only pair chr15-0172xxx & chr15-1032xxx which is found in the previous study [6] is presented in Figure 4d as a significant and sufficiently strong association in the context of heat tolerance.

The hypothesis that the unselected pool contains independent loci is not correct as it is shown in Figures 1 and 2. This fact was the basis for the the usage of correlation change analysis rather than simple correlation analysis.

Three loci become constant in the heat selected pool: chr02-0522xxx, chr04-0488xxx and chr05-0196xxx. Loci chr02-0522xxx and chr05-0196xxx contain NA markers only. Locus chr04-0488xxx contains WA markers only.

We exclude constant loci and keep for analysis only one locus from any pair of loci in strong linkage disequilibrium. After that, we apply two multi testing approaches: calculation of $q$-value to estimate FDR and BToRIG to find pairs with significant highest RIGs. The FDR approach identifies 23 significantly dependent pairs of loci and BToRIG identifies 29 correlated pairs of loci, which include 22 of pairs identified by FDR. Removing weak correlations and pairs with small change of correlations in the heat selected pool in comparison with the unselected pool decreases number of pairs of loci associated with heat tolerance to four pairs: chr15-0172xxx & chr15-1032xxx, chr07-0859xxx & chr15-0172xxx, chr01-0119xxx & chr13-0910xxx and chr10-0420xxx & chr15-0172xxx.

We also find that distribution of chr04-0461xxx & chr04-0496xxx in the two pools are very similar but distribution of three adjacent loci chr04-0461xxx, chr04-0488xxx and chr04-0496xxx are drastically changed. The association between chr04-0461xxx and chr04-0496xxx in selected and in unselected pools is the same and formally we exclude it from Figure 4d. Nevertheless, the situation in chromosome 4 is very special: the combination NA, NA, WA, NA in loci chr04-0454xxx, chr04-0461xxx, chr04-0488xxx and chr04-0496xxx occurs in 31% of genotypes in the heat selected pool and in 0% of genotypes in the unselected pool (Table 1). Therefore, this association should be considered as a result of selection.

The application of statistical methods developed for other purposes are proving useful in the determination of interactions among QTLs of complex traits, bringing us closer to understanding the entire heritability of traits.

The biology of the interactions found will require further experimentation. In particular the linked set of QTLs with heat tolerance promoting alleles coming from both parents, and the interaction within the set is of interest. The genetic architecture of this region could be the result of adaptation to heat tolerance in both populations.

**Acknowledgements:** This work was in part supported by the Wellcome Trust Institutional Strategic Support Fund WT097828/Z/11/Z (RM33G0255 and RM33G0335).

## References

1. Liti, G., & Louis, E. J. (2012). Advances in quantitative trait analysis in yeast. *PLoS genetics*, *8*(8), e1002912.

2. Ehrenreich IM, Torabi N, Jia Y, Kent J, Martis S, Shapiro JA, Gresham D, Caudy AA, Kruglyak L. 2010. Dissection of genetically complex traits with extremely large pools of yeast segregants. *Nature* **464**(7291): 1039-1042.
3. Schwartz K, Wenger JW, Dunn B, Sherlock G. 2012. APJ1 and GRE3 homologs work in concert to allow growth in xylose in a natural Saccharomyces sensu stricto hybrid yeast. *Genetics* **191**(2): 621-632.
4. Swinnen S, Schaerlaekens K, Pais T, Claesen J, Hubmann G, Yang Y, Demeke M, Foulquie-Moreno MR, Goovaerts A, Souvereyns K et al. 2012. Identification of novel causative genes determining the complex trait of high ethanol tolerance in yeast using pooled-segregant whole-genome sequence analysis. *Genome Res* **22**(5): 975-984.
5. Hubmann G, Mathe L, Foulquie-Moreno MR, Duitama J, Nevoigt E, Thevelein JM. 2013. Identification of multiple interacting alleles conferring low glycerol and high ethanol yield in Saccharomyces cerevisiae ethanolic fermentation. *Biotechnol Biofuels* **6**(1): 87.
6. Parts, L., Cubillos, F. A., Warringer, J., Jain, K., Salinas, F., Bumpstead, S. J., Molin, M., Zia, A., Simpson, J. T., Quail, M. A., Moses, A., Louis, E. J., Durbin, R., & Liti, G. (2011). Revealing the genetic structure of a trait by sequencing a population under selection. *Genome research*, *21*(7), 1131-1138.
7. Cubillos, F. A., Parts, L., Salinas, F., Bergström, A., Scovacricchi, E., Zia, A., Illingworth, C. J. R., Mustonen, V., Ibstedt, S., Warringer, J., Louis, E. J., Durbin, R., & Liti, G. (2013). High-resolution mapping of complex traits with a four-parent advanced intercross yeast population. *Genetics*, *195*(3), 1141-1155.

8. Cubillos, F. A., Billi, E., Zörgö, E., Parts, L., Fargier, P., Omholt, S., Blomberg, A., Warringer, J., Louis, E. J., & Liti, G. (2011). Assessing the complex architecture of polygenic traits in diverged yeast populations. *Molecular ecology*, *20*(7), 1401-1413.
9. Bloom, J. S., Ehrenreich, I. M., Loo, W. T., Lite, T. L. V., & Kruglyak, L. (2013). Finding the sources of missing heritability in a yeast cross. *Nature*,*494*(7436), 234-237.
10. Steinmetz, L. M., Sinha, H., Richards, D. R., Spiegelman, J. I., Oefner, P. J., McCusker, J. H., & Davis, R. W. (2002). Dissecting the architecture of a quantitative trait locus in yeast. *Nature*, *416*(6878), 326-330.
11. Sinha H, Nicholson BP, Steinmetz LM, McCusker JH. 2006. Complex genetic interactions in a quantitative trait locus. *PLoS Genet* **2**(2): e13.
12. Ben-Ari G, Zenvirth D, Sherman A, David L, Klutstein M, Lavi U, Hillel J, Simchen G. 2006. Four linked genes participate in controlling sporulation efficiency in budding yeast. *PLoS Genet* **2**(11): e195.
13. Sinha, H., David, L., Pascon, R. C., Clauder-Münster, S., Krishnakumar, S., Nguyen, M., Shi, G., Dean, J., Davis, R. W., Oefner, P. J., McCusker, J. H., & Steinmetz, L. M. (2008). Sequential elimination of major-effect contributors identifies additional quantitative trait loci conditioning high-temperature growth in yeast. *Genetics*, *180*(3), 1661-1670.
14. Ehrenreich IM, Bloom J, Torabi N, Wang X, Jia Y, Kruglyak L. 2012. Genetic architecture of highly complex chemical resistance traits across four yeast strains. *PLoS Genet* **8**(3): e1002570.
15. Mitchell, T. M. (1997). Machine learning. 1997. *Burr Ridge, IL: McGraw Hill*, *NY*.

16. Storey, J. D., & Tibshirani, R. (2003). Statistical significance for genome wide studies. *Proceedings of the National Academy of Sciences*, *100*(16), 9440-9445.
17. Greenwood, P. E. & Nikulin, M. S. (1996). A Guide to Chi-squared Testing. New York: Wiley
18. Hill, W. G., & Robertson, A. (1968). Linkage disequilibrium in finite populations.*Theoretical and Applied Genetics*, *38*(6), 226-231.
19. Lempel, A., & Greenberger, H. (1974). Families of sequences with optimal Hamming-correlation properties. *Information Theory, IEEE Transactions on*,*20*(1), 90-94.
20. Lewontin, R. C., & Kojima, K. I. (1960). The evolutionary dynamics of complex polymorphisms. *Evolution*, 458-472.
21. Hirschhorn, J. N., & Daly, M. J. (2005). Genome-wide association studies for common diseases and complex traits. *Nature Reviews Genetics*, *6*(2), 95-108.
22. Verhoeven, K. J., Simonsen, K. L., & McIntyre, L. M. (2005). Implementing false discovery rate control: increasing your power. *Oikos*, *108*(3), 643-647.
23. Benjamini, Y., & Hochberg, Y. (1995). Controlling the false discovery rate: a practical and powerful approach to multiple testing. *Journal of the Royal Statistical Society. Series B (Methodological)*, 289-300.
24. Kullback, S., & Leibler, R. A. (1951). On information and sufficiency. *The Annals of Mathematical Statistics*, 79-86.
25. Gorban, A. N., Smirnova, E. V., & Tyukina, T. A. (2010). Correlations, risk and crisis: From physiology to finance. *Physica A*, 389(16), 3193-3217.
26. Censi, F., Giuliani, A., Bartolini, P., & Calcagnini, G. (2011). A multiscale graph theoretical approach to gene regulation networks: a case study in atrial fibrillation. *Biomedical Engineering, IEEE Transactions on*, 58(10), 2943-2946.

**Table S1. Heat selected genotypes (Data from the suplementary materials for the paper [Parts, L., Cubillos, F. A., Warringer, J., Jain, K., Salinas, F., Bumpstead, S. J., Molin, M., Zia, A., Simpson, J. T., Quail, M. A., Moses, A., Louis, E. J., Durbin, R., and Liti, G. (2011). Revealing the genetic structure of a trait by sequencing a population under selection. Genome research, 21(7), 1131-1138])**

| ID | chr01-0040xxx | chr01-0119xxx | chr02-0472xxx | chr02-0517xxx | chr02-0522xxx | chr04-0454xxx | chr04-0461xxx | chr04-0488xxx | chr04-0496xxx | chr04-1313xxx | chr05-0196xxx | chr07-0131xxx | chr07-0859xxx | chr09-0292xxx | chr10-0234xxx | chr10-0235xxx | chr10-0420xxx | chr12-0140xxx | chr12-0730xxx | chr13-0893xxx | chr13-0910xxx | chr14-0441xxx | chr15-0172xxx | chr15-0179xxx | chr15-1032xxx |
|---|---|---|---|---|---|---|---|---|---|---|---|---|---|---|---|---|---|---|---|---|---|---|---|---|---|
| 1 | WA | NA | NA | NA | NA | WA | WA | WA | WA | WA | NA | WA | NA | WA | NA | NA | WA | WA | NA | NA | NA | NA | NA | NA | NA |
| 2 | WA | WA | WA | NA | NA | WA | WA | WA | WA | NA | NA | NA | NA | NA | WA | WA | WA | NA | WA | NA | NA | WA | NA | NA | NA |
| 3 | WA | NA | NA | NA | NA | WA | WA | WA | WA | NA | NA | NA | NA | NA | NA | NA | WA | WA | NA | NA | NA | NA | NA | NA | NA |
| 4 | WA | NA | WA | NA | NA | NA | NA | WA | NA | NA | NA | WA | NA | WA | WA | WA | NA | NA | NA | NA | NA | WA | NA | NA | NA |
| 5 | WA | NA | NA | NA | NA | WA | WA | WA | WA | NA | NA | WA | NA | NA | NA | NA | NA | WA | NA | NA | NA | WA | NA | NA | NA |
| 6 | WA | NA | NA | NA | NA | NA | NA | WA | NA | NA | NA | NA | NA | NA | NA | NA | NA | NA | NA | NA | NA | NA | WA | WA | NA |
| 7 | - | - | - | - | - | - | - | - | - | - | - | - | - | - | - | - | - | - | - | - | - | - | - | - | - |
| 8 | WA | WA | NA | NA | NA | WA | WA | WA | WA | WA | NA | NA | NA | NA | NA | NA | NA | WA | NA | NA | WA | WA | NA | NA | WA |
| 9 | WA | NA | NA | NA | NA | WA | WA | WA | WA | WA | NA | NA | NA | NA | NA | NA | NA | WA | WA | NA | NA | WA | NA | NA | NA |
| 10 | WA | NA | WA | NA | NA | NA | NA | WA | NA | NA | NA | WA | NA | WA | WA | WA | NA | NA | NA | NA | NA | NA | NA | NA | NA |
| 11 | WA | WA | WA | NA | NA | WA | WA | WA | WA | WA | NA | NA | NA | NA | NA | NA | WA | WA | NA | NA | NA | NA | NA | NA | NA |
| 12 | WA | NA | WA | NA | NA | NA | NA | WA | NA | NA | NA | WA | NA | WA | WA | WA | WA | NA | WA | NA | NA | NA | NA | NA | NA |
| 13 | WA | NA | WA | NA | NA | NA | NA | WA | NA | NA | NA | NA | NA | NA | NA | NA | WA | NA | NA | NA | WA | NA | NA | NA | WA |
| 14 | WA | NA | NA | NA | NA | WA | WA | WA | WA | NA | NA | NA | NA | NA | NA | NA | NA | NA | NA | NA | NA | NA | WA | WA | NA |
| 15 | WA | NA | NA | NA | NA | WA | WA | WA | WA | NA | NA | NA | NA | NA | WA | WA | WA | WA | WA | NA | NA | NA | NA | NA | NA |
| 16 | WA | NA | NA | NA | NA | WA | WA | WA | WA | NA | NA | NA | NA | NA | NA | NA | NA | NA | NA | NA | NA | NA | NA | NA | NA |
| 17 | WA | NA | NA | NA | NA | NA | NA | WA | WA | WA | NA | WA | NA | NA | NA | NA | NA | WA | WA | NA | WA | WA | NA | NA | NA |
| 18 | NA | NA | WA | NA | NA | WA | WA | WA | WA | NA | NA | WA | NA | NA | WA | NA | NA | WA | NA | NA | NA | WA | NA | NA | NA |
| 19 | WA | NA | WA | NA | NA | WA | WA | WA | WA | NA | NA | WA | NA | WA | NA | NA | NA | WA | WA | NA | NA | WA | NA | NA | NA |
| 20 | WA | NA | WA | NA | NA | NA | NA | WA | NA | WA | NA | WA | NA | NA | NA | NA | NA | NA | WA | NA | NA | WA | NA | NA | NA |
| 21 | WA | NA | NA | NA | NA | NA | NA | WA | NA | NA | NA | NA | NA | NA | NA | NA | NA | NA | NA | NA | NA | NA | NA | NA | NA |
| 22 | WA | NA | NA | NA | NA | WA | WA | WA | WA | NA | NA | WA | NA | NA | NA | NA | NA | WA | NA | NA | NA | WA | NA | NA | NA |
| 23 | WA | WA | WA | NA | NA | WA | WA | WA | WA | NA | NA | WA | NA | NA | WA | WA | NA | NA | NA | NA | NA | NA | NA | NA | NA |
| 24 | WA | NA | NA | NA | NA | NA | NA | WA | NA | NA | NA | WA | NA | NA | NA | NA | NA | NA | NA | NA | NA | NA | NA | NA | NA |
| 25 | WA | NA | NA | NA | NA | WA | WA | WA | WA | NA | NA | NA | NA | NA | NA | NA | NA | WA | NA | NA | NA | WA | NA | NA | NA |
| 26 | WA | NA | NA | NA | NA | NA | NA | WA | WA | WA | NA | NA | NA | NA | NA | NA | WA | WA | WA | NA | NA | NA | NA | NA | NA |
| 27 | WA | NA | NA | NA | NA | WA | WA | WA | WA | NA | NA | NA | NA | NA | WA | WA | WA | NA | NA | NA | NA | NA | NA | NA | NA |
| 28 | WA | NA | NA | NA | NA | WA | WA | WA | WA | WA | NA | NA | NA | NA | NA | NA | NA | WA | WA | NA | NA | WA | NA | NA | WA |
| 29 | WA | NA | WA | NA | NA | NA | - | WA | NA | NA | NA | NA | NA | WA | NA | NA | NA | NA | NA | NA | NA | WA | NA | NA | NA |
| 30 | WA | NA | WA | NA | NA | WA | WA | WA | WA | NA | NA | WA | NA | NA | NA | NA | WA | NA | WA | NA | NA | NA | NA | NA | NA |
| 31 | WA | NA | WA | NA | NA | WA | WA | WA | WA | NA | NA | NA | NA | NA | NA | NA | WA | NA | NA | NA | NA | WA | NA | NA | NA |
| 32 | WA | WA | NA | NA | NA | WA | WA | WA | WA | NA | NA | NA | NA | NA | WA | WA | WA | NA | NA | NA | NA | WA | NA | NA | NA |
| 33 | WA | NA | NA | NA | NA | NA | NA | WA | NA | WA | NA | NA | NA | NA | WA | WA | WA | WA | NA | NA | NA | NA | NA | NA | NA |
| 34 | WA | WA | NA | NA | NA | WA | WA | WA | WA | NA | NA | NA | NA | NA | WA | WA | WA | WA | NA | NA | NA | NA | NA | NA | NA |
| 35 | WA | NA | NA | NA | NA | NA | NA | WA | NA | NA | NA | NA | NA | WA | NA | NA | NA | WA | WA | NA | NA | WA | NA | NA | WA |
| 36 | WA | NA | WA | NA | NA | NA | NA | WA | NA | NA | NA | WA | NA | NA | WA | WA | WA | NA | NA | NA | NA | NA | NA | NA | NA |
| 37 | WA | NA | WA | NA | NA | WA | WA | WA | WA | NA | NA | WA | NA | NA | NA | NA | NA | NA | NA | NA | NA | WA | NA | NA | NA |
| 38 | WA | NA | NA | NA | NA | NA | NA | WA | NA | NA | NA | NA | NA | WA | NA | NA | WA | WA | NA | NA | NA | NA | NA | NA | NA |
| 39 | WA | NA | WA | NA | NA | NA | NA | WA | NA | NA | NA | NA | NA | NA | NA | NA | NA | WA | NA | NA | NA | WA | NA | NA | WA |
| 40 | WA | NA | NA | NA | NA | WA | WA | WA | WA | NA | NA | NA | NA | WA | NA | NA | WA | NA | NA | NA | NA | WA | NA | NA | NA |
| 41 | WA | NA | NA | NA | NA | NA | NA | WA | NA | NA | NA | WA | NA | NA | NA | NA | WA | NA | NA | NA | NA | WA | NA | NA | NA |
| 42 | WA | NA | NA | NA | NA | WA | WA | WA | WA | WA | NA | WA | NA | NA | NA | NA | NA | NA | WA | NA | NA | WA | NA | NA | WA |
| 43 | WA | NA | NA | NA | NA | WA | WA | WA | WA | NA | NA | WA | NA | WA | NA | NA | WA | WA | NA | NA | NA | WA | NA | NA | NA |
| 44 | WA | NA | WA | NA | NA | WA | WA | WA | WA | NA | NA | WA | NA | WA | NA | NA | WA | NA | NA | NA | NA | WA | NA | NA | NA |
| 45 | WA | NA | NA | NA | NA | WA | WA | WA | WA | WA | NA | WA | NA | NA | NA | NA | NA | NA | WA | NA | NA | NA | NA | NA | WA |
| 46 | WA | NA | NA | NA | NA | NA | NA | WA | NA | NA | NA | WA | NA | WA | NA | NA | NA | NA | WA | NA | NA | WA | NA | NA | NA |
| 47 | WA | NA | WA | NA | NA | NA | NA | WA | NA | WA | NA | NA | NA | NA | WA | WA | NA | NA | NA | NA | WA | NA | NA | NA | NA |
| 48 | WA | NA | NA | NA | NA | WA | WA | WA | WA | NA | NA | NA | NA | NA | WA | WA | NA | WA | WA | NA | NA | NA | NA | NA | NA |
| 49 | WA | NA | NA | NA | NA | NA | NA | WA | NA | WA | NA | NA | NA | WA | NA | NA | NA | NA | WA | NA | NA | NA | WA | NA | NA |
| 50 | WA | WA | WA | NA | NA | WA | WA | WA | WA | NA | NA | NA | NA | WA | NA | NA | WA | WA | NA | NA | NA | NA | NA | NA | NA |
| 51 | WA | NA | WA | NA | NA | WA | WA | WA | NA | WA | NA | WA | NA | WA | NA | NA | WA | WA | NA | NA | NA | NA | NA | NA | NA |
| 52 | WA | NA | NA | NA | NA | WA | WA | WA | WA | WA | NA | WA | NA | NA | NA | NA | NA | NA | NA | NA | NA | WA | WA | WA | NA |
| 53 | WA | NA | NA | NA | NA | NA | NA | WA | NA | NA | NA | WA | NA | NA | NA | NA | NA | NA | WA | NA | NA | NA | NA | WA | NA |
| 54 | WA | WA | NA | NA | NA | WA | WA | WA | WA | NA | NA | NA | NA | NA | NA | NA | WA | WA | WA | NA | NA | WA | NA | NA | NA |
| 55 | WA | NA | NA | NA | NA | WA | WA | WA | WA | NA | NA | WA | NA | WA | NA | NA | WA | NA | NA | NA | NA | WA | NA | NA | NA |
| 56 | WA | - | - | NA | NA | - | - | WA | - | - | - | - | - | - | - | - | - | - | - | - | - | - | NA | - | - |
| 57 | WA | WA | NA | NA | NA | WA | WA | WA | WA | NA | NA | NA | NA | NA | NA | NA | NA | WA | NA | NA | NA | NA | NA | NA | NA |
| 58 | WA | NA | NA | NA | NA | WA | WA | WA | WA | NA | NA | WA | NA | WA | NA | NA | WA | NA | WA | NA | WA | WA | NA | NA | NA |
| 59 | WA | WA | NA | NA | NA | NA | NA | WA | NA | NA | NA | NA | WA | WA | WA | NA | NA | WA | WA | NA | NA | NA | NA | NA | NA |
| 60 | WA | WA | NA | NA | NA | NA | NA | WA | NA | NA | NA | NA | WA | NA | NA | NA | WA | WA | WA | NA | WA | NA | NA | NA | NA |
| 61 | WA | NA | WA | NA | NA | WA | WA | WA | WA | NA | NA | NA | NA | WA | NA | NA | WA | WA | NA | NA | NA | WA | NA | NA | WA |
| 62 | WA | WA | WA | NA | NA | WA | WA | WA | WA | NA | NA | NA | WA | WA | NA | NA | WA | WA | NA | NA | NA | WA | NA | NA | NA |
| 63 | WA | NA | WA | NA | NA | NA | NA | WA | NA | NA | NA | WA | NA | WA | NA | NA | WA | NA | WA | NA | NA | NA | WA | WA | NA |
| 64 | WA | NA | NA | NA | NA | NA | NA | WA | WA | NA | NA | WA | NA | NA | WA | WA | NA | NA | NA | NA | NA | NA | WA | WA | NA |
| 65 | WA | NA | WA | NA | NA | WA | WA | WA | WA | WA | NA | WA | NA | NA | NA | NA | NA | WA | NA | NA | NA | NA | NA | NA | NA |
| 66 | WA | NA | NA | NA | NA | WA | WA | WA | WA | WA | NA | NA | NA | NA | NA | NA | WA | NA | NA | NA | NA | WA | NA | NA | WA |
| 67 | WA | NA | NA | NA | NA | WA | WA | WA | WA | NA | NA | WA | NA | WA | WA | WA | NA | WA | NA | NA | WA | NA | NA | NA | NA |
| 68 | WA | NA | NA | NA | NA | WA | WA | WA | WA | WA | NA | NA | WA | NA | WA | WA | WA | NA | WA | NA | NA | WA | NA | NA | NA |
| 69 | WA | WA | NA | NA | NA | WA | WA | WA | WA | WA | NA | WA | NA | WA | WA | WA | NA | NA | NA | NA | NA | WA | NA | NA | NA |
| 70 | WA | NA | NA | NA | NA | WA | WA | WA | WA | WA | NA | NA | NA | NA | WA | NA | NA | WA | NA | NA | NA | NA | NA | NA | NA |
| 71 | - | - | - | - | NA | - | - | WA | - | - | - | - | - | - | NA | NA | - | - | - | - | - | - | NA | - | - |
| 72 | WA | NA | WA | NA | NA | WA | WA | WA | WA | NA | NA | WA | NA | NA | NA | NA | NA | NA | NA | NA | NA | NA | NA | NA | NA |
| 73 | WA | NA | WA | NA | NA | NA | NA | WA | NA | NA | NA | NA | NA | WA | NA | NA | NA | NA | NA | NA | NA | NA | WA | WA | NA |
| 74 | WA | NA | NA | NA | NA | WA | NA | WA | NA | NA | NA | NA | NA | WA | NA | NA | WA | NA | NA | NA | NA | WA | NA | NA | WA |
| 75 | WA | NA | NA | NA | NA | WA | WA | WA | WA | NA | NA | NA | NA | NA | NA | NA | WA | WA | NA | NA | NA | NA | NA | NA | NA |
| 76 | WA | NA | WA | NA | NA | NA | NA | WA | NA | NA | NA | NA | NA | WA | WA | WA | NA | NA | NA | NA | NA | WA | NA | NA | WA |
| 77 | WA | NA | WA | NA | NA | NA | NA | WA | NA | WA | NA | WA | WA | NA | WA | WA | NA | NA | NA | NA | NA | NA | NA | NA | NA |
| 78 | WA | WA | WA | NA | NA | WA | WA | WA | WA | WA | NA | NA | NA | NA | NA | NA | WA | WA | NA | NA | NA | NA | NA | NA | NA |
| 79 | WA | NA | NA | NA | NA | WA | WA | WA | WA | WA | NA | NA | NA | WA | NA | NA | NA | NA | NA | NA | NA | WA | NA | NA | NA |
| 80 | WA | NA | NA | NA | NA | NA | WA | WA | NA | NA | NA | NA | WA | NA | NA | NA | NA | NA | WA | NA | NA | WA | NA | WA | NA |
| 81 | WA | WA | NA | NA | NA | NA | NA | WA | NA | NA | NA | NA | NA | NA | NA | NA | NA | NA | NA | NA | NA | WA | NA | NA | NA |
| 82 | WA | NA | NA | NA | NA | WA | WA | WA | WA | NA | NA | WA | NA | WA | WA | WA | WA | NA | WA | NA | NA | WA | NA | NA | NA |
| 83 | WA | NA | NA | NA | NA | WA | WA | WA | WA | NA | NA | WA | NA | NA | NA | NA | NA | WA | NA | NA | NA | NA | NA | NA | WA |
| 84 | WA | NA | NA | NA | NA | WA | WA | WA | WA | WA | NA | NA | NA | WA | NA | NA | NA | NA | NA | NA | NA | WA | NA | NA | NA |
| 85 | WA | NA | NA | NA | NA | WA | WA | WA | WA | NA | NA | NA | NA | NA | WA | WA | NA | WA | NA | NA | NA | NA | NA | NA | NA |
| 86 | WA | NA | NA | NA | NA | WA | NA | WA | NA | NA | NA | NA | NA | WA | NA | NA | WA | NA | NA | NA | NA | WA | NA | NA | WA |
| 87 | WA | NA | NA | NA | NA | NA | NA | WA | NA | WA | NA | WA | NA | NA | NA | NA | NA | NA | NA | NA | NA | WA | NA | NA | NA |
| 88 | WA | NA | NA | NA | NA | NA | NA | WA | NA | NA | NA | WA | NA | NA | NA | NA | NA | WA | WA | NA | NA | WA | NA | NA | NA |
| 89 | WA | WA | NA | NA | NA | NA | NA | WA | WA | NA | NA | NA | NA | NA | WA | WA | WA | NA | NA | NA | NA | NA | NA | NA | NA |
| 90 | WA | NA | WA | NA | NA | NA | NA | WA | NA | NA | NA | NA | NA | WA | NA | WA | NA | NA | WA | NA | NA | WA | NA | NA | NA |
| 91 | WA | NA | NA | NA | NA | WA | WA | WA | WA | NA | NA | WA | NA | NA | NA | NA | WA | WA | NA | NA | NA | WA | NA | NA | NA |
| 92 | WA | NA | NA | NA | NA | WA | WA | WA | NA | NA | NA | NA | NA | NA | WA | WA | WA | WA | NA | NA | NA | WA | NA | NA | NA |
| 93 | WA | WA | NA | NA | NA | NA | NA | WA | NA | NA | NA | NA | NA | NA | NA | NA | WA | NA | NA | NA | NA | WA | NA | NA | NA |
| 94 | WA | NA | NA | NA | NA | WA | WA | WA | WA | NA | NA | NA | NA | NA | NA | NA | WA | NA | NA | NA | NA | WA | NA | NA | WA |
| 95 | WA | WA | NA | NA | NA | NA | NA | WA | WA | NA | NA | WA | NA | NA | NA | NA | NA | NA | NA | NA | NA | NA | NA | NA | NA |
| 96 | WA | - | - | NA | NA | - | - | WA | - | NA | NA | - | NA | - | - | - | - | - | - | NA | NA | NA | NA | NA | NA |
| 97 | WA | NA | NA | NA | NA | WA | WA | WA | WA | NA | NA | WA | WA | WA | NA | NA | NA | NA | NA | NA | NA | WA | NA | NA | WA |
| 98 | WA | NA | NA | NA | NA | WA | NA | WA | NA | WA | NA | NA | NA | WA | NA | NA | WA | WA | NA | NA | NA | NA | NA | NA | NA |
| 99 | WA | NA | NA | NA | NA | WA | WA | WA | NA | WA | NA | NA | NA | NA | WA | WA | WA | WA | NA | NA | NA | WA | WA | WA | NA |
| 100 | WA | WA | NA | NA | NA | WA | WA | WA | WA | NA | NA | NA | NA | NA | NA | NA | NA | NA | NA | NA | NA | WA | NA | NA | NA |
| 101 | WA | WA | NA | NA | NA | WA | WA | WA | WA | NA | NA | NA | NA | NA | NA | NA | NA | WA | NA | NA | NA | NA | NA | NA | NA |
| 102 | WA | NA | NA | NA | NA | WA | WA | WA | WA | NA | NA | NA | NA | WA | WA | WA | WA | WA | NA | NA | NA | NA | NA | NA | WA |
| 103 | WA | NA | NA | NA | NA | WA | WA | WA | WA | WA | NA | NA | NA | WA | WA | WA | WA | NA | WA | NA | NA | WA | NA | NA | NA |
| 104 | WA | WA | NA | NA | NA | WA | WA | WA | WA | WA | NA | NA | WA | NA | WA | WA | NA | NA | NA | NA | NA | WA | NA | NA | NA |
| 105 | WA | WA | WA | NA | NA | WA | WA | WA | WA | NA | NA | WA | NA | NA | NA | NA | WA | NA | WA | NA | NA | NA | NA | NA | WA |
| 106 | WA | NA | NA | NA | NA | NA | NA | WA | NA | NA | NA | WA | NA | NA | NA | NA | WA | NA | WA | NA | NA | NA | NA | NA | NA |
| 107 | WA | NA | NA | NA | NA | NA | NA | WA | NA | NA | NA | NA | NA | WA | WA | WA | NA | WA | NA | NA | NA | NA | NA | NA | WA |
| 108 | WA | NA | WA | NA | NA | NA | NA | WA | NA | NA | NA | WA | NA | NA | NA | NA | NA | WA | NA | NA | NA | WA | NA | WA | WA |
| 109 | WA | NA | NA | NA | NA | NA | NA | WA | NA | NA | NA | NA | NA | NA | WA | WA | WA | NA | NA | NA | NA | WA | NA | NA | NA |
| 110 | WA | NA | NA | NA | NA | WA | WA | WA | WA | NA | NA | NA | NA | WA | NA | NA | NA | NA | WA | NA | NA | NA | NA | NA | NA |
| 111 | WA | NA | WA | NA | NA | WA | WA | WA | WA | NA | NA | NA | NA | NA | NA | NA | WA | WA | NA | NA | NA | WA | NA | NA | NA |
| 112 | WA | NA | NA | NA | NA | NA | NA | WA | NA | NA | NA | WA | NA | NA | NA | NA | NA | NA | NA | NA | NA | WA | NA | NA | NA |
| 113 | WA | NA | NA | NA | NA | NA | NA | WA | WA | NA | NA | NA | NA | NA | WA | WA | WA | WA | NA | NA | WA | NA | NA | NA | NA |
| 114 | WA | NA | WA | NA | NA | NA | NA | WA | NA | WA | NA | NA | NA | WA | WA | WA | WA | NA | WA | NA | NA | NA | NA | NA | NA |
| 115 | WA | WA | NA | NA | NA | NA | NA | WA | NA | WA | NA | NA | NA | NA | NA | NA | NA | WA | WA | NA | NA | WA | NA | NA | NA |
| 116 | WA | NA | NA | NA | NA | NA | NA | WA | NA | NA | NA | NA | NA | WA | WA | NA | NA | WA | NA | NA | NA | NA | WA | WA | NA |
| 117 | WA | NA | WA | NA | NA | NA | NA | WA | NA | WA | NA | NA | NA | NA | NA | NA | WA | NA | NA | NA | NA | NA | NA | NA | NA |
| 118 | WA | WA | NA | NA | NA | WA | WA | WA | WA | WA | NA | NA | NA | NA | NA | NA | WA | WA | NA | NA | NA | WA | NA | NA | NA |
| 119 | WA | WA | NA | NA | NA | NA | NA | WA | NA | WA | NA | NA | NA | WA | NA | NA | WA | WA | NA | NA | NA | WA | NA | NA | WA |
| 120 | WA | NA | NA | NA | NA | WA | WA | WA | NA | NA | NA | WA | NA | NA | NA | NA | WA | WA | NA | NA | NA | NA | NA | NA | WA |
| 121 | WA | NA | NA | NA | NA | NA | NA | WA | NA | NA | NA | WA | NA | WA | NA | NA | NA | NA | NA | NA | NA | WA | NA | NA | NA |
| 122 | WA | NA | NA | NA | NA | NA | NA | WA | NA | WA | NA | WA | NA | WA | WA | WA | NA | WA | NA | NA | NA | NA | NA | NA | NA |
| 123 | WA | NA | NA | NA | NA | NA | NA | WA | NA | NA | NA | NA | NA | NA | WA | WA | WA | NA | NA | NA | NA | WA | NA | NA | NA |
| 124 | WA | WA | NA | NA | NA | NA | NA | WA | NA | WA | NA | WA | NA | WA | WA | WA | NA | WA | WA | NA | NA | WA | NA | NA | NA |
| 125 | WA | WA | WA | NA | NA | WA | WA | WA | WA | WA | NA | NA | NA | NA | WA | WA | WA | NA | WA | NA | NA | NA | NA | NA | NA |
| 126 | WA | NA | NA | NA | NA | NA | NA | WA | NA | NA | NA | WA | NA | NA | WA | WA | NA | NA | WA | NA | NA | NA | WA | WA | NA |
| 127 | WA | NA | NA | NA | NA | WA | WA | WA | WA | WA | NA | WA | NA | NA | NA | NA | NA | NA | NA | NA | NA | WA | NA | NA | NA |
| 128 | WA | NA | WA | NA | NA | WA | WA | WA | NA | NA | NA | NA | WA | NA | NA | NA | NA | NA | WA | NA | WA | WA | NA | NA | WA |
| 129 | WA | WA | WA | NA | NA | WA | WA | WA | WA | NA | NA | NA | NA | NA | NA | NA | WA | NA | NA | NA | NA | NA | WA | WA | NA |
| 130 | WA | NA | NA | NA | NA | WA | WA | WA | WA | WA | NA | WA | NA | NA | NA | NA | WA | WA | NA | NA | NA | WA | NA | NA | NA |
| 131 | NA | WA | WA | NA | NA | WA | WA | WA | WA | NA | NA | WA | NA | WA | NA | NA | NA | WA | NA | NA | NA | WA | WA | WA | NA |
| 132 | WA | WA | NA | NA | NA | WA | WA | WA | WA | WA | NA | NA | NA | NA | WA | NA | WA | WA | NA | NA | NA | WA | NA | NA | NA |
| 133 | WA | NA | WA | NA | NA | NA | NA | WA | NA | WA | NA | NA | NA | NA | NA | NA | NA | WA | NA | NA | NA | NA | NA | NA | NA |
| 134 | WA | NA | NA | NA | NA | NA | NA | WA | WA | NA | NA | NA | NA | NA | WA | WA | NA | WA | NA | NA | NA | WA | NA | NA | NA |
| 135 | WA | NA | WA | NA | NA | WA | WA | WA | WA | NA | NA | NA | NA | NA | NA | NA | NA | NA | WA | NA | NA | WA | NA | NA | NA |
| 136 | WA | NA | WA | NA | NA | WA | WA | WA | WA | NA | NA | WA | NA | WA | NA | NA | WA | WA | NA | NA | NA | WA | NA | NA | NA |
| 137 | WA | NA | NA | NA | NA | WA | WA | WA | WA | NA | NA | NA | NA | NA | NA | NA | NA | WA | NA | NA | NA | NA | WA | NA | NA |
| 138 | WA | WA | WA | NA | NA | NA | NA | WA | NA | NA | NA | WA | WA | NA | NA | NA | WA | NA | NA | NA | NA | NA | NA | NA | WA |
| 139 | WA | WA | WA | NA | NA | NA | NA | WA | NA | NA | NA | WA | NA | NA | NA | NA | WA | NA | NA | NA | NA | WA | NA | NA | NA |
| 140 | WA | NA | WA | NA | NA | NA | NA | WA | NA | WA | NA | WA | NA | NA | NA | NA | NA | NA | WA | NA | NA | NA | NA | NA | NA |
| 141 | WA | NA | NA | NA | NA | NA | NA | WA | NA | WA | NA | WA | NA | NA | WA | WA | NA | NA | NA | NA | NA | NA | WA | WA | NA |
| 142 | WA | NA | WA | NA | NA | WA | WA | WA | WA | WA | NA | NA | NA | NA | NA | NA | NA | WA | NA | NA | NA | NA | NA | NA | NA |
| 143 | WA | NA | NA | NA | NA | NA | NA | WA | NA | NA | NA | WA | WA | NA | NA | NA | WA | NA | NA | NA | NA | WA | NA | NA | WA |
| 144 | WA | NA | NA | NA | NA | WA | WA | WA | WA | NA | NA | NA | NA | NA | NA | NA | NA | NA | NA | NA | NA | WA | NA | NA | NA |
| 145 | WA | NA | WA | NA | NA | WA | WA | WA | WA | WA | NA | WA | NA | NA | NA | NA | WA | WA | NA | NA | NA | WA | NA | NA | NA |
| 146 | WA | NA | NA | NA | NA | NA | NA | WA | WA | WA | NA | WA | NA | NA | NA | NA | NA | WA | WA | NA | NA | WA | NA | NA | WA |
| 147 | WA | WA | WA | NA | NA | WA | WA | WA | WA | WA | NA | WA | NA | WA | WA | WA | NA | NA | WA | NA | NA | NA | NA | NA | NA |
| 148 | WA | WA | NA | NA | NA | WA | WA | WA | WA | NA | NA | WA | NA | NA | NA | NA | NA | WA | NA | NA | NA | NA | NA | NA | NA |
| 149 | WA | NA | NA | NA | NA | NA | WA | WA | WA | WA | NA | NA | NA | NA | NA | NA | WA | WA | WA | NA | NA | NA | NA | NA | NA |
| 150 | WA | NA | NA | NA | NA | WA | WA | WA | WA | WA | NA | NA | NA | NA | NA | NA | WA | NA | NA | NA | NA | NA | NA | NA | NA |
| 151 | WA | NA | NA | NA | NA | WA | WA | WA | WA | NA | NA | NA | NA | WA | WA | WA | WA | WA | WA | NA | NA | WA | NA | NA | NA |
| 152 | WA | NA | WA | NA | NA | WA | WA | WA | WA | NA | NA | WA | NA | WA | WA | WA | WA | NA | WA | NA | NA | NA | NA | NA | NA |
| 153 | WA | NA | NA | NA | NA | NA | NA | WA | NA | NA | NA | NA | NA | NA | NA | NA | WA | NA | NA | NA | NA | NA | NA | NA | WA |
| 154 | WA | NA | NA | NA | NA | WA | WA | WA | WA | NA | NA | NA | NA | WA | NA | NA | NA | WA | NA | NA | NA | WA | NA | NA | NA |
| 155 | WA | NA | NA | NA | NA | WA | WA | WA | WA | NA | NA | WA | NA | WA | WA | NA | WA | NA | NA | NA | NA | NA | NA | NA | NA |
| 156 | WA | NA | WA | NA | NA | NA | NA | WA | WA | NA | NA | WA | NA | NA | WA | WA | NA | WA | NA | NA | NA | NA | WA | WA | NA |
| 157 | WA | NA | NA | NA | NA | WA | WA | WA | WA | NA | NA | WA | NA | WA | WA | WA | WA | WA | NA | NA | NA | NA | NA | NA | NA |
| 158 | WA | NA | NA | NA | NA | WA | WA | WA | NA | WA | NA | NA | WA | NA | NA | NA | NA | WA | NA | NA | NA | NA | NA | NA | WA |
| 159 | WA | NA | NA | NA | NA | WA | WA | WA | WA | NA | NA | NA | NA | WA | NA | NA | NA | NA | WA | NA | NA | WA | NA | NA | NA |
| 160 | WA | NA | NA | NA | NA | WA | WA | WA | WA | WA | NA | WA | NA | NA | NA | NA | WA | NA | WA | NA | NA | WA | NA | NA | WA |
| 161 | WA | NA | WA | NA | NA | NA | NA | WA | NA | NA | NA | WA | NA | WA | NA | NA | NA | WA | NA | NA | NA | NA | NA | NA | NA |
| 162 | WA | NA | WA | NA | NA | WA | WA | WA | WA | WA | NA | WA | NA | NA | NA | NA | WA | NA | NA | NA | NA | WA | NA | NA | NA |
| 163 | WA | WA | NA | NA | NA | NA | NA | WA | NA | NA | NA | WA | NA | NA | NA | NA | NA | NA | NA | NA | NA | WA | NA | NA | NA |
| 164 | WA | NA | NA | NA | NA | WA | WA | WA | WA | WA | NA | NA | WA | WA | WA | WA | WA | NA | NA | NA | NA | NA | NA | NA | NA |
| 165 | WA | WA | NA | NA | NA | WA | WA | WA | NA | NA | NA | NA | NA | NA | WA | WA | WA | WA | WA | NA | NA | NA | NA | NA | NA |
| 166 | WA | NA | NA | NA | NA | NA | NA | WA | NA | NA | NA | NA | NA | WA | NA | NA | WA | NA | NA | NA | NA | NA | NA | NA | NA |
| 167 | WA | NA | WA | NA | NA | WA | WA | WA | WA | WA | NA | NA | NA | NA | WA | WA | NA | NA | WA | NA | NA | WA | NA | NA | NA |
| 168 | WA | NA | WA | NA | NA | WA | WA | WA | WA | WA | NA | NA | NA | NA | NA | NA | WA | NA | NA | NA | NA | NA | NA | NA | WA |
| 169 | WA | - | - | NA | NA | - | - | WA | - | - | NA | - | - | - | - | - | - | - | NA | - | - | - | NA | NA | NA |
| 170 | WA | NA | WA | NA | NA | WA | WA | WA | WA | WA | NA | NA | NA | NA | WA | WA | NA | NA | NA | NA | NA | WA | NA | NA | NA |
| 171 | WA | NA | NA | NA | NA | NA | NA | WA | NA | NA | NA | NA | NA | NA | NA | NA | WA | NA | NA | NA | NA | WA | NA | NA | NA |
| 172 | WA | WA | WA | NA | NA | WA | WA | WA | WA | NA | NA | NA | NA | NA | NA | NA | NA | WA | NA | NA | NA | NA | NA | NA | NA |
| 173 | WA | NA | WA | NA | NA | NA | NA | WA | NA | NA | NA | WA | NA | NA | WA | WA | WA | NA | NA | NA | NA | NA | NA | NA | NA |
| 174 | WA | NA | NA | NA | NA | WA | WA | WA | WA | WA | NA | NA | NA | NA | NA | NA | WA | NA | NA | NA | NA | NA | NA | NA | NA |
| 175 | WA | WA | WA | NA | NA | WA | WA | WA | NA | NA | NA | WA | NA | NA | WA | WA | NA | NA | WA | NA | NA | WA | NA | NA | WA |
| 176 | WA | NA | WA | NA | NA | WA | WA | WA | NA | NA | NA | WA | NA | NA | NA | NA | NA | NA | NA | NA | NA | NA | NA | NA | NA |
| 177 | WA | NA | NA | NA | NA | NA | NA | WA | NA | WA | NA | WA | NA | NA | WA | WA | WA | WA | NA | NA | NA | NA | NA | NA | NA |
| 178 | WA | WA | NA | NA | NA | NA | NA | WA | WA | WA | NA | WA | NA | NA | NA | NA | NA | NA | NA | NA | NA | WA | NA | NA | WA |
| 179 | WA | NA | NA | NA | NA | NA | NA | WA | NA | WA | NA | WA | NA | WA | NA | NA | NA | NA | WA | NA | NA | NA | NA | NA | NA |
| 180 | WA | NA | WA | NA | NA | WA | WA | WA | WA | NA | NA | WA | NA | NA | NA | NA | NA | NA | NA | NA | NA | NA | NA | NA | NA |
| 181 | WA | NA | NA | NA | NA | WA | WA | WA | WA | NA | NA | NA | NA | NA | WA | WA | NA | NA | WA | NA | NA | NA | NA | NA | WA |
| 182 | WA | NA | NA | NA | NA | WA | WA | WA | WA | NA | NA | NA | NA | NA | NA | NA | NA | WA | NA | NA | NA | WA | NA | NA | NA |
| 183 | WA | NA | NA | NA | NA | NA | NA | WA | NA | WA | NA | NA | NA | NA | NA | NA | WA | NA | WA | NA | NA | WA | NA | NA | NA |
| 184 | WA | NA | NA | NA | NA | NA | NA | WA | NA | NA | NA | NA | WA | WA | NA | NA | NA | NA | WA | NA | NA | NA | NA | NA | NA |
| 185 | WA | WA | NA | NA | NA | NA | NA | WA | NA | NA | NA | NA | NA | NA | WA | WA | NA | NA | NA | NA | NA | NA | NA | NA | WA |
| 186 | WA | NA | WA | NA | NA | WA | WA | WA | WA | NA | NA | NA | NA | NA | NA | NA | NA | WA | NA | NA | NA | WA | NA | NA | WA |
| 187 | WA | NA | WA | NA | NA | WA | WA | WA | WA | NA | NA | WA | NA | NA | WA | WA | NA | NA | WA | NA | NA | NA | WA | WA | NA |
| 188 | WA | NA | WA | NA | NA | NA | NA | WA | NA | NA | NA | WA | NA | NA | NA | NA | NA | NA | NA | NA | NA | NA | NA | NA | WA |
| 189 | WA | WA | WA | NA | NA | WA | WA | WA | WA | NA | NA | WA | NA | WA | NA | NA | WA | NA | NA | NA | NA | NA | NA | NA | NA |
| 190 | WA | NA | NA | NA | NA | WA | WA | WA | WA | WA | NA | NA | NA | NA | NA | NA | WA | WA | NA | NA | WA | NA | NA | NA | NA |
| 191 | WA | NA | WA | NA | NA | NA | NA | WA | NA | NA | NA | WA | NA | WA | NA | NA | WA | NA | WA | NA | NA | NA | WA | WA | NA |
| 192 | WA | NA | NA | NA | NA | NA | NA | WA | NA | NA | NA | WA | NA | NA | NA | NA | WA | WA | NA | NA | WA | NA | NA | NA | NA |
| 193 | WA | WA | WA | NA | NA | WA | WA | WA | WA | NA | NA | WA | NA | WA | NA | NA | NA | NA | NA | NA | WA | NA | NA | NA | NA |
| 194 | WA | NA | NA | NA | NA | NA | NA | WA | NA | NA | NA | NA | NA | NA | NA | NA | NA | NA | NA | NA | NA | NA | NA | NA | WA |
| 195 | WA | WA | WA | NA | NA | WA | WA | WA | WA | NA | NA | NA | NA | NA | NA | NA | WA | WA | WA | NA | NA | NA | NA | NA | WA |
| 196 | WA | NA | NA | NA | NA | WA | WA | WA | WA | NA | NA | WA | NA | NA | NA | NA | WA | WA | WA | NA | WA | WA | NA | NA | NA |
| 197 | WA | WA | NA | NA | NA | WA | WA | WA | WA | NA | NA | NA | NA | WA | WA | WA | NA | NA | WA | NA | NA | WA | NA | NA | NA |
| 198 | WA | NA | WA | NA | NA | WA | WA | WA | WA | NA | NA | WA | WA | WA | NA | NA | NA | WA | NA | NA | NA | NA | NA | NA | NA |
| 199 | WA | WA | WA | NA | NA | WA | WA | WA | WA | NA | NA | WA | NA | NA | WA | WA | WA | NA | WA | NA | NA | WA | NA | NA | NA |
| 200 | WA | NA | NA | NA | NA | WA | WA | WA | WA | NA | NA | NA | NA | NA | NA | NA | WA | NA | NA | NA | NA | NA | WA | WA | NA |
| 201 | WA | NA | WA | NA | NA | WA | WA | WA | WA | NA | NA | WA | WA | NA | NA | NA | WA | NA | WA | NA | NA | WA | NA | NA | NA |
| 202 | WA | WA | NA | NA | NA | NA | NA | WA | NA | NA | NA | NA | NA | NA | NA | NA | NA | WA | NA | NA | NA | NA | NA | NA | NA |
| 203 | WA | NA | NA | NA | NA | NA | NA | WA | NA | NA | NA | WA | NA | NA | NA | NA | WA | NA | NA | NA | WA | WA | NA | NA | NA |
| 204 | WA | NA | NA | NA | NA | WA | WA | WA | WA | NA | NA | NA | NA | NA | NA | NA | WA | WA | NA | NA | NA | WA | NA | NA | NA |
| 205 | WA | NA | NA | NA | NA | WA | WA | WA | WA | NA | NA | NA | NA | NA | NA | NA | NA | WA | NA | NA | NA | NA | NA | NA | WA |
| 206 | WA | NA | NA | NA | NA | NA | NA | WA | NA | NA | NA | NA | NA | NA | NA | NA | WA | WA | NA | NA | NA | WA | NA | NA | NA |
| 207 | WA | NA | WA | NA | NA | NA | NA | WA | WA | NA | NA | WA | NA | NA | NA | NA | NA | NA | NA | NA | NA | WA | NA | NA | WA |
| 208 | WA | NA | NA | NA | NA | NA | NA | WA | NA | NA | NA | NA | NA | NA | NA | NA | WA | WA | NA | NA | NA | NA | NA | NA | NA |
| 209 | WA | NA | NA | NA | NA | WA | WA | WA | WA | NA | NA | WA | NA | NA | WA | WA | NA | WA | NA | NA | NA | NA | NA | NA | NA |
| 210 | WA | WA | NA | WA | NA | WA | WA | WA | WA | NA | NA | WA | NA | NA | NA | NA | WA | NA | NA | NA | NA | WA | NA | NA | NA |
| 211 | WA | WA | WA | NA | NA | WA | WA | WA | WA | WA | NA | NA | NA | WA | NA | NA | NA | WA | NA | NA | NA | NA | NA | NA | NA |
| 212 | WA | WA | NA | NA | NA | WA | WA | WA | WA | WA | NA | WA | NA | NA | NA | NA | WA | WA | WA | NA | NA | NA | NA | NA | NA |
| 213 | WA | NA | WA | NA | NA | NA | NA | WA | WA | NA | NA | NA | NA | NA | WA | WA | NA | NA | NA | NA | NA | WA | NA | NA | NA |
| 214 | WA | WA | NA | NA | NA | WA | WA | WA | WA | NA | NA | NA | WA | NA | WA | WA | NA | NA | NA | NA | NA | WA | NA | NA | NA |
| 215 | WA | WA | NA | NA | NA | WA | WA | WA | WA | WA | NA | NA | NA | NA | NA | NA | NA | NA | NA | NA | NA | NA | NA | NA | NA |
| 216 | WA | WA | NA | NA | NA | WA | WA | WA | NA | WA | NA | NA | NA | NA | WA | WA | WA | NA | NA | NA | NA | NA | NA | NA | WA |
| 217 | WA | NA | NA | NA | NA | NA | NA | WA | WA | NA | NA | WA | NA | NA | WA | WA | WA | WA | NA | NA | NA | WA | NA | NA | NA |
| 218 | WA | NA | NA | NA | NA | WA | WA | WA | WA | NA | NA | WA | NA | NA | NA | NA | NA | WA | WA | NA | NA | NA | NA | NA | WA |
| 219 | WA | NA | WA | NA | NA | WA | WA | WA | WA | WA | NA | NA | NA | NA | NA | NA | WA | NA | NA | NA | NA | NA | NA | NA | NA |
| 220 | WA | NA | NA | NA | NA | NA | NA | WA | NA | NA | NA | NA | NA | WA | NA | NA | WA | NA | WA | NA | NA | NA | NA | NA | NA |
| 221 | WA | NA | NA | NA | NA | WA | WA | WA | WA | NA | NA | NA | WA | WA | NA | NA | NA | NA | NA | NA | NA | WA | NA | NA | NA |
| 222 | WA | WA | NA | NA | NA | WA | WA | WA | NA | WA | NA | WA | WA | WA | WA | WA | NA | NA | NA | NA | NA | NA | NA | NA | NA |
| 223 | WA | WA | NA | NA | NA | WA | WA | WA | WA | NA | NA | WA | NA | WA | NA | NA | NA | WA | NA | NA | NA | WA | WA | WA | NA |
| 224 | WA | NA | NA | NA | NA | WA | WA | WA | WA | NA | NA | WA | NA | WA | WA | WA | NA | NA | WA | NA | NA | WA | NA | NA | NA |
| 225 | WA | NA | NA | NA | NA | WA | WA | WA | WA | NA | NA | WA | NA | WA | WA | WA | WA | WA | NA | NA | NA | WA | NA | NA | NA |
| 226 | WA | NA | NA | NA | NA | NA | NA | WA | NA | NA | NA | WA | NA | NA | WA | WA | WA | NA | NA | NA | NA | NA | NA | NA | NA |
| 227 | WA | NA | NA | NA | NA | WA | WA | WA | WA | NA | NA | NA | NA | NA | NA | NA | NA | WA | NA | NA | NA | NA | NA | NA | WA |
| 228 | WA | WA | NA | NA | NA | WA | WA | WA | WA | WA | NA | WA | NA | NA | NA | NA | WA | WA | WA | NA | NA | WA | NA | NA | NA |
| 229 | WA | NA | NA | NA | NA | WA | WA | WA | WA | NA | NA | WA | NA | NA | NA | NA | NA | WA | NA | NA | NA | WA | NA | NA | NA |
| 230 | WA | NA | NA | NA | NA | NA | NA | WA | NA | WA | NA | WA | NA | WA | WA | WA | NA | WA | WA | NA | NA | NA | NA | NA | NA |
| 231 | WA | NA | WA | NA | NA | NA | NA | WA | NA | WA | NA | NA | NA | WA | NA | NA | WA | WA | NA | NA | NA | WA | NA | NA | NA |
| 232 | WA | NA | NA | NA | NA | NA | NA | WA | NA | NA | NA | NA | NA | NA | NA | NA | NA | WA | NA | NA | NA | NA | NA | NA | NA |
| 233 | WA | NA | WA | NA | NA | NA | NA | WA | NA | WA | NA | WA | NA | NA | NA | NA | NA | NA | NA | NA | NA | WA | NA | NA | NA |
| 234 | WA | NA | WA | NA | NA | WA | WA | WA | WA | NA | NA | WA | NA | NA | NA | NA | NA | NA | NA | NA | NA | NA | NA | NA | WA |
| 235 | WA | NA | NA | NA | NA | WA | WA | WA | WA | NA | NA | WA | NA | WA | NA | NA | WA | NA | NA | NA | NA | NA | NA | NA | NA |
| 236 | WA | NA | NA | NA | NA | NA | NA | WA | NA | NA | NA | WA | NA | WA | WA | WA | NA | NA | WA | NA | NA | WA | NA | NA | NA |
| 237 | WA | NA | NA | NA | NA | NA | NA | WA | NA | NA | NA | NA | WA | NA | WA | WA | NA | NA | NA | NA | NA | WA | NA | NA | NA |
| 238 | WA | NA | NA | NA | NA | NA | NA | WA | NA | NA | NA | WA | NA | WA | WA | WA | NA | WA | WA | NA | NA | NA | NA | NA | NA |
| 239 | WA | NA | WA | NA | NA | NA | NA | WA | NA | WA | NA | WA | NA | NA | NA | NA | WA | WA | NA | NA | NA | NA | NA | NA | WA |
| 240 | WA | NA | NA | NA | NA | WA | WA | WA | WA | WA | NA | NA | NA | WA | NA | NA | WA | WA | NA | NA | WA | WA | NA | NA | NA |
| 241 | WA | WA | NA | NA | NA | WA | WA | WA | WA | NA | NA | NA | NA | WA | NA | NA | NA | WA | NA | NA | NA | NA | NA | NA | NA |
| 242 | WA | NA | NA | NA | NA | WA | WA | WA | NA | WA | NA | WA | NA | WA | NA | NA | WA | WA | NA | NA | NA | NA | NA | NA | NA |
| 243 | WA | NA | NA | NA | NA | NA | NA | WA | NA | NA | NA | WA | NA | WA | WA | WA | NA | NA | NA | NA | NA | WA | NA | NA | NA |
| 244 | WA | NA | WA | NA | NA | WA | WA | WA | WA | WA | NA | WA | NA | NA | NA | NA | WA | WA | NA | NA | NA | WA | NA | NA | NA |
| 245 | WA | NA | WA | NA | NA | WA | NA | WA | WA | WA | NA | NA | WA | NA | NA | NA | NA | NA | NA | NA | NA | WA | NA | NA | NA |
| 246 | WA | NA | NA | NA | NA | NA | NA | WA | NA | NA | NA | NA | NA | NA | NA | NA | WA | NA | NA | NA | NA | NA | NA | NA | NA |
| 247 | WA | NA | NA | NA | NA | WA | WA | WA | WA | WA | NA | NA | NA | WA | WA | WA | NA | NA | NA | NA | NA | WA | NA | NA | NA |
| 248 | WA | NA | NA | NA | NA | NA | NA | WA | NA | NA | NA | WA | NA | WA | NA | NA | NA | WA | NA | NA | NA | WA | NA | NA | NA |
| 249 | WA | WA | NA | NA | NA | WA | WA | WA | WA | NA | NA | NA | NA | NA | NA | NA | WA | WA | NA | NA | NA | NA | NA | NA | WA |
| 250 | WA | NA | NA | NA | NA | WA | WA | WA | WA | NA | NA | WA | NA | NA | NA | NA | NA | WA | WA | NA | NA | NA | NA | NA | NA |
| 251 | WA | NA | NA | NA | NA | NA | NA | WA | NA | WA | NA | NA | NA | NA | NA | NA | WA | WA | WA | NA | WA | NA | WA | WA | NA |
| 252 | WA | NA | NA | NA | NA | NA | NA | WA | NA | WA | NA | NA | NA | NA | NA | NA | WA | WA | NA | NA | NA | NA | NA | NA | WA |
| 253 | WA | NA | NA | NA | NA | NA | NA | WA | NA | NA | NA | WA | NA | NA | WA | WA | NA | NA | WA | NA | NA | WA | NA | NA | NA |
| 254 | WA | NA | NA | NA | NA | NA | NA | WA | NA | NA | NA | NA | NA | WA | NA | NA | NA | WA | WA | NA | NA | WA | NA | NA | NA |
| 255 | WA | NA | WA | NA | NA | WA | WA | WA | WA | WA | NA | NA | NA | NA | NA | NA | WA | WA | WA | NA | NA | NA | NA | NA | NA |
| 256 | WA | WA | WA | NA | NA | WA | WA | WA | WA | NA | NA | NA | NA | NA | NA | NA | NA | WA | WA | NA | NA | NA | NA | NA | NA |
| 257 | WA | WA | NA | NA | NA | NA | NA | WA | NA | WA | NA | NA | NA | NA | NA | NA | WA | WA | NA | NA | NA | NA | NA | NA | NA |
| 258 | WA | NA | NA | NA | NA | NA | NA | WA | NA | NA | NA | NA | NA | WA | WA | WA | NA | WA | NA | NA | NA | NA | NA | NA | WA |
| 259 | WA | NA | WA | NA | NA | WA | WA | WA | WA | NA | NA | NA | NA | WA | WA | WA | NA | WA | WA | NA | NA | NA | NA | NA | NA |
| 260 | WA | WA | WA | NA | NA | WA | WA | WA | WA | WA | NA | NA | NA | NA | NA | NA | NA | WA | NA | NA | NA | WA | NA | NA | NA |
| 261 | WA | NA | WA | NA | NA | WA | WA | WA | WA | WA | NA | WA | NA | NA | NA | NA | NA | WA | NA | NA | NA | NA | NA | NA | NA |
| 262 | WA | NA | WA | NA | NA | NA | NA | WA | NA | WA | NA | NA | NA | WA | WA | WA | NA | NA | NA | NA | NA | WA | WA | WA | NA |
| 263 | WA | NA | WA | NA | NA | NA | NA | WA | WA | NA | NA | NA | NA | NA | NA | NA | WA | WA | NA | NA | NA | NA | NA | NA | NA |
| 264 | WA | NA | NA | NA | NA | WA | WA | WA | WA | NA | NA | WA | NA | NA | NA | NA | WA | NA | NA | NA | NA | WA | NA | NA | NA |
| 265 | WA | WA | NA | NA | NA | NA | NA | WA | WA | WA | NA | WA | NA | WA | NA | NA | NA | WA | WA | NA | NA | WA | NA | NA | WA |
| 266 | WA | WA | WA | NA | NA | WA | WA | WA | WA | NA | NA | NA | WA | NA | WA | NA | WA | NA | NA | NA | NA | NA | NA | NA | NA |
| 267 | WA | NA | WA | NA | NA | WA | WA | WA | WA | NA | NA | WA | NA | WA | WA | WA | NA | WA | NA | NA | NA | NA | NA | NA | WA |
| 268 | WA | NA | WA | NA | NA | WA | WA | WA | WA | NA | NA | NA | NA | WA | WA | WA | WA | NA | NA | NA | NA | NA | NA | NA | WA |
| 269 | WA | NA | WA | NA | NA | NA | NA | WA | NA | NA | NA | WA | NA | WA | NA | NA | WA | WA | NA | NA | NA | NA | NA | NA | NA |
| 270 | WA | NA | NA | NA | NA | WA | WA | WA | WA | WA | NA | NA | NA | NA | WA | WA | NA | WA | NA | NA | NA | WA | NA | NA | NA |
| 271 | WA | WA | WA | NA | NA | NA | NA | WA | NA | WA | NA | NA | NA | NA | NA | NA | WA | WA | NA | NA | NA | NA | NA | NA | NA |
| 272 | WA | NA | NA | NA | NA | NA | NA | WA | NA | NA | NA | WA | NA | NA | WA | WA | WA | WA | NA | NA | NA | NA | NA | NA | NA |
| 273 | WA | NA | WA | NA | NA | WA | WA | WA | WA | NA | NA | WA | NA | NA | NA | NA | NA | WA | NA | NA | NA | NA | NA | NA | NA |
| 274 | WA | NA | WA | NA | NA | WA | WA | WA | WA | NA | NA | NA | NA | WA | NA | NA | NA | NA | NA | NA | NA | NA | NA | NA | NA |
| 275 | WA | NA | WA | NA | NA | WA | WA | WA | WA | WA | NA | NA | NA | WA | NA | NA | NA | WA | NA | NA | NA | NA | NA | NA | WA |
| 276 | WA | WA | NA | NA | NA | WA | WA | WA | WA | NA | NA | WA | NA | NA | WA | WA | WA | NA | NA | NA | NA | NA | NA | NA | NA |
| 277 | WA | NA | NA | NA | NA | WA | WA | WA | WA | NA | NA | NA | WA | NA | NA | NA | NA | WA | NA | NA | NA | WA | NA | NA | WA |
| 278 | WA | WA | WA | NA | NA | WA | WA | WA | WA | NA | NA | NA | NA | NA | NA | NA | WA | NA | NA | NA | NA | NA | NA | NA | NA |
| 279 | WA | NA | NA | NA | NA | WA | WA | WA | WA | NA | NA | NA | NA | NA | NA | NA | WA | NA | WA | NA | NA | NA | NA | NA | NA |
| 280 | WA | NA | NA | NA | NA | WA | WA | WA | WA | NA | NA | NA | NA | NA | WA | WA | WA | WA | NA | NA | NA | NA | WA | WA | NA |
| 281 | WA | WA | NA | NA | NA | WA | WA | WA | WA | NA | NA | WA | NA | NA | NA | NA | WA | WA | NA | NA | NA | NA | NA | NA | NA |
| 282 | WA | NA | WA | NA | NA | WA | WA | WA | WA | NA | NA | WA | NA | WA | WA | WA | WA | NA | WA | NA | NA | NA | NA | NA | NA |
| 283 | WA | NA | NA | NA | NA | WA | WA | WA | WA | WA | NA | NA | NA | NA | NA | NA | WA | WA | NA | NA | NA | WA | NA | NA | NA |
| 284 | WA | WA | NA | NA | NA | WA | WA | WA | WA | NA | NA | NA | NA | WA | WA | WA | NA | NA | WA | NA | NA | NA | NA | NA | NA |
| 285 | WA | NA | WA | NA | NA | WA | WA | WA | WA | WA | NA | WA | NA | WA | WA | WA | WA | WA | WA | NA | NA | WA | NA | NA | NA |
| 286 | WA | WA | NA | NA | NA | NA | NA | WA | NA | NA | NA | WA | NA | NA | NA | NA | NA | NA | WA | NA | NA | WA | NA | NA | NA |
| 287 | WA | NA | WA | NA | NA | NA | NA | WA | WA | NA | NA | NA | NA | NA | WA | WA | NA | NA | WA | NA | NA | NA | NA | NA | NA |
| 288 | WA | NA | WA | NA | NA | NA | NA | WA | NA | NA | NA | NA | NA | NA | WA | WA | NA | NA | WA | NA | NA | WA | NA | NA | NA |
| 289 | WA | NA | NA | NA | NA | WA | WA | WA | WA | NA | NA | WA | NA | NA | WA | WA | WA | WA | NA | NA | NA | NA | NA | NA | NA |
| 290 | WA | NA | NA | NA | NA | NA | NA | WA | NA | WA | NA | WA | NA | WA | NA | NA | WA | NA | NA | NA | NA | NA | NA | NA | NA |
| 291 | WA | - | NA | NA | NA | WA | WA | WA | WA | - | NA | NA | NA | - | - | - | NA | NA | NA | NA | NA | NA | NA | NA | - |
| 292 | WA | NA | WA | NA | NA | WA | WA | WA | WA | NA | NA | WA | NA | NA | NA | NA | WA | WA | NA | NA | NA | WA | NA | NA | NA |
| 293 | WA | WA | WA | NA | NA | WA | WA | WA | - | WA | NA | WA | NA | - | - | WA | - | NA | WA | NA | NA | WA | NA | NA | NA |
| 294 | WA | NA | NA | NA | NA | NA | NA | WA | NA | NA | NA | NA | NA | NA | NA | NA | WA | WA | NA | NA | NA | NA | NA | NA | NA |
| 295 | WA | WA | WA | NA | NA | WA | WA | WA | WA | WA | NA | NA | NA | NA | WA | WA | NA | NA | WA | NA | NA | NA | NA | NA | NA |
| 296 | WA | NA | NA | NA | NA | NA | NA | WA | NA | NA | NA | NA | NA | NA | WA | WA | NA | WA | WA | NA | NA | WA | WA | WA | NA |
| 297 | WA | NA | WA | NA | NA | NA | WA | WA | NA | NA | NA | NA | NA | WA | NA | NA | NA | WA | WA | NA | NA | NA | NA | NA | NA |
| 298 | WA | NA | NA | NA | NA | NA | NA | WA | NA | NA | NA | WA | WA | WA | NA | NA | NA | NA | WA | NA | NA | WA | NA | NA | NA |
| 299 | WA | NA | WA | NA | NA | NA | NA | WA | NA | NA | NA | WA | NA | NA | WA | WA | NA | WA | NA | NA | NA | NA | NA | NA | NA |
| 300 | WA | NA | NA | NA | NA | WA | WA | WA | WA | NA | NA | NA | NA | NA | NA | NA | NA | WA | NA | NA | NA | NA | WA | WA | NA |
| 301 | WA | NA | WA | NA | NA | WA | WA | WA | WA | NA | NA | WA | NA | WA | NA | NA | NA | WA | NA | NA | NA | WA | NA | NA | NA |
| 302 | WA | NA | WA | NA | NA | NA | NA | WA | NA | NA | NA | WA | NA | NA | NA | NA | WA | WA | NA | NA | NA | WA | NA | NA | NA |
| 303 | WA | NA | NA | NA | NA | WA | WA | WA | WA | NA | NA | WA | NA | WA | WA | WA | NA | NA | WA | NA | NA | NA | NA | NA | NA |
| 304 | WA | WA | WA | NA | NA | WA | WA | WA | WA | WA | NA | WA | NA | WA | NA | NA | WA | NA | NA | NA | NA | WA | NA | NA | NA |
| 305 | WA | NA | WA | NA | NA | WA | WA | WA | WA | NA | NA | WA | NA | NA | NA | NA | WA | NA | NA | NA | NA | WA | NA | NA | NA |
| 306 | WA | WA | WA | NA | NA | WA | WA | WA | WA | NA | NA | WA | NA | NA | NA | NA | WA | WA | WA | NA | NA | WA | NA | NA | NA |
| 307 | WA | NA | WA | NA | NA | WA | WA | WA | WA | NA | NA | WA | NA | NA | NA | NA | WA | WA | NA | NA | NA | WA | NA | NA | NA |
| 308 | WA | NA | NA | NA | NA | WA | WA | WA | WA | NA | NA | NA | NA | WA | NA | NA | NA | NA | NA | NA | NA | WA | NA | NA | NA |
| 309 | WA | NA | WA | NA | NA | NA | NA | WA | NA | WA | NA | NA | NA | WA | NA | NA | WA | NA | NA | NA | NA | NA | NA | NA | NA |
| 310 | WA | WA | NA | NA | NA | NA | NA | WA | NA | WA | NA | NA | NA | NA | NA | NA | WA | WA | NA | NA | NA | WA | NA | NA | NA |
| 311 | WA | NA | NA | NA | NA | NA | NA | WA | NA | NA | NA | NA | NA | WA | NA | NA | NA | NA | NA | NA | NA | NA | NA | NA | NA |
| 312 | WA | NA | WA | NA | NA | NA | NA | WA | NA | WA | NA | WA | NA | NA | NA | NA | WA | NA | WA | NA | WA | WA | NA | NA | WA |
| 313 | WA | NA | NA | NA | NA | WA | WA | WA | WA | WA | NA | WA | NA | NA | WA | WA | WA | NA | NA | NA | WA | NA | NA | NA | NA |
| 314 | WA | WA | NA | NA | NA | WA | WA | WA | WA | NA | NA | NA | NA | WA | NA | NA | NA | NA | NA | NA | NA | NA | NA | NA | NA |
| 315 | WA | WA | NA | NA | NA | WA | WA | WA | WA | NA | NA | NA | NA | NA | NA | NA | NA | WA | NA | NA | NA | NA | NA | NA | NA |
| 316 | WA | WA | NA | NA | NA | NA | WA | WA | WA | NA | NA | WA | NA | NA | NA | NA | WA | NA | NA | NA | NA | NA | NA | NA | NA |
| 317 | WA | NA | WA | NA | NA | NA | NA | WA | NA | NA | NA | WA | NA | NA | NA | NA | WA | WA | NA | NA | NA | NA | NA | NA | NA |
| 318 | WA | WA | NA | NA | NA | WA | WA | WA | WA | NA | NA | NA | NA | WA | NA | NA | WA | WA | NA | NA | NA | WA | WA | NA | NA |
| 319 | WA | NA | NA | NA | NA | WA | WA | WA | WA | WA | NA | NA | NA | WA | NA | NA | NA | NA | WA | NA | NA | NA | NA | NA | NA |
| 320 | WA | WA | NA | NA | NA | WA | WA | WA | WA | WA | NA | NA | NA | WA | NA | NA | WA | NA | WA | NA | NA | WA | NA | NA | NA |
| 321 | WA | NA | WA | NA | NA | WA | WA | WA | WA | NA | NA | NA | NA | WA | NA | NA | WA | NA | NA | NA | NA | WA | NA | NA | NA |
| 322 | WA | NA | WA | NA | NA | NA | NA | WA | NA | NA | NA | WA | NA | WA | WA | WA | NA | NA | NA | NA | NA | NA | NA | WA | NA |
| 323 | WA | NA | WA | NA | NA | NA | NA | WA | NA | NA | NA | WA | NA | NA | WA | WA | WA | WA | WA | NA | NA | NA | NA | NA | NA |
| 324 | WA | NA | NA | NA | NA | WA | WA | WA | NA | NA | NA | NA | NA | NA | NA | NA | WA | NA | WA | NA | WA | NA | NA | NA | NA |
| 325 | WA | NA | WA | NA | NA | NA | NA | WA | NA | NA | NA | WA | NA | WA | NA | NA | NA | WA | NA | NA | NA | WA | NA | NA | NA |
| 326 | WA | NA | NA | NA | NA | NA | NA | WA | NA | WA | NA | WA | NA | NA | NA | NA | NA | WA | NA | NA | NA | WA | NA | NA | NA |
| 327 | WA | NA | NA | NA | NA | NA | NA | WA | NA | NA | NA | NA | NA | WA | NA | NA | NA | WA | NA | NA | NA | NA | NA | NA | NA |
| 328 | WA | NA | WA | NA | NA | NA | NA | WA | NA | NA | NA | WA | NA | WA | NA | NA | NA | WA | NA | NA | NA | WA | NA | NA | NA |
| 329 | WA | WA | NA | NA | NA | NA | NA | WA | WA | NA | NA | NA | NA | NA | WA | WA | NA | NA | NA | NA | WA | NA | NA | NA | NA |
| 330 | WA | NA | NA | NA | NA | WA | WA | WA | WA | WA | NA | WA | NA | NA | NA | NA | WA | WA | WA | NA | NA | WA | NA | NA | NA |
| 331 | WA | WA | NA | NA | NA | WA | WA | WA | WA | NA | NA | NA | NA | NA | WA | WA | NA | NA | NA | NA | NA | NA | NA | NA | NA |
| 332 | WA | WA | WA | NA | NA | WA | WA | WA | WA | NA | NA | WA | NA | WA | NA | NA | WA | NA | NA | NA | NA | NA | NA | NA | NA |
| 333 | WA | WA | NA | NA | NA | WA | WA | WA | WA | WA | NA | WA | NA | NA | NA | NA | WA | NA | NA | NA | NA | NA | WA | WA | NA |
| 334 | WA | WA | NA | NA | NA | WA | WA | WA | WA | NA | NA | NA | NA | NA | NA | NA | WA | NA | NA | NA | NA | NA | NA | NA | WA |
| 335 | WA | NA | NA | NA | NA | NA | NA | WA | NA | NA | NA | NA | NA | NA | NA | NA | NA | NA | NA | NA | NA | NA | NA | NA | WA |
| 336 | WA | NA | NA | NA | NA | WA | WA | WA | WA | NA | NA | WA | NA | WA | WA | WA | NA | WA | NA | NA | NA | NA | NA | NA | NA |
| 337 | WA | NA | WA | NA | NA | WA | WA | WA | WA | WA | NA | WA | NA | WA | NA | NA | NA | WA | NA | NA | NA | WA | NA | NA | NA |
| 338 | WA | NA | WA | NA | NA | WA | WA | WA | WA | NA | NA | NA | NA | NA | NA | NA | NA | NA | NA | NA | NA | NA | NA | NA | WA |
| 339 | WA | NA | WA | NA | NA | WA | WA | WA | WA | NA | NA | NA | NA | WA | NA | NA | WA | WA | WA | NA | NA | WA | NA | NA | WA |
| 340 | WA | WA | NA | NA | NA | WA | WA | WA | WA | NA | NA | NA | NA | WA | NA | NA | NA | WA | NA | NA | NA | NA | NA | NA | NA |
| 341 | WA | WA | NA | NA | NA | WA | WA | WA | WA | NA | NA | WA | NA | NA | NA | NA | NA | WA | WA | NA | WA | NA | NA | NA | NA |
| 342 | WA | NA | WA | NA | NA | WA | WA | WA | WA | NA | NA | WA | NA | NA | NA | NA | WA | WA | NA | NA | NA | WA | NA | NA | NA |
| 343 | WA | WA | NA | NA | NA | WA | WA | WA | WA | NA | NA | WA | NA | NA | NA | NA | WA | WA | NA | NA | NA | NA | NA | NA | NA |
| 344 | WA | WA | NA | NA | NA | NA | NA | WA | WA | WA | NA | WA | NA | WA | WA | WA | NA | NA | NA | NA | NA | WA | NA | NA | NA |
| 345 | WA | NA | NA | NA | NA | WA | WA | WA | WA | NA | NA | WA | NA | NA | NA | NA | WA | WA | NA | NA | NA | WA | NA | NA | NA |
| 346 | WA | WA | WA | NA | NA | WA | WA | WA | WA | NA | NA | WA | NA | NA | NA | NA | NA | WA | NA | NA | NA | NA | NA | NA | NA |
| 347 | WA | NA | NA | NA | NA | NA | NA | WA | NA | WA | NA | WA | NA | WA | NA | NA | NA | NA | NA | NA | NA | NA | NA | NA | NA |
| 348 | WA | NA | NA | NA | NA | WA | WA | WA | WA | WA | NA | NA | NA | NA | NA | NA | NA | NA | NA | NA | NA | NA | NA | NA | NA |
| 349 | WA | NA | NA | NA | NA | WA | WA | WA | WA | WA | NA | WA | NA | NA | WA | WA | WA | NA | WA | NA | WA | NA | NA | NA | NA |
| 350 | WA | NA | WA | NA | NA | WA | WA | WA | WA | NA | NA | NA | NA | NA | WA | WA | WA | NA | NA | NA | NA | NA | NA | NA | WA |
| 351 | WA | WA | NA | NA | NA | NA | NA | WA | WA | NA | NA | NA | NA | NA | NA | NA | NA | NA | WA | NA | NA | NA | NA | NA | NA |
| 352 | WA | NA | NA | NA | NA | WA | WA | WA | NA | NA | NA | NA | NA | NA | WA | WA | WA | WA | NA | NA | NA | WA | NA | NA | NA |
| 353 | WA | NA | NA | NA | NA | NA | NA | WA | NA | NA | NA | WA | NA | WA | NA | NA | NA | NA | NA | NA | NA | WA | NA | NA | NA |
| 354 | WA | NA | NA | NA | NA | NA | NA | WA | NA | NA | NA | NA | NA | NA | NA | NA | WA | WA | WA | NA | NA | NA | NA | NA | NA |
| 355 | WA | WA | NA | NA | NA | NA | NA | WA | WA | NA | NA | WA | NA | WA | NA | NA | NA | WA | NA | NA | NA | NA | WA | NA | NA |
| 356 | WA | NA | NA | NA | NA | WA | WA | WA | WA | NA | NA | NA | NA | NA | NA | NA | WA | NA | NA | NA | NA | NA | NA | NA | NA |
| 357 | WA | WA | NA | NA | NA | NA | NA | WA | WA | NA | NA | NA | NA | WA | NA | NA | NA | NA | NA | NA | NA | NA | WA | NA | WA |
| 358 | WA | WA | NA | NA | NA | NA | NA | WA | NA | NA | NA | WA | NA | WA | NA | NA | NA | NA | NA | NA | NA | WA | NA | NA | WA |
| 359 | WA | WA | NA | NA | NA | WA | WA | WA | WA | WA | NA | WA | NA | NA | WA | WA | WA | NA | NA | NA | NA | WA | NA | NA | WA |

| | | | | | | | | | | | | | | | | | | | | | | | | |
|---|---|---|---|---|---|---|---|---|---|---|---|---|---|---|---|---|---|---|---|---|---|---|---|---|
| 360 WA | NA | NA | NA | NA | NA | NA | WA | WA | NA | NA | NA | NA | WA | WA | WA | WA | WA | NA | NA | WA | NA | NA | NA | WA |
| 361 WA | NA | WA | NA | NA | WA | WA | WA | WA | NA | NA | NA | NA | NA | NA | NA | WA | NA | NA | NA | NA | WA | NA | NA | NA |
| 362 WA | NA | WA | NA | NA | NA | WA | WA | NA | NA | NA | WA | NA | NA | NA | NA | NA | NA | NA | NA | NA | NA | WA | WA | NA |
| 363 WA | WA | NA | NA | NA | WA | WA | WA | WA | NA | NA | WA | NA | NA | NA | NA | NA | NA | NA | NA | NA | NA | NA | NA | NA |
| 364 WA | NA | NA | NA | NA | WA | WA | WA | WA | NA | NA | WA | NA | NA | NA | NA | NA | NA | NA | NA | NA | WA | NA | NA | NA |
| 365 WA | NA | WA | NA | NA | NA | NA | WA | NA | WA | NA | NA | NA | NA | WA | WA | NA | WA | NA | NA | NA | NA | NA | NA | NA |
| 366 WA | NA | NA | NA | NA | WA | WA | WA | WA | NA | NA | NA | NA | NA | NA | NA | WA | NA | WA | NA | NA | NA | NA | NA | NA |
| 367 WA | NA | NA | NA | NA | WA | WA | WA | WA | WA | NA | NA | NA | WA | WA | WA | WA | NA | WA | WA | NA | WA | NA | NA | NA |
| 368 WA | WA | NA | NA | NA | NA | NA | WA | WA | WA | NA | NA | NA | NA | NA | NA | WA | WA | NA | NA | NA | WA | NA | NA | NA |
| 369 WA | NA | WA | NA | NA | WA | WA | WA | WA | NA | NA | WA | NA | NA | NA | NA | WA | NA | NA | NA | NA | NA | NA | NA | NA |
| 370 WA | NA | NA | NA | NA | WA | WA | WA | WA | NA | NA | WA | NA | NA | NA | NA | NA | WA | NA | NA | NA | NA | NA | NA | NA |
| 371 WA | NA | WA | NA | NA | WA | WA | WA | WA | WA | NA | WA | NA | NA | NA | NA | WA | WA | NA | NA | NA | NA | NA | NA | WA |
| 372 WA | NA | WA | NA | NA | NA | NA | WA | NA | WA | NA | NA | NA | NA | WA | WA | NA | WA | NA | NA | NA | NA | NA | NA | NA |
| 373 WA | WA | NA | NA | NA | WA | WA | WA | WA | WA | NA | NA | NA | WA | NA | NA | NA | NA | NA | NA | NA | WA | NA | NA | NA |
| 374 WA | WA | WA | NA | NA | WA | WA | WA | WA | NA | NA | WA | NA | NA | NA | NA | NA | NA | NA | NA | WA | NA | WA | WA | NA |
| 375 WA | NA | NA | NA | NA | WA | WA | WA | WA | WA | NA | NA | NA | NA | WA | WA | WA | NA | NA | NA | NA | NA | NA | NA | NA |
| 376 WA | NA | NA | NA | NA | NA | NA | WA | WA | NA | NA | WA | NA | WA | WA | WA | NA | NA | WA | NA | NA | WA | NA | NA | NA |
| 377 WA | NA | WA | NA | NA | NA | NA | WA | NA | NA | NA | NA | WA | NA | WA | WA | NA | WA | NA | NA | NA | NA | NA | NA | NA |
| 378 WA | NA | NA | NA | NA | WA | WA | WA | WA | NA | NA | NA | NA | WA | NA | NA | WA | WA | NA | NA | NA | WA | NA | NA | NA |
| 379 WA | NA | WA | NA | NA | WA | WA | WA | WA | NA | NA | NA | NA | NA | WA | WA | WA | WA | NA | NA | NA | NA | NA | NA | NA |
| 380 WA | NA | NA | NA | NA | WA | WA | WA | WA | NA | NA | NA | NA | NA | NA | NA | WA | NA | WA | NA | NA | WA | NA | NA | NA |
| 381 WA | NA | WA | NA | NA | NA | NA | WA | WA | NA | NA | NA | NA | NA | WA | WA | WA | WA | NA | NA | NA | NA | NA | NA | NA |
| 382 WA | NA | WA | NA | NA | NA | NA | WA | NA | NA | NA | WA | NA | WA | NA | NA | WA | NA | NA | NA | WA | WA | NA | NA | NA |
| 383 WA | NA | NA | NA | NA | NA | NA | WA | NA | NA | NA | NA | NA | WA | WA | WA | WA | WA | NA | NA | NA | NA | WA | WA | NA |
| 384 WA | NA | NA | NA | NA | NA | NA | WA | NA | NA | NA | WA | NA | WA | NA | NA | NA | WA | WA | NA | NA | NA | NA | NA | NA |
| 385 WA | NA | WA | NA | NA | NA | NA | WA | NA | NA | NA | WA | NA | WA | NA | NA | NA | WA | WA | NA | NA | NA | NA | NA | NA |
| 386 WA | NA | NA | NA | NA | NA | NA | WA | NA | WA | NA | WA | NA | NA | NA | NA | NA | NA | WA | NA | NA | WA | NA | NA | NA |
| 387 WA | WA | WA | NA | NA | WA | WA | WA | WA | NA | NA | WA | NA | NA | WA | WA | NA | NA | NA | NA | NA | NA | NA | NA | WA |
| 388 WA | NA | NA | NA | NA | NA | NA | WA | WA | NA | NA | NA | NA | NA | NA | NA | WA | WA | NA | NA | NA | WA | WA | WA | NA |
| 389 WA | NA | NA | NA | NA | WA | WA | WA | WA | NA | NA | NA | NA | NA | WA | WA | WA | NA | WA | NA | NA | NA | NA | NA | NA |
| 390 WA | WA | WA | NA | NA | WA | WA | WA | WA | NA | NA | WA | NA | NA | NA | NA | NA | WA | NA | NA | NA | WA | NA | NA | NA |
| 391 WA | NA | NA | NA | NA | WA | WA | WA | WA | NA | NA | WA | NA | NA | NA | NA | WA | NA | NA | NA | NA | WA | NA | NA | NA |
| 392 WA | NA | NA | NA | NA | NA | NA | WA | NA | NA | NA | NA | NA | WA | WA | WA | WA | WA | WA | NA | NA | NA | NA | NA | NA |
| 393 WA | WA | WA | NA | NA | WA | WA | WA | WA | NA | NA | WA | NA | WA | NA | NA | WA | WA | NA | WA | WA | NA | NA | NA | NA |
| 394 WA | NA | WA | NA | NA | WA | WA | WA | WA | NA | NA | WA | NA | WA | NA | NA | WA | WA | NA | NA | WA | WA | WA | WA | NA |
| 395 WA | WA | NA | NA | NA | WA | WA | WA | WA | WA | NA | NA | NA | WA | NA | NA | NA | NA | NA | NA | NA | NA | NA | NA | NA |
| 396 WA | NA | NA | NA | NA | WA | WA | WA | WA | NA | NA | NA | NA | NA | NA | NA | NA | NA | WA | NA | NA | NA | NA | NA | WA |
| 397 WA | NA | NA | NA | NA | WA | WA | WA | WA | WA | NA | WA | WA | WA | WA | WA | NA | NA | WA | NA | NA | WA | WA | WA | NA |
| 398 WA | NA | NA | NA | NA | WA | WA | WA | WA | NA | NA | WA | NA | NA | WA | WA | WA | WA | NA | NA | NA | NA | NA | NA | NA |
| 399 WA | NA | NA | NA | NA | WA | WA | WA | WA | NA | NA | WA | WA | NA | NA | NA | NA | WA | NA | NA | NA | NA | NA | NA | NA |
| 400 WA | WA | NA | NA | NA | WA | WA | WA | WA | NA | NA | WA | NA | NA | NA | NA | WA | WA | NA | NA | NA | WA | WA | WA | NA |
| 401 WA | NA | NA | NA | NA | WA | WA | WA | WA | NA | NA | NA | NA | WA | NA | NA | NA | WA | NA | NA | WA | WA | NA | NA | WA |
| 402 WA | NA | WA | NA | NA | WA | WA | WA | WA | NA | NA | WA | NA | NA | NA | NA | WA | WA | NA | NA | NA | NA | NA | NA | WA |
| 403 WA | NA | NA | WA | NA | NA | NA | WA | NA | NA | NA | WA | NA | NA | NA | NA | NA | WA | WA | NA | NA | WA | WA | WA | WA |
| 404 WA | WA | NA | NA | NA | NA | NA | WA | NA | NA | NA | NA | NA | WA | WA | WA | NA | NA | NA | NA | NA | NA | NA | NA | WA |
| 405 WA | NA | NA | NA | NA | WA | WA | WA | WA | WA | NA | NA | NA | NA | WA | WA | NA | WA | WA | NA | NA | WA | WA | WA | NA |
| 406 WA | NA | NA | NA | NA | NA | NA | WA | WA | NA | NA | NA | NA | NA | NA | NA | NA | NA | WA | NA | WA | NA | NA | NA | NA |
| 407 WA | NA | NA | NA | NA | NA | NA | WA | NA | WA | NA | WA | NA | NA | WA | WA | NA | WA | NA | NA | NA | NA | NA | NA | NA |
| 408 WA | NA | NA | NA | NA | NA | NA | WA | NA | WA | NA | NA | NA | NA | NA | NA | WA | NA | NA | NA | NA | WA | NA | NA | NA |
| 409 WA | NA | NA | NA | NA | WA | WA | WA | WA | NA | NA | WA | NA | NA | WA | WA | NA | WA | NA | NA | NA | NA | WA | NA | NA |
| 410 WA | WA | NA | NA | NA | NA | NA | WA | NA | NA | NA | NA | NA | WA | WA | WA | NA | NA | NA | NA | NA | WA | NA | NA | NA |
| 411 WA | NA | NA | NA | NA | NA | NA | WA | NA | WA | NA | WA | NA | NA | WA | WA | WA | NA | NA | NA | NA | NA | NA | NA | NA |
| 412 WA | WA | NA | NA | NA | NA | NA | WA | NA | WA | NA | NA | NA | NA | WA | WA | NA | NA | NA | NA | NA | WA | NA | NA | NA |
| 413 WA | NA | NA | NA | NA | NA | NA | WA | NA | NA | NA | WA | NA | NA | WA | WA | NA | NA | WA | NA | NA | NA | NA | NA | NA |
| 414 WA | NA | NA | NA | NA | NA | NA | WA | WA | NA | NA | NA | NA | NA | NA | NA | WA | WA | NA | NA | NA | WA | WA | WA | NA |
| 415 WA | NA | NA | NA | NA | WA | WA | WA | WA | NA | NA | NA | WA | NA | NA | NA | WA | NA | WA | NA | NA | NA | NA | NA | WA |
| 416 WA | WA | NA | NA | NA | WA | WA | WA | WA | WA | NA | NA | NA | NA | NA | NA | WA | NA | NA | NA | NA | WA | NA | NA | NA |
| 417 WA | NA | WA | NA | NA | NA | NA | WA | WA | NA | NA | NA | NA | NA | NA | NA | WA | WA | WA | NA | NA | WA | WA | WA | NA |
| 418 WA | NA | WA | NA | NA | WA | WA | WA | WA | NA | NA | NA | NA | NA | WA | WA | WA | WA | NA | NA | WA | WA | NA | NA | NA |
| 419 WA | NA | NA | NA | NA | NA | NA | WA | NA | NA | NA | NA | NA | NA | WA | WA | WA | NA | WA | NA | NA | NA | NA | NA | WA |
| 420 WA | NA | WA | NA | NA | WA | WA | WA | WA | NA | NA | WA | NA | WA | NA | NA | WA | NA | NA | NA | NA | WA | NA | NA | NA |
| 421 WA | WA | WA | NA | NA | NA | NA | WA | NA | WA | NA | WA | NA | NA | NA | NA | NA | NA | NA | NA | NA | WA | WA | WA | NA |
| 422 WA | NA | NA | NA | NA | NA | NA | WA | NA | NA | NA | WA | NA | WA | NA | NA | WA | NA | WA | NA | NA | WA | NA | NA | NA |
| 423 WA | NA | WA | NA | NA | WA | WA | WA | WA | WA | NA | NA | NA | NA | WA | WA | NA | WA | NA | NA | NA | NA | NA | NA | NA |
| 424 WA | NA | NA | NA | NA | NA | NA | WA | NA | NA | NA | WA | NA | WA | WA | WA | WA | NA | NA | NA | NA | NA | NA | NA | WA |
| 425 WA | WA | NA | NA | NA | WA | WA | WA | WA | NA | NA | NA | NA | NA | NA | NA | NA | NA | NA | NA | NA | NA | NA | NA | NA |
| 426 WA | WA | WA | NA | NA | NA | NA | WA | NA | WA | NA | NA | NA | NA | NA | NA | NA | WA | NA | NA | NA | WA | NA | NA | NA |
| 427 WA | NA | NA | NA | NA | NA | NA | WA | NA | NA | NA | NA | NA | NA | NA | NA | NA | WA | NA | NA | NA | WA | NA | NA | NA |
| 428 WA | NA | WA | NA | NA | WA | WA | WA | WA | NA | NA | NA | NA | NA | NA | NA | WA | NA | NA | NA | NA | WA | NA | NA | NA |
| 429 WA | NA | WA | NA | NA | WA | WA | WA | WA | NA | NA | NA | NA | NA | NA | NA | WA | NA | NA | NA | NA | WA | NA | NA | NA |
| 430 WA | WA | NA | NA | NA | NA | NA | WA | NA | NA | NA | NA | NA | NA | NA | NA | WA | WA | NA | NA | NA | NA | WA | WA | NA |
| 431 WA | WA | NA | NA | NA | NA | NA | WA | NA | NA | NA | WA | NA | WA | NA | NA | WA | NA | NA | NA | NA | NA | NA | NA | NA |
| 432 WA | NA | WA | NA | NA | WA | WA | WA | WA | NA | NA | NA | NA | WA | WA | WA | WA | WA | WA | NA | NA | WA | NA | NA | NA |
| 433 WA | NA | WA | NA | NA | NA | NA | WA | NA | NA | NA | WA | NA | NA | NA | NA | WA | NA | NA | NA | NA | NA | NA | NA | NA |
| 434 WA | NA | NA | NA | NA | NA | NA | WA | NA | NA | NA | WA | NA | WA | NA | NA | WA | NA | NA | NA | WA | NA | NA | NA | NA |
| 435 WA | NA | NA | NA | NA | WA | WA | WA | WA | NA | NA | WA | NA | WA | NA | NA | WA | NA | NA | NA | NA | WA | NA | NA | NA |
| 436 WA | NA | NA | NA | NA | WA | WA | WA | WA | NA | NA | NA | NA | WA | NA | NA | NA | NA | NA | NA | NA | NA | NA | NA | WA |
| 437 WA | NA | NA | NA | NA | WA | WA | WA | WA | NA | NA | NA | NA | WA | NA | NA | WA | WA | NA | NA | NA | NA | NA | NA | WA |
| 438 WA | NA | WA | NA | NA | NA | NA | WA | NA | NA | NA | NA | NA | NA | NA | NA | WA | NA | NA | NA | WA | NA | NA | NA | WA |
| 439 WA | NA | NA | NA | NA | NA | NA | WA | WA | WA | NA | WA | NA | NA | NA | NA | WA | NA | WA | NA | NA | WA | NA | NA | NA |
| 440 WA | NA | WA | NA | NA | NA | NA | WA | WA | NA | NA | WA | NA | NA | NA | NA | WA | NA | NA | NA | NA | WA | NA | NA | WA |
| 441 WA | NA | NA | NA | NA | WA | WA | WA | NA | WA | NA | NA | NA | NA | NA | NA | WA | WA | NA | NA | NA | NA | NA | NA | NA |
| 442 WA | NA | NA | NA | NA | WA | WA | WA | NA | NA | NA | WA | NA | NA | NA | NA | NA | NA | NA | NA | NA | NA | NA | NA | NA |
| 443 WA | NA | NA | NA | NA | NA | NA | WA | WA | WA | NA | NA | NA | WA | NA | NA | WA | NA | NA | NA | NA | WA | NA | NA | NA |
| 444 WA | NA | NA | NA | NA | NA | NA | WA | NA | WA | NA | NA | NA | NA | WA | WA | NA | WA | WA | NA | NA | NA | NA | NA | NA |
| 445 WA | WA | WA | NA | NA | WA | WA | WA | WA | NA | NA | WA | NA | NA | NA | NA | WA | NA | NA | NA | NA | NA | NA | NA | NA |
| 446 WA | NA | WA | NA | NA | NA | NA | WA | NA | NA | NA | NA | NA | NA | WA | WA | NA | WA | WA | NA | NA | WA | NA | NA | NA |
| 447 WA | NA | WA | NA | NA | NA | NA | WA | NA | NA | NA | WA | NA | WA | NA | NA | WA | NA | WA | NA | NA | WA | WA | WA | NA |
| 448 WA | NA | WA | NA | NA | WA | WA | WA | WA | NA | NA | NA | NA | NA | NA | NA | NA | WA | NA | NA | NA | NA | NA | NA | WA |
| 449 WA | NA | NA | NA | NA | WA | WA | WA | WA | NA | NA | WA | NA | NA | WA | WA | NA | WA | WA | NA | NA | NA | NA | NA | NA |
| 450 WA | NA | NA | NA | NA | WA | WA | WA | WA | NA | NA | WA | NA | NA | NA | NA | NA | NA | WA | NA | WA | NA | NA | NA | NA |
| 451 WA | NA | NA | NA | NA | WA | WA | WA | WA | NA | NA | WA | WA | WA | WA | WA | NA | NA | NA | NA | NA | WA | NA | NA | WA |
| 452 WA | NA | NA | NA | NA | WA | WA | WA | WA | NA | NA | WA | NA | NA | WA | WA | WA | WA | WA | NA | NA | NA | NA | NA | NA |
| 453 WA | WA | NA | NA | NA | WA | WA | WA | NA | NA | NA | NA | NA | NA | WA | WA | WA | WA | WA | NA | NA | NA | NA | NA | NA |
| 454 WA | NA | NA | NA | NA | NA | NA | WA | NA | NA | NA | WA | NA | NA | NA | NA | NA | NA | NA | NA | NA | NA | NA | NA | NA |
| 455 WA | NA | WA | NA | NA | WA | WA | WA | WA | NA | NA | WA | NA | WA | WA | WA | NA | NA | WA | NA | NA | WA | NA | NA | NA |
| 456 WA | NA | WA | NA | NA | WA | WA | WA | WA | WA | NA | NA | NA | WA | NA | NA | NA | WA | WA | NA | NA | NA | NA | NA | NA |
| 457 WA | WA | NA | NA | NA | WA | WA | WA | WA | NA | NA | WA | NA | NA | NA | NA | NA | WA | NA | NA | NA | NA | NA | NA | NA |
| 458 WA | NA | NA | NA | NA | WA | WA | WA | - | WA | NA | NA | NA | NA | - | - | WA | WA | - | NA | NA | WA | NA | NA | NA |
| 459 WA | NA | NA | NA | NA | WA | WA | WA | NA | NA | NA | NA | NA | NA | WA | WA | WA | WA | NA | NA | NA | WA | NA | NA | NA |
| 460 WA | NA | NA | NA | NA | WA | WA | WA | WA | WA | NA | WA | NA | NA | WA | WA | WA | WA | NA | NA | NA | WA | NA | NA | NA |
| 461 WA | NA | WA | NA | NA | WA | WA | WA | WA | NA | NA | NA | NA | NA | NA | NA | WA | NA | NA | NA | NA | WA | NA | NA | NA |
| 462 WA | NA | NA | NA | NA | WA | WA | WA | WA | WA | NA | NA | NA | NA | WA | NA | WA | NA | WA | NA | NA | NA | NA | NA | NA |
| 463 WA | NA | WA | NA | NA | WA | WA | WA | WA | NA | NA | WA | WA | WA | WA | WA | NA | NA | WA | NA | NA | NA | NA | NA | NA |
| 464 WA | NA | NA | NA | NA | WA | WA | WA | WA | WA | NA | NA | NA | NA | NA | NA | WA | NA | WA | NA | NA | WA | NA | NA | NA |
| 465 WA | NA | NA | NA | NA | NA | NA | WA | NA | WA | NA | NA | NA | NA | WA | WA | NA | NA | NA | NA | NA | NA | NA | NA | NA |
| 466 WA | WA | NA | NA | NA | WA | WA | WA | WA | NA | NA | WA | NA | NA | NA | NA | NA | NA | NA | NA | NA | NA | NA | NA | NA |
| 467 WA | NA | NA | NA | NA | WA | WA | WA | WA | WA | NA | WA | NA | WA | NA | NA | NA | NA | NA | NA | NA | WA | NA | NA | NA |
| 468 WA | NA | NA | NA | NA | WA | WA | WA | WA | NA | NA | WA | NA | WA | NA | NA | WA | NA | NA | NA | NA | WA | NA | NA | NA |
| 469 WA | NA | WA | NA | NA | WA | WA | WA | WA | WA | NA | NA | NA | NA | NA | NA | WA | NA | WA | NA | NA | WA | NA | NA | NA |
| 470 WA | WA | NA | NA | NA | WA | WA | WA | WA | WA | NA | NA | NA | WA | WA | WA | WA | NA | NA | NA | NA | NA | NA | NA | NA |
| 471 WA | NA | NA | NA | NA | WA | WA | WA | WA | WA | NA | NA | NA | NA | NA | NA | WA | NA | NA | NA | NA | NA | NA | NA | NA |
| 472 WA | NA | NA | NA | NA | WA | WA | WA | WA | NA | NA | NA | WA | WA | NA | NA | NA | WA | NA | NA | NA | - | NA | NA | NA |
| 473 WA | NA | WA | NA | NA | WA | WA | WA | WA | NA | NA | WA | NA | NA | NA | NA | NA | NA | NA | NA | NA | WA | NA | NA | WA |
| 474 WA | NA | NA | NA | NA | NA | NA | WA | NA | NA | NA | WA | NA | NA | NA | NA | WA | WA | WA | NA | NA | WA | NA | NA | NA |
| 475 WA | NA | WA | NA | NA | WA | WA | WA | NA | NA | NA | NA | NA | WA | WA | WA | NA | NA | NA | NA | NA | NA | WA | WA | NA |
| 476 WA | WA | NA | NA | NA | WA | WA | WA | WA | NA | NA | WA | NA | WA | WA | WA | NA | WA | NA | NA | NA | NA | NA | NA | WA |
| 477 WA | NA | NA | NA | NA | NA | NA | WA | NA | NA | NA | NA | NA | WA | NA | NA | NA | NA | NA | NA | NA | NA | NA | NA | NA |
| 478 WA | NA | NA | NA | NA | WA | WA | WA | WA | WA | NA | NA | NA | WA | NA | NA | WA | NA | NA | NA | NA | NA | WA | WA | NA |
| 479 WA | NA | WA | NA | NA | NA | NA | WA | NA | WA | NA | NA | NA | WA | NA | NA | WA | WA | NA | NA | NA | NA | NA | NA | NA |
| 480 WA | NA | NA | NA | NA | NA | NA | WA | NA | NA | NA | WA | NA | NA | NA | NA | WA | NA | WA | NA | NA | NA | WA | NA | NA |

| | | | | | | | | | | | | | | | | | | | | | | | | | |
|---|---|---|---|---|---|---|---|---|---|---|---|---|---|---|---|---|---|---|---|---|---|---|---|---|---|
| 481 | WA | WA | WA | NA | NA | WA | NA | WA | WA | WA | NA | WA | NA | WA | NA | NA | WA | NA | NA | NA | NA | NA | NA | NA | NA |
| 482 | WA | NA | WA | NA | NA | NA | NA | WA | WA | NA | NA | NA | NA | WA | NA | NA | NA | NA | NA | NA | NA | NA | NA | NA | NA |
| 483 | WA | NA | NA | NA | NA | WA | WA | WA | WA | NA | NA | WA | NA | NA | NA | WA | WA | WA | NA | NA | NA | NA | NA | NA | NA |
| 484 | WA | NA | NA | NA | NA | WA | WA | WA | WA | WA | NA | WA | NA | NA | WA | WA | WA | NA | NA | NA | WA | NA | NA | NA | NA |
| 485 | WA | WA | NA | NA | NA | NA | NA | WA | NA | WA | NA | WA | NA | WA | WA | WA | NA | NA | NA | NA | NA | WA | NA | NA | WA |
| 486 | WA | WA | NA | NA | NA | WA | WA | WA | WA | NA | NA | NA | NA | WA | WA | WA | WA | NA | NA | NA | NA | NA | NA | NA | NA |
| 487 | WA | NA | NA | NA | NA | NA | NA | WA | NA | WA | NA | NA | NA | NA | WA | WA | NA | WA | NA | NA | NA | NA | NA | NA | NA |
| 488 | WA | WA | NA | NA | NA | WA | WA | WA | WA | NA | NA | WA | NA | WA | WA | WA | NA | NA | NA | NA | NA | NA | NA | NA | NA |
| 489 | WA | NA | NA | NA | NA | NA | NA | WA | NA | WA | NA | NA | NA | WA | NA | NA | NA | NA | NA | NA | NA | NA | NA | NA | NA |
| 490 | WA | NA | NA | NA | NA | NA | NA | WA | NA | NA | NA | NA | NA | NA | NA | NA | WA | NA | NA | NA | NA | NA | NA | NA | NA |
| 491 | WA | NA | NA | NA | NA | WA | WA | WA | WA | NA | NA | WA | NA | NA | NA | NA | NA | NA | WA | NA | NA | WA | NA | NA | NA |
| 492 | WA | NA | WA | NA | NA | NA | NA | WA | NA | NA | NA | WA | NA | WA | WA | - | NA | NA | NA | NA | NA | - | NA | NA | NA |
| 493 | WA | NA | NA | NA | NA | NA | NA | WA | NA | NA | NA | NA | NA | NA | NA | NA | NA | WA | NA | NA | NA | WA | NA | NA | NA |
| 494 | WA | WA | NA | NA | NA | NA | NA | WA | NA | WA | NA | NA | NA | NA | NA | NA | NA | NA | NA | NA | NA | NA | NA | NA | NA |
| 495 | WA | NA | NA | NA | NA | WA | WA | WA | WA | NA | NA | WA | NA | NA | NA | NA | NA | NA | NA | NA | NA | NA | NA | NA | NA |
| 496 | WA | NA | NA | NA | NA | NA | NA | WA | NA | NA | NA | WA | NA | NA | NA | NA | NA | NA | NA | NA | WA | WA | NA | NA | WA |
| 497 | WA | WA | NA | NA | NA | WA | WA | WA | WA | WA | NA | NA | NA | NA | NA | NA | WA | WA | WA | NA | NA | NA | NA | NA | NA |
| 498 | WA | NA | NA | NA | NA | WA | WA | WA | WA | NA | NA | WA | NA | WA | WA | WA | NA | NA | WA | NA | NA | WA | NA | NA | NA |
| 499 | WA | WA | WA | NA | NA | NA | NA | WA | NA | WA | NA | NA | NA | NA | WA | WA | WA | NA | WA | NA | NA | NA | NA | NA | NA |
| 500 | WA | WA | NA | NA | NA | NA | NA | WA | NA | NA | NA | WA | NA | NA | NA | NA | WA | NA | NA | NA | NA | WA | NA | NA | NA |
| 501 | WA | WA | NA | NA | NA | WA | WA | WA | WA | NA | NA | NA | NA | WA | NA | NA | WA | WA | NA | NA | NA | NA | NA | NA | NA |
| 502 | WA | NA | NA | NA | NA | NA | NA | WA | NA | NA | NA | NA | NA | NA | NA | NA | NA | WA | NA | NA | NA | NA | NA | NA | NA |
| 503 | WA | NA | NA | NA | NA | WA | WA | WA | NA | NA | NA | NA | NA | NA | WA | WA | NA | NA | NA | NA | NA | WA | NA | NA | NA |
| 504 | WA | NA | WA | NA | NA | NA | NA | WA | NA | WA | NA | WA | NA | WA | NA | NA | NA | NA | NA | NA | WA | WA | NA | NA | NA |
| 505 | WA | NA | WA | NA | NA | WA | WA | WA | WA | NA | NA | WA | NA | NA | NA | NA | WA | WA | NA | NA | NA | NA | NA | NA | NA |
| 506 | WA | NA | WA | NA | NA | NA | NA | WA | NA | WA | NA | WA | NA | NA | WA | WA | WA | WA | WA | NA | NA | NA | NA | NA | WA |
| 507 | WA | NA | WA | NA | NA | WA | WA | WA | WA | WA | NA | WA | NA | NA | WA | WA | NA | NA | NA | NA | NA | NA | NA | NA | NA |
| 508 | WA | WA | NA | NA | NA | WA | WA | WA | WA | NA | NA | NA | WA | NA | WA | WA | WA | WA | NA | NA | NA | WA | NA | NA | NA |
| 509 | WA | NA | NA | NA | NA | WA | WA | WA | WA | WA | NA | WA | NA | WA | NA | NA | WA | WA | NA | NA | NA | WA | NA | NA | NA |
| 510 | WA | WA | NA | NA | NA | NA | NA | WA | WA | WA | NA | NA | NA | NA | NA | NA | NA | WA | NA | NA | NA | WA | NA | NA | NA |
| 511 | WA | NA | WA | NA | NA | WA | WA | WA | NA | NA | NA | NA | NA | WA | NA | NA | WA | WA | NA | NA | NA | WA | NA | NA | WA |
| 512 | WA | NA | NA | NA | NA | WA | WA | WA | WA | NA | NA | WA | NA | NA | WA | WA | WA | NA | WA | NA | NA | NA | NA | NA | NA |
| 513 | WA | NA | NA | NA | NA | WA | WA | WA | WA | WA | NA | WA | NA | WA | NA | NA | WA | WA | NA | NA | NA | WA | NA | NA | NA |
| 514 | WA | NA | NA | NA | NA | WA | WA | WA | WA | WA | NA | NA | NA | NA | WA | WA | NA | NA | WA | NA | NA | WA | NA | NA | NA |
| 515 | WA | NA | WA | NA | NA | NA | NA | WA | NA | NA | NA | WA | NA | NA | NA | NA | WA | NA | WA | NA | NA | WA | NA | NA | NA |
| 516 | WA | NA | NA | NA | NA | WA | WA | WA | WA | WA | NA | NA | NA | NA | WA | WA | NA | WA | WA | NA | NA | NA | NA | NA | NA |
| 517 | WA | WA | WA | NA | NA | NA | NA | WA | NA | NA | NA | NA | WA | NA | NA | NA | NA | NA | NA | NA | NA | WA | NA | NA | NA |
| 518 | WA | WA | NA | NA | NA | WA | WA | WA | WA | NA | NA | WA | NA | WA | WA | WA | NA | NA | NA | NA | NA | NA | NA | NA | NA |
| 519 | WA | NA | NA | NA | NA | WA | WA | WA | WA | NA | NA | NA | NA | NA | NA | NA | NA | NA | WA | NA | NA | WA | NA | NA | NA |
| 520 | WA | NA | NA | NA | NA | NA | NA | WA | NA | WA | NA | WA | NA | NA | NA | NA | NA | WA | NA | NA | NA | WA | NA | NA | NA |
| 521 | WA | NA | NA | NA | NA | WA | WA | WA | WA | NA | NA | WA | NA | NA | NA | NA | NA | NA | NA | NA | NA | WA | NA | NA | NA |
| 522 | WA | NA | NA | NA | NA | WA | WA | WA | WA | NA | NA | WA | NA | NA | NA | NA | NA | NA | WA | NA | NA | NA | NA | NA | NA |
| 523 | WA | NA | NA | NA | NA | NA | NA | WA | NA | NA | NA | WA | NA | WA | NA | NA | WA | NA | NA | NA | NA | WA | NA | NA | NA |
| 524 | WA | NA | NA | NA | NA | WA | WA | WA | WA | NA | NA | NA | WA | WA | NA | NA | WA | WA | NA | NA | NA | NA | NA | NA | NA |
| 525 | WA | NA | NA | NA | NA | NA | NA | WA | NA | WA | NA | WA | NA | WA | NA | NA | NA | NA | NA | NA | NA | NA | NA | NA | NA |
| 526 | WA | WA | NA | NA | NA | NA | NA | WA | WA | NA | NA | WA | NA | NA | WA | WA | NA | NA | WA | NA | NA | NA | NA | NA | NA |
| 527 | WA | WA | WA | NA | NA | WA | WA | WA | WA | WA | NA | NA | NA | NA | WA | WA | WA | NA | WA | NA | NA | WA | NA | NA | NA |
| 528 | WA | NA | WA | NA | NA | WA | WA | WA | WA | NA | NA | WA | NA | NA | NA | NA | WA | NA | NA | NA | WA | NA | NA | NA | NA |
| 529 | WA | WA | WA | NA | NA | WA | WA | WA | WA | NA | NA | NA | NA | NA | NA | NA | NA | WA | NA | NA | NA | NA | NA | NA | NA |
| 530 | WA | NA | NA | NA | NA | WA | WA | WA | WA | NA | NA | NA | NA | WA | NA | NA | WA | WA | NA | NA | NA | WA | NA | NA | WA |
| 531 | WA | NA | NA | NA | NA | WA | WA | WA | WA | NA | NA | NA | NA | WA | NA | NA | NA | NA | NA | NA | NA | WA | NA | NA | NA |
| 532 | WA | NA | NA | NA | NA | - | - | WA | - | NA | NA | WA | NA | - | - | - | - | NA | NA | NA | NA | NA | NA | NA | NA |
| 533 | WA | NA | NA | NA | NA | WA | WA | WA | WA | NA | NA | WA | NA | NA | NA | NA | NA | NA | NA | WA | NA | NA | WA | WA | WA |
| 534 | WA | NA | NA | NA | NA | WA | WA | WA | WA | NA | NA | WA | NA | WA | NA | NA | WA | NA | NA | NA | NA | NA | NA | NA | NA |
| 535 | WA | NA | NA | NA | NA | WA | WA | WA | WA | NA | NA | NA | NA | NA | WA | WA | NA | NA | NA | NA | WA | NA | WA | WA | NA |
| 536 | WA | NA | NA | NA | NA | WA | WA | WA | WA | WA | NA | NA | NA | WA | NA | NA | WA | WA | NA | NA | WA | NA | NA | NA | NA |
| 537 | WA | NA | WA | NA | NA | WA | WA | WA | WA | NA | NA | NA | NA | WA | NA | NA | WA | NA | NA | NA | NA | NA | NA | NA | WA |
| 538 | WA | NA | NA | NA | NA | NA | NA | WA | NA | NA | NA | WA | NA | WA | NA | NA | WA | NA | NA | NA | NA | WA | WA | WA | NA |
| 539 | WA | WA | NA | NA | NA | NA | NA | WA | WA | WA | NA | NA | WA | NA | NA | NA | WA | WA | NA | NA | NA | WA | NA | NA | NA |
| 540 | WA | NA | NA | NA | NA | NA | NA | WA | NA | NA | NA | NA | NA | NA | NA | NA | WA | WA | WA | NA | WA | WA | NA | NA | NA |
| 541 | WA | NA | NA | NA | NA | WA | WA | WA | WA | NA | NA | WA | NA | NA | WA | WA | NA | NA | NA | NA | NA | WA | NA | NA | NA |
| 542 | WA | NA | NA | NA | NA | WA | WA | WA | WA | NA | NA | WA | NA | NA | WA | WA | NA | NA | WA | NA | NA | WA | NA | NA | WA |
| 543 | WA | NA | NA | NA | NA | WA | WA | WA | WA | NA | NA | WA | NA | NA | NA | NA | NA | NA | WA | NA | NA | WA | NA | NA | NA |
| 544 | WA | NA | NA | NA | NA | NA | NA | WA | NA | NA | NA | WA | NA | NA | NA | NA | WA | NA | NA | NA | WA | WA | NA | NA | WA |
| 545 | WA | NA | WA | NA | NA | NA | NA | WA | WA | NA | NA | NA | NA | WA | NA | NA | WA | WA | NA | NA | NA | NA | NA | NA | NA |
| 546 | WA | NA | NA | NA | NA | WA | WA | WA | WA | NA | NA | NA | NA | NA | NA | NA | WA | WA | NA | NA | NA | NA | NA | NA | NA |
| 547 | WA | NA | WA | NA | NA | NA | NA | WA | NA | NA | NA | WA | NA | NA | NA | NA | WA | NA | NA | NA | NA | NA | NA | NA | WA |
| 548 | WA | NA | WA | NA | NA | WA | WA | WA | WA | WA | NA | NA | NA | WA | WA | WA | WA | NA | WA | NA | NA | WA | NA | NA | NA |
| 549 | WA | NA | NA | NA | NA | WA | WA | WA | WA | NA | NA | NA | NA | WA | WA | WA | NA | WA | NA | NA | NA | WA | WA | WA | NA |
| 550 | WA | NA | NA | NA | NA | WA | WA | WA | WA | WA | NA | NA | NA | WA | WA | WA | WA | NA | NA | NA | NA | WA | NA | NA | NA |
| 551 | WA | NA | NA | NA | NA | NA | NA | WA | NA | NA | NA | NA | NA | WA | NA | NA | NA | WA | NA | NA | NA | WA | NA | NA | WA |
| 552 | WA | NA | NA | NA | NA | NA | NA | WA | WA | NA | NA | NA | NA | WA | WA | WA | WA | WA | NA | NA | NA | NA | NA | NA | WA |
| 553 | WA | NA | WA | NA | NA | WA | WA | WA | NA | NA | NA | NA | NA | NA | WA | WA | NA | WA | NA | NA | NA | NA | NA | NA | WA |
| 554 | WA | NA | NA | NA | NA | WA | WA | WA | WA | NA | NA | NA | NA | NA | NA | NA | WA | WA | NA | NA | NA | WA | NA | NA | NA |
| 555 | WA | NA | WA | NA | NA | WA | WA | WA | WA | WA | NA | NA | NA | NA | NA | NA | WA | WA | WA | NA | NA | NA | NA | NA | NA |
| 556 | WA | NA | NA | NA | NA | NA | NA | WA | NA | NA | NA | NA | NA | NA | NA | NA | NA | WA | NA | NA | NA | WA | NA | NA | NA |
| 557 | WA | NA | NA | NA | NA | WA | WA | WA | WA | NA | NA | NA | NA | NA | WA | WA | WA | WA | WA | NA | NA | NA | NA | NA | WA |
| 558 | WA | NA | WA | NA | NA | WA | WA | WA | WA | NA | NA | NA | NA | NA | WA | NA | WA | NA | NA | NA | NA | NA | NA | NA | NA |
| 559 | WA | NA | WA | NA | NA | WA | WA | WA | WA | NA | NA | WA | NA | NA | WA | WA | NA | NA | WA | NA | NA | WA | NA | NA | WA |
| 560 | WA | NA | NA | NA | NA | NA | NA | WA | NA | NA | NA | NA | NA | NA | NA | NA | NA | WA | WA | WA | NA | NA | NA | NA | NA |
| 561 | WA | NA | WA | NA | NA | WA | WA | WA | WA | WA | NA | NA | NA | NA | WA | WA | NA | NA | NA | NA | WA | NA | NA | NA | NA |
| 562 | WA | NA | NA | NA | NA | WA | WA | WA | WA | WA | NA | WA | NA | NA | NA | NA | NA | WA | WA | NA | NA | WA | NA | NA | NA |
| 563 | WA | NA | WA | NA | NA | WA | WA | WA | WA | WA | NA | WA | NA | NA | NA | NA | WA | NA | NA | NA | NA | WA | NA | NA | NA |
| 564 | WA | NA | NA | NA | NA | WA | WA | WA | NA | NA | NA | NA | NA | - | NA | - | NA | NA | - | NA | NA | NA | NA | NA | WA |
| 565 | WA | NA | WA | NA | NA | NA | NA | WA | WA | NA | NA | NA | NA | WA | NA | NA | NA | WA | NA | NA | NA | NA | NA | NA | WA |
| 566 | WA | NA | NA | NA | NA | NA | NA | WA | NA | WA | NA | WA | NA | WA | WA | WA | NA | NA | WA | NA | NA | WA | NA | NA | NA |
| 567 | WA | WA | NA | NA | NA | WA | WA | WA | WA | NA | NA | WA | WA | NA | WA | WA | WA | NA | NA | NA | NA | NA | NA | NA | WA |
| 568 | WA | WA | WA | NA | NA | NA | NA | WA | NA | NA | NA | WA | NA | NA | NA | NA | WA | NA | NA | NA | NA | NA | WA | WA | NA |
| 569 | WA | WA | - | NA | NA | - | - | WA | - | WA | NA | - | NA | NA | NA | NA | NA | - | - | NA | NA | NA | NA | NA | - |
| 570 | WA | WA | WA | NA | NA | NA | NA | WA | NA | WA | NA | WA | NA | NA | NA | NA | NA | WA | NA | NA | NA | NA | NA | NA | NA |
| 571 | WA | NA | NA | NA | NA | WA | WA | WA | WA | WA | NA | NA | NA | NA | WA | WA | NA | WA | NA | NA | NA | NA | NA | NA | WA |
| 572 | WA | WA | NA | NA | NA | WA | WA | WA | WA | NA | NA | - | NA | NA | WA | - | NA | NA | WA | NA | NA | - | NA | NA | NA |
| 573 | WA | NA | WA | NA | NA | NA | NA | WA | NA | NA | NA | WA | NA | NA | NA | NA | WA | WA | WA | NA | NA | WA | NA | NA | WA |
| 574 | WA | NA | WA | NA | NA | NA | NA | WA | NA | WA | NA | NA | NA | NA | NA | NA | NA | NA | NA | NA | NA | NA | NA | NA | NA |
| 575 | WA | NA | NA | NA | NA | WA | WA | WA | WA | NA | NA | WA | NA | NA | NA | NA | NA | NA | - | NA | NA | NA | NA | NA | NA |
| 576 | WA | NA | NA | NA | NA | WA | WA | WA | - | NA | NA | - | NA | NA | NA | NA | NA | NA | WA | NA | NA | NA | NA | NA | NA |
| 577 | WA | NA | NA | NA | NA | NA | NA | WA | NA | NA | NA | NA | NA | NA | NA | NA | WA | NA | NA | NA | NA | NA | NA | NA | WA |
| 578 | WA | NA | NA | NA | NA | WA | WA | WA | WA | NA | NA | NA | NA | NA | WA | WA | NA | WA | NA | NA | NA | WA | NA | NA | NA |
| 579 | WA | WA | NA | NA | NA | WA | WA | WA | WA | NA | NA | NA | NA | NA | NA | NA | WA | NA | NA | NA | NA | NA | NA | NA | NA |
| 580 | WA | NA | WA | NA | NA | WA | WA | WA | WA | WA | NA | NA | NA | NA | NA | NA | WA | NA | NA | NA | NA | WA | NA | NA | NA |
| 581 | WA | WA | NA | NA | NA | NA | NA | WA | WA | NA | NA | NA | NA | NA | NA | NA | WA | WA | NA | NA | NA | WA | NA | NA | NA |
| 582 | WA | NA | WA | NA | NA | NA | NA | WA | WA | NA | NA | NA | NA | NA | WA | WA | WA | NA | NA | NA | NA | NA | NA | NA | NA |
| 583 | WA | WA | NA | NA | NA | WA | WA | WA | NA | NA | NA | WA | NA | NA | NA | NA | NA | WA | WA | NA | NA | NA | NA | NA | NA |
| 584 | WA | NA | NA | NA | NA | WA | WA | WA | WA | WA | NA | NA | NA | NA | NA | NA | NA | WA | NA | NA | WA | NA | NA | NA | WA |
| 585 | WA | NA | WA | NA | NA | WA | WA | WA | WA | WA | NA | WA | NA | NA | NA | NA | NA | WA | NA | NA | NA | NA | NA | NA | NA |
| 586 | WA | NA | WA | NA | NA | WA | WA | WA | NA | NA | NA | WA | NA | NA | NA | NA | NA | NA | NA | NA | NA | NA | NA | NA | NA |
| 587 | WA | NA | WA | NA | NA | WA | WA | WA | WA | NA | NA | NA | NA | NA | NA | NA | WA | WA | WA | NA | NA | NA | NA | NA | NA |
| 588 | WA | NA | WA | NA | NA | NA | NA | WA | NA | WA | NA | NA | NA | NA | NA | NA | NA | NA | NA | NA | NA | NA | NA | NA | NA |
| 589 | WA | NA | WA | NA | NA | NA | NA | WA | NA | WA | NA | NA | NA | NA | NA | NA | NA | NA | NA | NA | NA | NA | NA | NA | NA |
| 590 | WA | WA | NA | NA | NA | WA | WA | WA | WA | NA | NA | NA | NA | WA | NA | NA | NA | WA | NA | NA | NA | NA | NA | NA | NA |
| 591 | WA | NA | WA | NA | NA | NA | WA | WA | WA | NA | NA | NA | NA | WA | NA | NA | NA | NA | NA | NA | NA | NA | NA | NA | NA |
| 592 | WA | NA | NA | NA | NA | WA | WA | WA | - | - | NA | - | NA | NA | - | - | WA | NA | - | NA | WA | NA | NA | NA | NA |
| 593 | WA | - | NA | NA | NA | NA | NA | WA | NA | WA | NA | - | NA | - | - | - | NA | NA | NA | NA | NA | WA | NA | NA | - |
| 594 | WA | NA | WA | NA | NA | NA | NA | WA | NA | WA | NA | NA | NA | NA | NA | NA | NA | NA | WA | NA | NA | NA | NA | NA | NA |
| 595 | WA | NA | NA | NA | NA | NA | NA | WA | NA | WA | NA | NA | NA | NA | WA | WA | NA | WA | NA | NA | NA | NA | NA | NA | NA |
| 596 | WA | - | - | NA | NA | NA | NA | WA | - | NA | NA | WA | NA | WA | - | - | - | NA | NA | NA | NA | WA | NA | NA | NA |
| 597 | WA | - | - | NA | NA | - | - | WA | - | NA | NA | - | NA | - | NA | NA | NA | WA | NA | NA | NA | NA | NA | NA | - |
| 598 | WA | WA | NA | NA | NA | NA | NA | WA | WA | NA | NA | WA | NA | NA | NA | NA | NA | NA | NA | NA | NA | WA | WA | WA | NA |
| 599 | WA | NA | WA | NA | NA | WA | WA | WA | WA | NA | NA | WA | NA | NA | NA | NA | NA | NA | NA | NA | NA | NA | NA | NA | WA |
| 600 | WA | NA | WA | NA | NA | NA | NA | WA | WA | NA | NA | WA | NA | NA | WA | WA | NA | NA | NA | NA | NA | NA | NA | NA | NA |
| 601 | WA | NA | WA | NA | NA | WA | WA | WA | NA | NA | NA | WA | NA | NA | NA | NA | NA | WA | NA | NA | NA | WA | NA | NA | WA |

| | | | | | | | | | | | | | | | | | | | | | | | | |
|---|---|---|---|---|---|---|---|---|---|---|---|---|---|---|---|---|---|---|---|---|---|---|---|---|
| 602 WA | NA | NA | NA | NA | WA | WA | WA | WA | NA | NA | NA | NA | WA | WA | WA | WA | NA | NA | NA | WA | NA | NA | NA | NA |
| 603 WA | WA | NA | NA | NA | NA | NA | WA | WA | WA | NA | NA | WA | NA | NA | NA | WA | WA | NA | NA | NA | NA | NA | NA | NA |
| 604 WA | NA | WA | NA | NA | WA | WA | WA | WA | NA | NA | WA | NA | NA | WA | WA | NA | NA | NA | NA | WA | NA | NA | NA | NA |
| 605 WA | WA | NA | NA | NA | - | - | WA | WA | WA | NA | - | NA | NA | NA | NA | - | WA | - | NA | NA | - | NA | NA | NA |
| 606 WA | NA | WA | NA | NA | WA | WA | WA | WA | NA | NA | NA | NA | WA | NA | NA | WA | WA | NA | NA | NA | NA | NA | NA | NA |
| 607 WA | NA | NA | NA | NA | NA | NA | WA | - | - | NA | - | NA | NA | - | - | WA | NA | NA | NA | - | NA | NA | NA | NA |
| 608 WA | WA | NA | NA | NA | NA | NA | WA | NA | NA | NA | WA | NA | NA | NA | NA | WA | NA | NA | NA | NA | WA | NA | NA | NA |
| 609 WA | - | - | NA | NA | WA | WA | WA | WA | - | NA | NA | NA | - | - | - | WA | WA | NA | NA | NA | NA | NA | NA | NA |
| 610 WA | WA | NA | NA | NA | WA | WA | WA | WA | WA | NA | NA | NA | NA | WA | WA | NA | NA | NA | NA | NA | NA | WA | NA | NA |
| 611 WA | NA | NA | NA | NA | NA | NA | WA | NA | NA | NA | NA | NA | NA | NA | NA | WA | NA | NA | NA | NA | WA | NA | NA | NA |
| 612 WA | WA | NA | NA | NA | WA | WA | WA | WA | WA | NA | WA | NA | WA | WA | WA | NA | NA | WA | NA | NA | NA | NA | NA | NA |
| 613 WA | NA | NA | NA | NA | WA | WA | WA | WA | WA | NA | NA | WA | NA | NA | NA | NA | WA | NA | NA | NA | NA | NA | NA | NA |
| 614 WA | WA | NA | NA | NA | WA | WA | WA | WA | NA | NA | NA | NA | NA | WA | WA | WA | NA | WA | NA | NA | - | NA | NA | NA |
| 615 WA | NA | NA | NA | NA | NA | NA | WA | NA | NA | NA | NA | NA | WA | WA | WA | NA | NA | NA | NA | NA | WA | NA | NA | WA |
| 616 WA | NA | NA | NA | NA | WA | WA | WA | WA | NA | NA | WA | WA | NA | NA | NA | WA | WA | WA | NA | NA | NA | NA | NA | NA |
| 617 WA | NA | NA | NA | NA | NA | NA | WA | NA | NA | NA | NA | NA | WA | NA | NA | NA | WA | NA | NA | NA | WA | NA | NA | NA |
| 618 WA | NA | WA | NA | NA | WA | WA | WA | WA | NA | NA | NA | NA | WA | NA | NA | WA | WA | NA | NA | NA | NA | NA | NA | NA |
| 619 WA | NA | WA | NA | NA | WA | WA | WA | WA | NA | NA | WA | NA | NA | WA | WA | NA | NA | WA | NA | NA | NA | NA | NA | NA |
| 620 WA | NA | WA | NA | NA | NA | NA | WA | WA | NA | NA | NA | NA | NA | NA | NA | WA | NA | NA | NA | NA | WA | NA | NA | NA |
| 621 WA | NA | NA | NA | NA | WA | WA | WA | WA | WA | NA | WA | NA | NA | NA | NA | NA | WA | WA | NA | NA | NA | WA | WA | NA |
| 622 WA | NA | WA | NA | NA | WA | WA | WA | WA | NA | NA | WA | NA | NA | NA | NA | WA | WA | WA | NA | NA | NA | NA | NA | NA |
| 623 WA | NA | WA | NA | NA | WA | WA | WA | NA | NA | NA | NA | NA | NA | NA | NA | WA | WA | NA | NA | NA | WA | WA | WA | NA |
| 624 WA | WA | NA | NA | NA | NA | NA | WA | NA | NA | NA | NA | NA | WA | NA | NA | NA | WA | NA | NA | NA | NA | WA | WA | NA |
| 625 WA | WA | - | NA | NA | NA | NA | WA | NA | NA | NA | NA | - | NA | NA | NA | NA | NA | NA | NA | NA | WA | NA | NA | NA |
| 626 WA | WA | NA | NA | NA | NA | NA | WA | NA | NA | NA | WA | NA | WA | WA | WA | NA | NA | WA | NA | NA | NA | WA | NA | NA |
| 627 WA | WA | NA | NA | NA | NA | NA | WA | NA | NA | NA | WA | NA | NA | NA | NA | NA | WA | WA | NA | NA | NA | NA | NA | NA |
| 628 WA | NA | NA | NA | NA | NA | NA | WA | NA | WA | NA | WA | NA | NA | NA | NA | NA | WA | NA | NA | NA | WA | NA | NA | NA |
| 629 WA | NA | NA | NA | NA | WA | WA | WA | WA | NA | NA | WA | NA | NA | NA | NA | NA | NA | NA | NA | NA | WA | NA | NA | NA |
| 630 WA | NA | NA | NA | NA | WA | WA | WA | WA | NA | NA | NA | NA | NA | WA | WA | WA | NA | NA | NA | NA | NA | NA | NA | NA |
| 631 WA | NA | NA | NA | NA | NA | NA | WA | NA | NA | NA | WA | NA | WA | NA | NA | WA | NA | NA | NA | NA | WA | NA | NA | NA |
| 632 WA | NA | - | NA | NA | WA | WA | WA | WA | WA | NA | NA | - | WA | NA | NA | - | - | NA | NA | NA | NA | NA | NA | - |
| 633 WA | NA | NA | NA | NA | NA | NA | WA | NA | WA | NA | WA | NA | WA | NA | NA | NA | NA | - | NA | NA | - | NA | NA | NA |
| 634 WA | NA | NA | NA | NA | WA | WA | WA | WA | NA | NA | WA | NA | NA | NA | NA | WA | WA | NA | NA | NA | NA | NA | NA | NA |
| 635 WA | NA | WA | NA | NA | WA | WA | WA | WA | WA | NA | NA | NA | WA | NA | NA | NA | NA | NA | NA | WA | WA | WA | NA | NA |
| 636 WA | NA | NA | NA | NA | NA | NA | WA | NA | NA | NA | NA | NA | WA | NA | NA | WA | WA | WA | NA | NA | NA | NA | NA | NA |
| 637 WA | NA | NA | NA | NA | NA | NA | WA | NA | WA | NA | WA | NA | WA | NA | NA | NA | WA | NA | NA | NA | NA | NA | NA | NA |
| 638 WA | NA | NA | NA | NA | WA | WA | WA | WA | NA | NA | NA | NA | WA | NA | NA | WA | WA | NA | NA | NA | WA | NA | NA | WA |
| 639 WA | NA | NA | NA | NA | WA | WA | WA | WA | WA | NA | NA | NA | NA | NA | NA | NA | NA | NA | NA | NA | WA | NA | NA | NA |
| 640 WA | NA | - | NA | NA | WA | WA | WA | WA | NA | NA | - | NA | WA | NA | NA | NA | NA | NA | NA | NA | WA | NA | NA | NA |
| 641 WA | NA | NA | NA | NA | WA | WA | WA | WA | NA | NA | NA | NA | NA | WA | WA | WA | NA | NA | NA | NA | WA | NA | NA | WA |
| 642 WA | NA | WA | NA | NA | WA | WA | WA | WA | WA | NA | NA | WA | NA | NA | NA | NA | NA | NA | NA | NA | WA | NA | NA | WA |
| 643 WA | NA | NA | NA | NA | WA | WA | WA | WA | NA | NA | NA | NA | NA | WA | WA | NA | NA | NA | NA | WA | NA | WA | WA | NA |
| 644 WA | NA | NA | NA | NA | WA | WA | WA | WA | WA | NA | NA | NA | WA | NA | NA | WA | WA | NA | NA | WA | NA | NA | NA | NA |
| 645 WA | NA | WA | NA | NA | WA | WA | WA | WA | NA | NA | NA | NA | WA | NA | NA | WA | NA | NA | NA | NA | - | - | NA | WA |
| 646 WA | NA | NA | NA | NA | NA | NA | WA | - | NA | NA | - | NA | - | NA | NA | - | NA | NA | NA | NA | WA | - | - | NA |
| 647 WA | NA | WA | NA | NA | NA | NA | WA | NA | NA | NA | NA | NA | WA | NA | NA | NA | NA | NA | NA | NA | WA | NA | NA | NA |
| 648 WA | WA | WA | NA | NA | WA | WA | WA | WA | NA | NA | NA | NA | NA | NA | NA | NA | NA | NA | NA | NA | NA | NA | NA | NA |
| 649 WA | NA | NA | NA | NA | NA | NA | WA | NA | NA | NA | WA | NA | NA | NA | NA | NA | NA | NA | NA | NA | NA | NA | NA | WA |
| 650 WA | WA | NA | NA | NA | WA | WA | WA | WA | NA | NA | NA | NA | WA | NA | NA | WA | WA | NA | NA | NA | WA | NA | NA | WA |
| 651 WA | NA | NA | NA | NA | WA | WA | WA | WA | NA | NA | NA | WA | NA | NA | NA | WA | NA | NA | NA | NA | NA | NA | NA | NA |
| 652 WA | WA | NA | NA | NA | WA | WA | WA | WA | WA | NA | NA | NA | NA | NA | NA | NA | WA | NA | NA | NA | NA | NA | NA | NA |
| 653 WA | NA | NA | NA | NA | WA | WA | WA | WA | NA | NA | NA | NA | NA | WA | WA | WA | NA | NA | NA | NA | WA | NA | NA | NA |
| 654 WA | WA | WA | NA | NA | - | - | WA | NA | NA | NA | WA | NA | NA | WA | WA | WA | NA | NA | NA | NA | - | NA | NA | NA |
| 655 WA | WA | NA | NA | NA | NA | NA | WA | NA | NA | NA | WA | NA | NA | NA | NA | WA | NA | NA | NA | NA | NA | NA | NA | NA |
| 656 WA | NA | NA | NA | NA | NA | NA | WA | NA | NA | NA | WA | NA | NA | NA | NA | WA | NA | WA | NA | NA | NA | NA | NA | NA |
| 657 WA | WA | NA | NA | NA | WA | WA | WA | WA | NA | NA | NA | NA | WA | NA | NA | WA | NA | - | NA | WA | NA | NA | NA | NA |
| 658 WA | NA | NA | NA | NA | WA | WA | WA | WA | NA | NA | NA | NA | NA | WA | WA | NA | NA | WA | NA | NA | WA | NA | NA | NA |
| 659 WA | NA | NA | NA | NA | WA | WA | WA | WA | WA | NA | WA | NA | NA | WA | WA | WA | NA | NA | NA | NA | WA | NA | NA | NA |
| 660 WA | NA | NA | NA | NA | WA | WA | WA | WA | WA | NA | WA | NA | NA | WA | WA | NA | NA | NA | NA | NA | WA | NA | NA | NA |
| 661 WA | NA | NA | NA | NA | WA | WA | WA | WA | NA | NA | NA | WA | NA | NA | NA | WA | NA | WA | NA | NA | WA | NA | NA | NA |
| 662 WA | NA | WA | NA | NA | WA | WA | WA | WA | NA | NA | WA | NA | WA | NA | NA | WA | WA | NA | NA | NA | WA | NA | NA | WA |
| 663 WA | NA | WA | NA | NA | WA | WA | WA | WA | WA | NA | NA | NA | NA | NA | NA | WA | NA | NA | NA | NA | NA | WA | WA | NA |
| 664 WA | NA | NA | NA | NA | NA | NA | WA | NA | NA | NA | NA | WA | WA | WA | WA | NA | WA | NA | NA | NA | NA | NA | NA | NA |
| 665 WA | NA | NA | NA | NA | WA | WA | WA | WA | WA | NA | NA | NA | NA | NA | NA | WA | WA | NA | NA | NA | WA | NA | NA | NA |
| 666 WA | NA | WA | NA | NA | WA | WA | WA | WA | WA | NA | WA | NA | NA | NA | NA | WA | NA | NA | NA | NA | NA | NA | NA | WA |
| 667 WA | NA | NA | NA | NA | NA | NA | WA | NA | WA | NA | NA | NA | WA | NA | NA | WA | WA | WA | NA | NA | WA | NA | NA | NA |
| 668 WA | NA | NA | NA | NA | NA | NA | WA | NA | WA | NA | NA | NA | NA | WA | WA | NA | NA | WA | NA | NA | NA | NA | NA | NA |
| 669 WA | NA | WA | NA | NA | NA | NA | WA | NA | WA | NA | WA | NA | WA | NA | NA | NA | NA | NA | NA | NA | WA | NA | NA | NA |
| 670 WA | - | NA | NA | NA | WA | WA | WA | WA | - | NA | NA | NA | - | - | - | WA | NA | NA | NA | - | - | - | NA | NA |
| 671 WA | NA | WA | NA | NA | WA | WA | WA | WA | WA | NA | NA | NA | NA | NA | NA | NA | WA | WA | NA | NA | NA | NA | NA | NA |
| 672 WA | NA | - | NA | NA | WA | WA | WA | WA | WA | NA | WA | NA | NA | WA | WA | - | - | - | NA | NA | WA | NA | NA | NA |
| 673 WA | NA | NA | NA | NA | WA | WA | WA | WA | WA | NA | NA | NA | WA | NA | NA | WA | WA | NA | NA | WA | NA | NA | NA | NA |
| 674 WA | WA | NA | NA | NA | WA | WA | WA | WA | NA | NA | NA | NA | WA | NA | NA | NA | WA | NA | NA | WA | WA | NA | NA | NA |
| 675 WA | NA | NA | NA | NA | NA | NA | WA | NA | NA | NA | NA | NA | NA | WA | WA | NA | NA | NA | NA | NA | WA | NA | NA | WA |
| 676 WA | WA | WA | NA | NA | NA | NA | WA | NA | WA | NA | NA | NA | NA | NA | NA | WA | WA | NA | NA | NA | NA | NA | NA | NA |
| 677 WA | NA | WA | NA | NA | NA | NA | WA | NA | NA | NA | WA | NA | NA | WA | WA | WA | WA | WA | NA | WA | NA | NA | NA | NA |
| 678 WA | NA | WA | NA | NA | WA | WA | WA | WA | WA | NA | WA | NA | NA | NA | NA | WA | NA | NA | NA | NA | WA | NA | NA | NA |
| 679 WA | NA | NA | NA | NA | WA | WA | WA | WA | WA | NA | NA | NA | NA | NA | NA | NA | WA | NA | NA | NA | NA | NA | NA | WA |
| 680 WA | WA | WA | NA | NA | WA | WA | WA | WA | NA | NA | NA | NA | WA | NA | NA | WA | NA | WA | NA | NA | NA | NA | NA | NA |
| 681 WA | NA | WA | NA | NA | WA | WA | WA | WA | NA | NA | NA | NA | WA | NA | NA | NA | NA | WA | NA | NA | NA | NA | NA | NA |
| 682 WA | NA | WA | NA | NA | WA | WA | WA | WA | WA | NA | NA | NA | WA | NA | NA | NA | NA | NA | NA | NA | WA | NA | NA | NA |
| 683 WA | NA | NA | NA | NA | WA | WA | WA | WA | WA | NA | WA | NA | WA | WA | WA | WA | NA | NA | NA | NA | NA | NA | NA | NA |
| 684 WA | NA | NA | NA | NA | WA | WA | WA | WA | NA | NA | WA | NA | NA | WA | WA | NA | WA | WA | NA | WA | WA | NA | NA | NA |
| 685 WA | NA | NA | NA | NA | WA | WA | WA | WA | WA | NA | WA | WA | NA | NA | NA | NA | NA | NA | NA | NA | WA | NA | NA | NA |
| 686 WA | WA | WA | NA | NA | WA | WA | WA | WA | NA | NA | NA | NA | NA | WA | WA | NA | WA | NA | NA | NA | WA | NA | NA | NA |
| 687 WA | NA | NA | NA | NA | WA | WA | WA | WA | WA | NA | WA | WA | NA | WA | WA | WA | WA | WA | NA | NA | WA | NA | NA | NA |
| 688 WA | NA | WA | NA | NA | NA | NA | WA | NA | NA | NA | WA | NA | NA | NA | NA | NA | NA | NA | NA | NA | NA | NA | NA | NA |
| 689 WA | NA | NA | NA | NA | WA | WA | WA | WA | NA | NA | WA | NA | NA | NA | NA | WA | WA | NA | NA | NA | NA | NA | NA | NA |
| 690 WA | NA | WA | NA | NA | WA | WA | WA | WA | NA | NA | WA | WA | NA | NA | NA | NA | NA | NA | NA | NA | NA | NA | NA | NA |
| 691 WA | NA | NA | NA | NA | WA | WA | WA | WA | WA | NA | WA | NA | WA | NA | NA | NA | NA | NA | NA | NA | WA | NA | NA | NA |
| 692 WA | WA | WA | NA | NA | WA | WA | WA | WA | NA | NA | NA | NA | NA | WA | WA | WA | WA | NA | NA | NA | WA | NA | NA | NA |
| 693 WA | WA | WA | NA | NA | NA | NA | WA | NA | WA | NA | WA | NA | NA | NA | NA | NA | NA | WA | NA | NA | NA | NA | NA | NA |
| 694 WA | NA | NA | NA | NA | WA | WA | WA | WA | NA | NA | NA | NA | NA | NA | NA | NA | WA | NA | NA | NA | NA | WA | WA | NA |
| 695 WA | NA | NA | NA | NA | WA | WA | WA | WA | NA | NA | NA | NA | NA | NA | NA | NA | WA | WA | NA | NA | WA | NA | NA | NA |
| 696 WA | NA | NA | NA | NA | NA | NA | WA | NA | WA | NA | WA | NA | WA | WA | WA | WA | NA | NA | NA | NA | NA | NA | NA | WA |
| 697 WA | NA | NA | NA | NA | WA | WA | WA | WA | NA | NA | NA | NA | NA | NA | NA | NA | WA | NA | NA | NA | WA | NA | NA | NA |
| 698 WA | NA | NA | NA | NA | WA | WA | WA | WA | WA | NA | NA | NA | NA | WA | WA | WA | NA | NA | NA | NA | NA | NA | NA | NA |
| 699 WA | NA | NA | NA | NA | WA | WA | WA | WA | NA | NA | NA | NA | WA | NA | NA | WA | WA | NA | NA | NA | WA | NA | NA | NA |
| 700 WA | WA | WA | NA | NA | WA | WA | WA | WA | WA | NA | NA | NA | WA | NA | NA | NA | NA | NA | NA | NA | WA | NA | NA | NA |
| 701 WA | NA | NA | NA | NA | WA | WA | WA | WA | WA | NA | WA | NA | NA | NA | NA | WA | NA | NA | NA | NA | NA | NA | NA | NA |
| 702 WA | NA | WA | NA | NA | WA | WA | WA | NA | WA | NA | WA | NA | WA | WA | WA | NA | NA | WA | NA | NA | NA | NA | NA | NA |
| 703 WA | NA | NA | NA | NA | WA | WA | WA | WA | NA | NA | NA | NA | NA | NA | NA | NA | WA | WA | NA | NA | WA | NA | NA | WA |
| 704 WA | NA | WA | NA | NA | WA | WA | WA | WA | NA | NA | WA | WA | WA | WA | WA | WA | WA | WA | NA | NA | WA | NA | NA | NA |
| 705 WA | NA | WA | NA | NA | NA | NA | WA | NA | NA | NA | NA | WA | WA | NA | NA | WA | NA | NA | NA | NA | NA | NA | NA | NA |
| 706 WA | NA | NA | NA | NA | WA | NA | WA | WA | NA | NA | NA | NA | NA | WA | WA | WA | NA | NA | WA | WA | NA | NA | NA | NA |
| 707 WA | NA | NA | NA | NA | WA | WA | WA | WA | NA | NA | NA | NA | NA | WA | WA | NA | WA | NA | NA | NA | WA | NA | NA | NA |
| 708 WA | NA | WA | NA | NA | WA | WA | WA | WA | NA | NA | NA | NA | NA | WA | WA | NA | NA | NA | NA | NA | WA | NA | NA | NA |
| 709 WA | NA | WA | NA | NA | WA | WA | WA | WA | NA | NA | NA | NA | NA | NA | NA | WA | NA | NA | NA | NA | WA | NA | NA | NA |
| 710 WA | NA | WA | NA | NA | NA | NA | WA | NA | NA | NA | NA | NA | WA | NA | NA | NA | NA | WA | NA | NA | NA | NA | NA | WA |
| 711 WA | NA | NA | NA | NA | WA | WA | WA | WA | WA | NA | WA | NA | NA | NA | NA | WA | NA | NA | NA | NA | NA | NA | NA | WA |
| 712 WA | NA | NA | NA | NA | WA | WA | WA | WA | - | NA | WA | NA | NA | - | - | - | NA | NA | NA | NA | NA | NA | NA | NA |
| 713 WA | NA | WA | NA | NA | WA | WA | WA | WA | WA | NA | NA | NA | NA | NA | NA | WA | WA | NA | NA | NA | NA | NA | NA | NA |
| 714 WA | WA | WA | NA | NA | WA | WA | WA | WA | NA | NA | NA | NA | NA | NA | NA | WA | WA | NA | NA | NA | NA | NA | NA | NA |
| 715 WA | WA | WA | NA | NA | WA | WA | WA | WA | NA | NA | - | NA | WA | NA | NA | NA | WA | NA | NA | NA | NA | NA | NA | NA |
| 716 WA | NA | NA | NA | NA | WA | WA | WA | WA | WA | NA | NA | NA | NA | NA | NA | WA | WA | NA | NA | NA | NA | NA | NA | WA |
| 717 WA | NA | NA | NA | NA | WA | WA | WA | WA | NA | NA | NA | NA | NA | NA | NA | WA | NA | NA | NA | NA | WA | NA | NA | WA |
| 718 WA | NA | NA | NA | NA | WA | WA | WA | WA | NA | NA | NA | NA | WA | NA | NA | NA | WA | NA | NA | NA | NA | NA | NA | NA |
| 719 WA | NA | NA | NA | NA | WA | WA | WA | WA | WA | NA | NA | NA | NA | - | - | WA | - | - | NA | NA | NA | NA | NA | NA |
| 720 WA | WA | - | - | NA | WA | - | WA | - | NA | NA | - | NA | NA | - | - | NA | - | - | NA | - | NA | NA | - | NA |
| 721 WA | NA | NA | NA | NA | WA | WA | WA | WA | NA | NA | NA | NA | WA | NA | NA | NA | NA | NA | NA | NA | NA | WA | NA | NA |
| 722 WA | NA | WA | NA | NA | WA | WA | WA | WA | WA | NA | NA | NA | WA | NA | NA | NA | WA | NA | WA | WA | WA | NA | NA | NA |

| | | | | | | | | | | | | | | | | | | | | | | | | | |
|---|---|---|---|---|---|---|---|---|---|---|---|---|---|---|---|---|---|---|---|---|---|---|---|---|---|
| 723 | WA | NA | WA | NA | NA | WA | WA | WA | WA | NA | NA | NA | NA | WA | NA | NA | NA | NA | NA | NA | NA | NA | NA | NA | NA |
| 724 | WA | WA | NA | NA | NA | WA | WA | WA | WA | WA | NA | - | NA | NA | NA | NA | NA | NA | NA | NA | WA | WA | NA | NA | NA |
| 725 | WA | NA | WA | NA | NA | NA | NA | WA | WA | WA | NA | NA | NA | NA | NA | NA | NA | NA | NA | WA | NA | WA | NA | NA | NA |
| 726 | WA | NA | NA | NA | NA | WA | WA | WA | WA | NA | NA | NA | NA | NA | NA | NA | WA | NA | NA | WA | NA | WA | NA | NA | WA |
| 727 | WA | WA | NA | NA | NA | WA | WA | WA | WA | WA | NA | NA | NA | NA | NA | NA | WA | NA | WA | NA | NA | WA | NA | NA | NA |
| 728 | WA | NA | NA | NA | NA | WA | WA | WA | WA | NA | NA | WA | NA | NA | NA | NA | NA | NA | NA | NA | NA | NA | NA | NA | NA |
| 729 | WA | WA | NA | NA | NA | WA | WA | WA | WA | WA | NA | WA | NA | NA | NA | NA | NA | WA | WA | NA | NA | NA | NA | NA | NA |
| 730 | WA | NA | NA | NA | NA | WA | WA | WA | WA | NA | NA | NA | NA | NA | WA | WA | WA | NA | NA | NA | NA | NA | NA | NA | NA |
| 731 | WA | NA | NA | NA | NA | WA | WA | WA | WA | NA | NA | NA | NA | WA | NA | NA | WA | WA | NA | NA | NA | NA | NA | NA | NA |
| 732 | WA | NA | NA | NA | NA | WA | WA | WA | WA | NA | NA | WA | NA | NA | WA | WA | NA | WA | NA | NA | NA | NA | NA | NA | WA |
| 733 | WA | NA | NA | NA | NA | NA | NA | WA | WA | NA | NA | WA | NA | WA | WA | NA | NA | NA | NA | NA | NA | WA | NA | NA | NA |
| 734 | WA | NA | WA | NA | NA | NA | NA | WA | NA | WA | NA | NA | NA | NA | NA | NA | NA | NA | NA | NA | NA | NA | NA | NA | NA |
| 735 | WA | NA | NA | NA | NA | WA | WA | WA | WA | NA | NA | WA | WA | NA | WA | WA | WA | NA | NA | NA | NA | NA | NA | NA | NA |
| 736 | WA | NA | NA | NA | NA | WA | WA | WA | WA | WA | NA | NA | NA | WA | NA | NA | NA | NA | NA | NA | NA | WA | NA | NA | NA |
| 737 | WA | NA | NA | NA | NA | NA | NA | WA | NA | NA | NA | WA | NA | NA | WA | WA | NA | NA | NA | NA | NA | WA | NA | NA | NA |
| 738 | WA | NA | NA | NA | NA | NA | NA | WA | NA | WA | NA | NA | NA | WA | WA | WA | NA | NA | NA | NA | NA | WA | NA | NA | NA |
| 739 | WA | NA | NA | NA | NA | NA | NA | WA | NA | NA | NA | WA | NA | NA | NA | NA | NA | NA | NA | NA | NA | NA | NA | NA | NA |
| 740 | WA | WA | NA | NA | NA | NA | NA | WA | NA | NA | NA | NA | NA | WA | NA | NA | NA | NA | NA | NA | NA | NA | NA | NA | NA |
| 741 | WA | NA | NA | NA | NA | NA | NA | WA | NA | NA | NA | WA | NA | NA | NA | NA | WA | NA | WA | NA | NA | NA | NA | NA | NA |
| 742 | WA | NA | NA | NA | NA | WA | WA | WA | WA | NA | NA | NA | NA | NA | NA | NA | NA | WA | NA | NA | NA | WA | WA | WA | NA |
| 743 | WA | NA | NA | NA | NA | WA | WA | WA | WA | NA | NA | NA | NA | WA | WA | WA | WA | WA | NA | NA | NA | NA | NA | NA | WA |
| 744 | WA | NA | NA | NA | NA | NA | NA | WA | NA | NA | NA | WA | WA | NA | NA | NA | WA | WA | WA | WA | WA | WA | NA | NA | NA |
| 745 | WA | NA | NA | NA | NA | WA | WA | WA | NA | WA | NA | WA | NA | NA | WA | WA | WA | WA | NA | NA | NA | NA | NA | NA | NA |
| 746 | WA | WA | NA | NA | NA | WA | WA | WA | WA | NA | NA | NA | NA | WA | NA | NA | NA | NA | NA | NA | NA | WA | NA | NA | NA |
| 747 | WA | WA | NA | NA | NA | WA | WA | WA | WA | WA | NA | WA | NA | WA | NA | NA | NA | WA | NA | NA | NA | WA | NA | NA | NA |
| 748 | WA | NA | WA | NA | NA | WA | WA | WA | WA | NA | NA | NA | NA | NA | WA | WA | NA | WA | WA | NA | NA | NA | NA | NA | NA |
| 749 | WA | NA | NA | NA | NA | NA | NA | WA | NA | WA | NA | NA | NA | WA | WA | WA | NA | WA | WA | NA | NA | WA | NA | NA | NA |
| 750 | WA | NA | NA | NA | NA | WA | WA | WA | WA | NA | NA | WA | NA | NA | WA | WA | NA | NA | NA | NA | NA | WA | NA | NA | NA |
| 751 | WA | NA | NA | NA | NA | WA | WA | WA | WA | NA | NA | WA | NA | NA | WA | WA | WA | NA | WA | NA | NA | WA | NA | NA | NA |
| 752 | WA | WA | NA | NA | NA | WA | WA | WA | WA | NA | NA | NA | NA | NA | WA | WA | NA | NA | NA | NA | NA | WA | NA | NA | NA |
| 753 | WA | NA | NA | NA | NA | WA | WA | WA | WA | WA | NA | NA | NA | WA | NA | NA | NA | NA | WA | NA | NA | WA | NA | NA | NA |
| 754 | WA | NA | NA | NA | NA | WA | WA | WA | WA | NA | NA | NA | NA | NA | NA | WA | NA | WA | NA | NA | NA | WA | NA | NA | NA |
| 755 | WA | WA | NA | NA | NA | WA | WA | WA | WA | NA | NA | WA | NA | NA | NA | NA | NA | WA | NA | NA | NA | NA | NA | NA | NA |
| 756 | WA | NA | NA | NA | NA | WA | WA | WA | WA | NA | NA | WA | NA | WA | NA | NA | NA | WA | NA | NA | NA | NA | NA | NA | NA |
| 757 | WA | NA | WA | NA | NA | NA | WA | WA | WA | WA | NA | WA | NA | NA | WA | WA | NA | NA | NA | WA | NA | WA | NA | NA | NA |
| 758 | WA | WA | WA | NA | NA | NA | NA | WA | NA | WA | NA | NA | NA | NA | NA | NA | NA | WA | WA | NA | NA | NA | NA | NA | NA |
| 759 | WA | NA | NA | NA | NA | WA | WA | WA | WA | NA | NA | WA | NA | NA | NA | NA | WA | WA | NA | NA | NA | NA | NA | NA | WA |
| 760 | WA | WA | NA | NA | NA | NA | NA | WA | NA | NA | NA | WA | NA | NA | NA | NA | NA | NA | WA | NA | NA | NA | NA | NA | NA |
| 761 | WA | NA | WA | NA | NA | WA | WA | WA | WA | WA | NA | WA | NA | WA | WA | WA | WA | WA | WA | NA | NA | WA | NA | NA | NA |
| 762 | WA | NA | NA | NA | NA | WA | WA | WA | WA | NA | NA | WA | NA | NA | NA | NA | WA | WA | NA | NA | NA | WA | WA | WA | NA |
| 763 | WA | NA | WA | NA | NA | NA | NA | WA | NA | NA | NA | NA | NA | NA | NA | NA | WA | NA | NA | NA | WA | NA | NA | NA | WA |
| 764 | WA | NA | NA | NA | NA | WA | WA | WA | WA | NA | NA | NA | NA | WA | NA | NA | NA | NA | WA | NA | NA | NA | NA | NA | WA |
| 765 | WA | WA | WA | NA | NA | WA | WA | WA | WA | WA | NA | WA | NA | NA | WA | WA | WA | WA | NA | NA | NA | WA | NA | NA | NA |
| 766 | WA | NA | NA | NA | NA | WA | WA | WA | WA | NA | NA | NA | WA | NA | WA | WA | NA | NA | WA | NA | NA | NA | NA | NA | NA |
| 767 | WA | NA | NA | NA | NA | WA | WA | WA | WA | NA | NA | NA | NA | NA | WA | WA | WA | WA | WA | WA | NA | NA | NA | NA | WA |
| 768 | WA | WA | WA | NA | NA | NA | NA | WA | NA | WA | NA | NA | NA | NA | NA | NA | NA | WA | WA | NA | NA | NA | NA | NA | NA |
| 769 | WA | NA | WA | NA | NA | WA | WA | WA | WA | NA | NA | NA | NA | WA | NA | NA | NA | NA | NA | NA | NA | NA | NA | NA | WA |
| 770 | WA | NA | WA | NA | NA | WA | WA | WA | WA | NA | NA | WA | NA | WA | WA | WA | WA | NA | NA | NA | NA | NA | NA | NA | NA |
| 771 | WA | NA | NA | NA | NA | WA | WA | WA | WA | NA | NA | NA | NA | NA | NA | NA | WA | WA | NA | NA | NA | NA | NA | NA | NA |
| 772 | WA | NA | NA | NA | NA | WA | WA | WA | WA | NA | NA | NA | NA | NA | NA | NA | WA | WA | NA | NA | NA | NA | NA | NA | NA |
| 773 | WA | NA | NA | NA | NA | WA | WA | WA | WA | NA | NA | NA | NA | NA | NA | NA | NA | NA | NA | NA | NA | NA | NA | NA | WA |
| 774 | WA | NA | WA | NA | NA | WA | WA | WA | WA | NA | NA | NA | NA | WA | NA | NA | WA | NA | NA | NA | NA | NA | NA | NA | NA |
| 775 | WA | NA | NA | NA | NA | WA | WA | WA | WA | NA | NA | NA | NA | NA | NA | NA | NA | NA | NA | NA | NA | NA | NA | NA | WA |
| 776 | WA | NA | NA | NA | NA | WA | WA | WA | WA | NA | NA | NA | NA | WA | NA | NA | NA | NA | NA | NA | NA | NA | NA | NA | NA |
| 777 | WA | NA | NA | NA | NA | NA | NA | WA | NA | WA | NA | NA | NA | NA | NA | NA | NA | WA | WA | NA | NA | NA | NA | NA | WA |
| 778 | WA | NA | WA | NA | NA | NA | NA | WA | NA | NA | NA | NA | NA | WA | WA | WA | NA | NA | NA | NA | NA | WA | NA | NA | WA |
| 779 | WA | NA | WA | NA | NA | WA | WA | WA | WA | NA | NA | NA | NA | WA | NA | NA | WA | WA | WA | NA | NA | WA | NA | NA | NA |
| 780 | WA | NA | WA | NA | NA | NA | NA | WA | NA | WA | NA | WA | WA | NA | NA | NA | NA | WA | NA | NA | NA | WA | NA | NA | NA |
| 781 | WA | NA | NA | NA | NA | WA | WA | WA | WA | WA | NA | WA | WA | NA | NA | NA | NA | NA | WA | NA | NA | NA | NA | NA | NA |
| 782 | WA | NA | NA | NA | NA | NA | NA | WA | NA | NA | NA | WA | NA | WA | NA | NA | WA | NA | NA | NA | NA | WA | WA | WA | NA |
| 783 | WA | NA | WA | NA | NA | WA | WA | WA | WA | NA | NA | WA | WA | NA | NA | NA | NA | NA | NA | NA | NA | NA | NA | NA | NA |
| 784 | WA | NA | NA | NA | NA | WA | WA | WA | WA | WA | NA | NA | NA | NA | NA | NA | NA | NA | NA | NA | NA | NA | NA | NA | NA |
| 785 | WA | NA | NA | NA | NA | WA | WA | WA | WA | NA | NA | WA | NA | NA | WA | WA | NA | NA | WA | NA | NA | NA | NA | NA | NA |
| 786 | WA | NA | WA | NA | NA | NA | NA | WA | NA | NA | NA | WA | NA | NA | NA | NA | NA | WA | NA | NA | NA | NA | NA | NA | NA |
| 787 | WA | NA | WA | NA | NA | WA | WA | WA | WA | NA | NA | WA | NA | WA | NA | NA | NA | NA | WA | NA | NA | WA | WA | NA | NA |
| 788 | WA | NA | NA | NA | NA | WA | WA | WA | WA | NA | NA | NA | NA | WA | WA | WA | WA | NA | WA | NA | NA | WA | NA | NA | NA |
| 789 | WA | NA | NA | NA | NA | WA | WA | WA | WA | NA | NA | WA | NA | WA | NA | NA | WA | WA | NA | NA | NA | WA | NA | NA | NA |
| 790 | WA | NA | NA | NA | NA | WA | WA | WA | WA | NA | NA | NA | WA | NA | NA | NA | NA | WA | NA | NA | WA | NA | NA | NA | NA |
| 791 | WA | NA | WA | NA | NA | WA | WA | WA | WA | WA | NA | NA | NA | WA | WA | WA | WA | NA | NA | NA | NA | WA | NA | NA | NA |
| 792 | WA | NA | WA | NA | NA | NA | NA | WA | NA | NA | NA | WA | NA | NA | NA | NA | NA | NA | NA | NA | NA | WA | WA | WA | WA |
| 793 | WA | NA | NA | NA | NA | WA | WA | WA | WA | NA | NA | NA | NA | WA | WA | WA | NA | NA | NA | NA | NA | NA | NA | NA | NA |
| 794 | WA | NA | WA | NA | NA | NA | NA | WA | NA | WA | NA | NA | NA | WA | NA | NA | WA | WA | NA | NA | NA | NA | NA | NA | NA |
| 795 | WA | NA | WA | NA | NA | NA | NA | WA | NA | WA | NA | NA | NA | NA | WA | WA | NA | NA | NA | NA | NA | WA | NA | NA | NA |
| 796 | WA | NA | WA | NA | NA | WA | WA | WA | WA | NA | NA | NA | NA | NA | NA | NA | NA | WA | WA | WA | NA | WA | NA | NA | NA |
| 797 | WA | NA | NA | NA | NA | WA | WA | WA | WA | WA | NA | NA | NA | NA | NA | NA | NA | NA | NA | NA | NA | WA | NA | NA | NA |
| 798 | WA | NA | NA | NA | NA | WA | WA | WA | WA | NA | NA | NA | NA | WA | WA | WA | WA | NA | NA | NA | WA | NA | NA | NA | NA |
| 799 | WA | WA | WA | NA | NA | NA | NA | WA | NA | WA | NA | NA | NA | NA | NA | NA | WA | WA | NA | NA | NA | NA | NA | NA | NA |
| 800 | - | NA | NA | NA | NA | - | NA | WA | NA | - | NA | WA | NA | WA | NA | NA | WA | NA | - | NA | NA | NA | NA | - | NA |
| 801 | WA | WA | WA | NA | NA | WA | WA | WA | WA | WA | NA | NA | NA | WA | WA | WA | NA | WA | WA | NA | NA | NA | NA | NA | NA |
| 802 | WA | NA | WA | NA | NA | NA | NA | WA | NA | NA | NA | NA | NA | WA | NA | NA | NA | WA | WA | NA | NA | WA | NA | NA | NA |
| 803 | WA | NA | NA | NA | NA | NA | NA | WA | WA | WA | NA | WA | NA | WA | NA | NA | NA | NA | WA | NA | NA | NA | NA | NA | NA |
| 804 | WA | NA | NA | NA | NA | WA | WA | WA | WA | NA | NA | WA | NA | NA | NA | NA | NA | NA | WA | NA | NA | WA | NA | NA | NA |
| 805 | WA | NA | NA | NA | NA | WA | WA | WA | WA | NA | NA | WA | NA | WA | WA | WA | WA | NA | WA | NA | NA | WA | NA | NA | NA |
| 806 | WA | NA | NA | NA | NA | WA | WA | WA | WA | NA | NA | NA | NA | WA | WA | WA | NA | NA | WA | NA | NA | NA | NA | NA | NA |
| 807 | WA | NA | WA | NA | NA | WA | NA | WA | WA | NA | NA | WA | NA | NA | NA | NA | WA | WA | NA | NA | NA | WA | NA | NA | NA |
| 808 | WA | WA | NA | NA | NA | NA | NA | WA | NA | NA | NA | NA | NA | NA | NA | NA | NA | NA | NA | NA | NA | WA | NA | NA | NA |
| 809 | WA | NA | WA | NA | NA | NA | NA | WA | NA | NA | NA | NA | NA | NA | NA | NA | NA | NA | NA | NA | NA | WA | NA | NA | NA |
| 810 | WA | NA | NA | NA | NA | NA | NA | WA | NA | NA | NA | WA | NA | WA | NA | NA | NA | NA | NA | NA | NA | NA | WA | WA | NA |
| 811 | WA | WA | NA | NA | NA | NA | NA | WA | WA | WA | NA | - | NA | NA | NA | NA | NA | NA | NA | NA | NA | NA | NA | NA | NA |
| 812 | WA | NA | NA | NA | NA | NA | NA | WA | NA | WA | NA | WA | NA | NA | WA | WA | NA | WA | NA | NA | NA | WA | NA | NA | NA |
| 813 | WA | NA | NA | NA | NA | WA | WA | WA | WA | NA | NA | NA | NA | NA | WA | WA | WA | WA | NA | NA | NA | NA | NA | NA | NA |
| 814 | WA | WA | NA | NA | NA | NA | NA | WA | NA | NA | NA | NA | NA | NA | WA | WA | WA | WA | - | NA | NA | NA | NA | NA | WA |
| 815 | WA | NA | NA | NA | NA | NA | NA | WA | NA | NA | NA | WA | NA | NA | NA | NA | WA | NA | NA | NA | NA | NA | NA | NA | NA |
| 816 | WA | NA | NA | NA | NA | WA | WA | WA | WA | WA | NA | WA | WA | WA | WA | WA | NA | NA | NA | NA | NA | NA | NA | NA | NA |
| 817 | WA | WA | NA | NA | NA | WA | WA | WA | WA | NA | NA | NA | WA | WA | NA | NA | NA | NA | NA | NA | NA | WA | NA | NA | WA |
| 818 | WA | NA | WA | NA | NA | NA | NA | WA | NA | NA | NA | NA | NA | WA | WA | WA | NA | WA | WA | NA | NA | NA | WA | WA | NA |
| 819 | WA | NA | NA | NA | NA | WA | WA | WA | WA | NA | NA | WA | NA | WA | NA | NA | WA | WA | WA | NA | NA | WA | NA | NA | NA |
| 820 | WA | NA | NA | NA | NA | WA | WA | WA | WA | NA | NA | NA | NA | NA | NA | NA | NA | WA | NA | NA | NA | NA | WA | NA | NA |
| 821 | WA | NA | NA | NA | NA | NA | NA | WA | NA | NA | NA | WA | NA | NA | WA | WA | NA | NA | NA | NA | NA | NA | WA | WA | NA |
| 822 | WA | NA | WA | NA | NA | WA | WA | WA | WA | NA | NA | NA | NA | NA | NA | NA | NA | WA | NA | NA | NA | NA | NA | NA | NA |
| 823 | WA | NA | NA | NA | NA | WA | WA | WA | WA | WA | NA | WA | NA | NA | NA | NA | NA | WA | NA | NA | NA | WA | WA | WA | NA |
| 824 | WA | WA | WA | NA | NA | WA | WA | WA | WA | NA | NA | WA | NA | NA | WA | WA | WA | WA | NA | NA | NA | NA | WA | NA | NA |
| 825 | WA | NA | NA | NA | NA | WA | WA | WA | WA | NA | NA | NA | NA | NA | NA | NA | WA | WA | NA | NA | NA | NA | NA | NA | NA |
| 826 | WA | NA | NA | NA | NA | WA | WA | WA | WA | WA | NA | NA | NA | NA | NA | NA | NA | NA | NA | NA | NA | WA | NA | NA | NA |
| 827 | WA | WA | WA | NA | NA | WA | WA | WA | WA | WA | NA | WA | NA | NA | WA | WA | NA | WA | NA | NA | NA | NA | NA | NA | NA |
| 828 | WA | NA | WA | NA | NA | WA | WA | WA | WA | NA | NA | NA | NA | WA | WA | WA | WA | NA | WA | NA | NA | WA | WA | WA | NA |
| 829 | WA | NA | NA | NA | NA | NA | NA | WA | NA | NA | NA | NA | WA | WA | NA | NA | NA | NA | NA | NA | NA | NA | NA | NA | NA |
| 830 | WA | NA | NA | NA | NA | WA | WA | WA | WA | NA | NA | WA | NA | WA | NA | NA | WA | NA | NA | NA | NA | WA | NA | NA | NA |
| 831 | WA | NA | WA | NA | NA | WA | WA | WA | WA | NA | NA | WA | NA | NA | NA | NA | NA | NA | NA | NA | NA | NA | NA | NA | WA |
| 832 | WA | NA | NA | NA | NA | WA | WA | WA | WA | NA | NA | NA | NA | WA | WA | WA | WA | WA | NA | NA | NA | NA | NA | NA | NA |
| 833 | WA | WA | WA | NA | NA | NA | NA | WA | NA | WA | NA | NA | WA | NA | WA | WA | NA | NA | NA | NA | NA | NA | NA | NA | NA |
| 834 | WA | NA | NA | NA | NA | WA | WA | WA | NA | NA | NA | NA | NA | NA | NA | NA | WA | NA | WA | NA | WA | NA | NA | NA | NA |
| 835 | WA | WA | NA | NA | NA | NA | NA | WA | NA | NA | NA | NA | NA | NA | NA | NA | NA | WA | NA | NA | NA | NA | NA | NA | NA |
| 836 | WA | WA | NA | NA | NA | WA | WA | WA | WA | NA | NA | WA | NA | NA | NA | NA | NA | WA | NA | NA | NA | NA | NA | NA | NA |
| 837 | WA | WA | WA | NA | NA | WA | WA | WA | WA | NA | NA | NA | NA | NA | NA | NA | NA | WA | NA | NA | NA | NA | NA | NA | NA |
| 838 | WA | NA | NA | NA | NA | WA | WA | WA | WA | NA | NA | WA | NA | WA | NA | NA | WA | NA | NA | NA | NA | NA | NA | NA | NA |
| 839 | WA | WA | NA | NA | NA | NA | NA | WA | NA | NA | NA | WA | NA | WA | WA | WA | WA | NA | WA | NA | NA | NA | NA | NA | NA |
| 840 | WA | WA | NA | NA | NA | WA | WA | WA | WA | WA | NA | NA | NA | NA | NA | NA | NA | NA | NA | NA | NA | WA | NA | NA | NA |
| 841 | WA | WA | WA | NA | NA | NA | NA | WA | NA | NA | NA | NA | NA | WA | NA | NA | WA | WA | NA | NA | NA | NA | NA | NA | NA |
| 842 | WA | NA | NA | NA | NA | WA | WA | WA | WA | NA | NA | NA | NA | NA | NA | NA | WA | WA | NA | NA | NA | WA | NA | NA | NA |
| 843 | WA | NA | NA | NA | NA | WA | WA | WA | WA | NA | NA | NA | NA | WA | NA | NA | NA | WA | WA | NA | NA | NA | WA | NA | NA |
| 844 | WA | NA | NA | NA | NA | NA | NA | WA | NA | NA | NA | NA | NA | WA | NA | NA | WA | WA | NA | NA | NA | NA | NA | NA | NA |
| 845 | WA | NA | NA | NA | NA | WA | WA | WA | WA | NA | NA | WA | NA | NA | NA | NA | WA | NA | NA | NA | NA | NA | NA | NA | NA |
| 846 | WA | WA | WA | NA | NA | WA | WA | WA | WA | NA | NA | NA | NA | NA | WA | WA | NA | NA | NA | NA | NA | NA | NA | NA | NA |
| 847 | WA | WA | NA | NA | NA | NA | NA | WA | NA | WA | NA | NA | NA | NA | NA | NA | WA | WA | WA | NA | NA | NA | NA | NA | NA |
| 848 | NA | WA | WA | NA | NA | WA | WA | WA | WA | NA | NA | NA | NA | NA | NA | NA | NA | WA | NA | NA | NA | WA | NA | NA | NA |
| 849 | WA | WA | NA | NA | NA | WA | WA | WA | WA | NA | NA | NA | NA | NA | NA | NA | NA | NA | WA | NA | NA | NA | NA | NA | NA |
| 850 | WA | NA | WA | NA | NA | WA | WA | WA | WA | WA | NA | WA | NA | NA | NA | NA | WA | WA | NA | NA | NA | NA | NA | NA | WA |
| 851 | WA | WA | NA | NA | NA | WA | WA | WA | WA | WA | NA | WA | NA | NA | NA | NA | WA | NA | NA | NA | NA | NA | NA | NA | NA |
| 852 | WA | NA | WA | NA | NA | WA | WA | WA | WA | NA | NA | WA | NA | NA | NA | NA | WA | WA | WA | NA | NA | WA | NA | NA | NA |
| 853 | WA | NA | WA | NA | NA | NA | NA | WA | WA | WA | NA | WA | NA | NA | WA | WA | NA | WA | NA | NA | NA | NA | NA | NA | NA |
| 854 | WA | NA | NA | NA | NA | WA | WA | WA | WA | NA | NA | WA | NA | WA | NA | NA | WA | NA | NA | NA | NA | NA | NA | NA | NA |
| 855 | WA | NA | WA | NA | NA | NA | NA | WA | NA | NA | NA | NA | NA | NA | NA | NA | WA | NA | WA | NA | NA | NA | NA | NA | WA |
| 856 | WA | NA | NA | NA | NA | WA | WA | WA | WA | NA | NA | NA | NA | NA | NA | NA | WA | NA | WA | NA | NA | NA | NA | NA | WA |
| 857 | WA | NA | NA | NA | NA | NA | NA | WA | NA | NA | NA | NA | NA | NA | WA | WA | WA | NA | NA | NA | NA | NA | NA | NA | NA |
| 858 | WA | NA | NA | NA | NA | NA | NA | WA | NA | NA | NA | WA | NA | NA | WA | WA | NA | NA | NA | NA | NA | NA | NA | NA | WA |
| 859 | WA | NA | NA | NA | NA | WA | WA | WA | WA | NA | NA | WA | NA | NA | NA | NA | NA | WA | WA | NA | NA | WA | NA | NA | NA |
| 860 | WA | NA | WA | NA | NA | NA | NA | WA | NA | NA | NA | WA | NA | NA | NA | NA | WA | WA | NA | NA | NA | NA | NA | NA | NA |
| 861 | WA | WA | NA | NA | NA | WA | WA | WA | WA | NA | NA | NA | NA | WA | WA | NA | WA | NA | NA | NA | NA | WA | NA | NA | NA |
| 862 | WA | NA | NA | NA | NA | NA | NA | WA | NA | NA | NA | NA | NA | NA | WA | WA | NA | WA | NA | NA | NA | NA | NA | NA | NA |
| 863 | WA | NA | WA | NA | NA | WA | WA | WA | WA | WA | NA | NA | NA | NA | WA | WA | WA | WA | NA | NA | NA | NA | NA | NA | NA |
| 864 | WA | - | - | NA | NA | - | - | WA | - | - | NA | - | NA | - | - | - | - | - | - | NA | NA | - | NA | NA | NA |
| 865 | WA | NA | NA | NA | NA | WA | WA | WA | WA | WA | NA | NA | NA | NA | NA | NA | NA | WA | WA | NA | NA | WA | WA | NA | NA |
| 866 | WA | NA | - | NA | NA | - | - | WA | - | - | NA | - | NA | - | - | - | NA | - | NA | NA | NA | - | NA | NA | NA |
| 867 | WA | NA | - | NA | NA | - | - | WA | - | NA | NA | NA | NA | - | NA | - | NA | - | NA | NA | NA | - | NA | NA | - |
| 868 | WA | NA | NA | NA | NA | WA | WA | WA | WA | WA | NA | NA | NA | NA | NA | NA | NA | WA | WA | NA | NA | WA | NA | NA | WA |
| 869 | WA | WA | WA | NA | NA | NA | NA | WA | NA | WA | NA | WA | NA | NA | NA | NA | WA | NA | NA | NA | NA | NA | NA | NA | WA |
| 870 | WA | NA | NA | NA | NA | NA | NA | WA | NA | WA | NA | WA | NA | WA | NA | NA | NA | WA | NA | NA | NA | NA | NA | NA | NA |
| 871 | WA | NA | NA | NA | NA | WA | WA | WA | WA | NA | NA | NA | NA | NA | NA | NA | NA | WA | NA | NA | NA | NA | NA | NA | NA |
| 872 | WA | NA | - | NA | NA | - | - | WA | - | NA | NA | NA | NA | - | NA | NA | NA | - | - | - | NA | - | NA | NA | NA |
| 873 | WA | WA | - | NA | NA | - | - | WA | - | NA | - | - | - | - | - | - | - | - | - | NA | - | - | - | - | - |
| 874 | WA | - | - | NA | NA | NA | - | WA | NA | - | NA | NA | NA | WA | NA | NA | - | NA | NA | - | NA | WA | - | NA | NA |
| 875 | WA | NA | WA | NA | NA | NA | NA | WA | NA | NA | NA | NA | NA | - | NA | NA | - | - | NA | - | NA | WA | NA | NA | NA |
| 876 | NA | NA | NA | NA | NA | WA | WA | WA | WA | NA | NA | NA | NA | NA | NA | NA | WA | WA | NA | NA | NA | NA | WA | WA | NA |
| 877 | WA | NA | NA | NA | NA | NA | NA | WA | NA | NA | NA | NA | NA | NA | NA | NA | NA | NA | NA | NA | NA | WA | NA | NA | NA |
| 878 | WA | WA | WA | NA | NA | NA | NA | WA | NA | NA | NA | WA | NA | NA | NA | NA | WA | WA | NA | NA | NA | WA | NA | NA | NA |
| 879 | WA | NA | NA | NA | NA | NA | NA | WA | NA | WA | NA | NA | NA | NA | - | - | WA | NA | NA | NA | NA | - | NA | NA | NA |
| 880 | WA | WA | NA | NA | NA | NA | NA | WA | NA | WA | NA | NA | NA | NA | NA | NA | NA | WA | NA | NA | NA | NA | NA | NA | NA |
| 881 | WA | NA | NA | NA | NA | WA | WA | WA | WA | NA | NA | NA | NA | NA | WA | WA | WA | NA | NA | NA | NA | WA | NA | NA | NA |
| 882 | WA | NA | NA | NA | NA | WA | WA | WA | WA | NA | NA | WA | NA | NA | WA | WA | NA | NA | WA | NA | NA | WA | NA | NA | NA |
| 883 | WA | NA | WA | NA | NA | WA | WA | WA | WA | NA | NA | NA | NA | WA | NA | NA | NA | WA | WA | NA | NA | NA | NA | NA | NA |
| 884 | WA | NA | NA | NA | NA | WA | WA | WA | WA | NA | NA | WA | WA | WA | WA | WA | WA | NA | NA | WA | WA | WA | NA | NA | NA |
| 885 | WA | NA | WA | NA | NA | WA | WA | WA | WA | WA | NA | WA | NA | NA | NA | NA | WA | NA | NA | NA | NA | WA | NA | NA | NA |
| 886 | WA | WA | WA | NA | NA | NA | NA | WA | NA | NA | NA | NA | NA | NA | NA | NA | WA | NA | WA | NA | NA | WA | NA | NA | NA |
| 887 | WA | NA | NA | NA | NA | WA | WA | WA | WA | NA | NA | NA | NA | NA | NA | NA | WA | WA | NA | NA | NA | WA | NA | NA | WA |
| 888 | WA | NA | NA | NA | NA | WA | WA | WA | WA | NA | NA | NA | NA | WA | NA | NA | NA | WA | NA | NA | NA | WA | WA | NA | NA |
| 889 | WA | WA | NA | NA | NA | WA | WA | WA | WA | NA | NA | WA | NA | NA | NA | NA | NA | NA | WA | NA | NA | NA | NA | NA | NA |
| 890 | WA | NA | NA | NA | NA | WA | WA | WA | WA | NA | NA | WA | NA | NA | NA | NA | NA | WA | NA | NA | NA | NA | NA | NA | NA |
| 891 | WA | WA | NA | NA | NA | WA | WA | WA | WA | NA | NA | WA | NA | NA | WA | WA | WA | WA | WA | NA | NA | WA | NA | NA | NA |
| 892 | WA | NA | NA | NA | NA | WA | WA | WA | NA | NA | NA | WA | WA | WA | NA | NA | NA | WA | NA | NA | NA | WA | NA | NA | NA |
| 893 | - | - | - | - | NA | - | - | WA | - | - | - | - | - | - | NA | - | NA | - | - | - | - | - | NA | - | - |
| 894 | WA | NA | NA | NA | NA | WA | NA | WA | NA | WA | NA | NA | NA | NA | NA | NA | NA | WA | NA | NA | NA | WA | WA | WA | NA |
| 895 | WA | NA | NA | NA | NA | WA | WA | WA | WA | NA | NA | WA | NA | NA | WA | WA | NA | WA | NA | NA | NA | NA | WA | NA | NA |
| 896 | WA | NA | NA | NA | NA | WA | WA | WA | WA | NA | NA | NA | WA | WA | NA | NA | WA | WA | WA | NA | NA | WA | NA | NA | NA |
| 897 | WA | NA | NA | NA | NA | WA | WA | WA | WA | WA | NA | WA | NA | NA | NA | NA | NA | WA | WA | NA | NA | WA | NA | NA | WA |
| 898 | WA | WA | NA | NA | NA | NA | NA | WA | NA | WA | NA | NA | NA | NA | NA | NA | NA | WA | WA | NA | NA | NA | NA | NA | WA |
| 899 | WA | WA | NA | NA | NA | WA | WA | WA | WA | NA | NA | NA | NA | NA | WA | WA | NA | WA | NA | NA | NA | NA | NA | NA | NA |
| 900 | - | NA | - | NA | NA | - | - | WA | - | NA | NA | NA | NA | - | NA | - | - | - | NA | - | NA | - | - | - | NA |
| 901 | WA | NA | WA | NA | NA | WA | WA | WA | WA | NA | NA | NA | NA | NA | WA | WA | NA | NA | NA | NA | NA | WA | NA | NA | NA |
| 902 | WA | NA | - | NA | NA | - | - | WA | - | NA | NA | NA | - | - | - | - | NA | - | NA | NA | NA | - | NA | NA | NA |
| 903 | WA | WA | NA | NA | NA | WA | WA | WA | WA | NA | NA | WA | NA | NA | NA | NA | WA | NA | WA | NA | NA | NA | NA | NA | NA |
| 904 | WA | WA | NA | NA | NA | WA | WA | WA | WA | WA | NA | NA | NA | NA | NA | NA | NA | WA | NA | NA | NA | WA | NA | NA | NA |
| 905 | WA | NA | WA | NA | NA | NA | NA | WA | NA | WA | NA | NA | NA | NA | NA | NA | NA | NA | NA | NA | NA | NA | NA | NA | NA |
| 906 | WA | WA | NA | NA | NA | WA | WA | WA | WA | WA | NA | NA | NA | NA | NA | NA | NA | NA | NA | NA | NA | NA | NA | NA | NA |
| 907 | WA | WA | WA | NA | NA | NA | NA | WA | NA | NA | NA | WA | NA | WA | NA | NA | NA | NA | NA | NA | NA | WA | WA | WA | NA |
| 908 | WA | WA | NA | NA | NA | WA | WA | WA | NA | NA | NA | NA | NA | NA | NA | NA | NA | NA | WA | NA | NA | NA | WA | WA | NA |
| 909 | WA | NA | NA | NA | NA | NA | NA | WA | NA | NA | NA | NA | NA | NA | NA | NA | WA | NA | NA | NA | NA | WA | WA | WA | NA |
| 910 | WA | WA | NA | NA | NA | NA | NA | WA | NA | NA | NA | WA | NA | NA | WA | WA | NA | NA | NA | NA | NA | NA | NA | NA | NA |
| 911 | WA | NA | NA | NA | NA | NA | NA | WA | NA | NA | NA | NA | NA | NA | WA | WA | WA | NA | WA | NA | NA | NA | NA | NA | WA |
| 912 | WA | NA | - | NA | NA | - | - | WA | - | NA | NA | NA | NA | WA | - | - | - | NA | - | - | NA | - | - | - | NA |
| 913 | WA | NA | NA | NA | NA | WA | WA | WA | WA | NA | NA | NA | NA | NA | NA | NA | NA | WA | NA | NA | NA | NA | NA | NA | NA |
| 914 | WA | WA | WA | NA | NA | NA | NA | WA | NA | NA | NA | WA | NA | NA | WA | WA | NA | NA | NA | NA | NA | NA | NA | NA | WA |
| 915 | WA | - | NA | NA | NA | - | - | WA | - | NA | NA | NA | - | - | - | - | NA | - | NA | NA | NA | - | NA | NA | - |
| 916 | WA | NA | NA | NA | NA | WA | WA | WA | WA | WA | NA | NA | WA | WA | NA | NA | NA | NA | NA | NA | WA | NA | NA | NA | NA |
| 917 | WA | NA | NA | NA | NA | WA | WA | WA | WA | NA | NA | NA | NA | NA | NA | NA | NA | NA | NA | NA | NA | WA | NA | NA | NA |
| 918 | WA | NA | WA | NA | NA | WA | WA | WA | WA | NA | NA | WA | NA | NA | NA | NA | NA | NA | NA | NA | NA | WA | NA | NA | WA |
| 919 | WA | WA | NA | NA | NA | NA | NA | WA | NA | NA | NA | NA | NA | NA | NA | NA | NA | NA | NA | NA | NA | NA | NA | NA | WA |
| 920 | WA | NA | NA | NA | NA | NA | NA | WA | WA | NA | NA | WA | NA | WA | WA | WA | NA | NA | WA | NA | NA | WA | NA | NA | WA |
| 921 | WA | NA | NA | NA | NA | WA | WA | WA | WA | WA | NA | WA | NA | WA | NA | NA | NA | NA | WA | NA | NA | WA | NA | NA | NA |
| 922 | WA | NA | NA | NA | NA | NA | NA | WA | NA | WA | NA | NA | NA | NA | WA | WA | NA | WA | WA | NA | WA | WA | WA | WA | NA |
| 923 | WA | NA | WA | NA | NA | WA | WA | WA | WA | WA | NA | WA | NA | NA | NA | NA | NA | NA | NA | NA | NA | NA | NA | NA | NA |
| 924 | WA | NA | WA | NA | NA | NA | NA | WA | NA | WA | NA | WA | NA | WA | NA | NA | NA | NA | NA | WA | NA | WA | NA | NA | NA |
| 925 | WA | NA | WA | NA | NA | WA | WA | WA | WA | NA | NA | NA | NA | NA | NA | NA | WA | WA | NA | NA | NA | WA | NA | NA | NA |
| 926 | WA | - | NA | NA | NA | WA | WA | WA | - | NA | NA | NA | - | - | NA | NA | NA | - | - | NA | NA | WA | NA | NA | NA |
| 927 | WA | NA | NA | NA | NA | NA | NA | WA | WA | NA | NA | NA | WA | WA | NA | NA | WA | WA | NA | NA | NA | NA | NA | NA | NA |
| 928 | WA | NA | NA | NA | NA | WA | WA | WA | WA | NA | NA | NA | NA | NA | NA | NA | WA | NA | WA | NA | NA | NA | NA | NA | NA |
| 929 | WA | NA | WA | NA | NA | WA | WA | WA | WA | NA | NA | NA | NA | WA | NA | NA | NA | NA | NA | NA | NA | WA | NA | NA | WA |
| 930 | WA | NA | NA | NA | NA | NA | NA | WA | NA | NA | NA | WA | NA | NA | NA | NA | NA | NA | NA | NA | WA | NA | NA | NA | WA |
| 931 | WA | WA | WA | NA | NA | WA | WA | WA | WA | NA | NA | NA | NA | NA | NA | NA | WA | WA | NA | NA | NA | WA | NA | NA | NA |
| 932 | WA | NA | NA | NA | NA | NA | WA | WA | WA | WA | NA | WA | WA | NA | NA | NA | WA | NA | NA | NA | NA | WA | NA | NA | WA |
| 933 | WA | NA | NA | NA | NA | WA | WA | WA | WA | WA | NA | WA | NA | WA | NA | NA | NA | WA | NA | NA | NA | NA | NA | NA | NA |
| 934 | WA | NA | WA | NA | NA | WA | WA | WA | WA | WA | NA | NA | NA | WA | WA | WA | WA | NA | NA | NA | NA | NA | NA | NA | NA |
| 935 | WA | WA | NA | NA | NA | WA | WA | WA | WA | WA | NA | NA | NA | WA | NA | NA | WA | NA | NA | NA | NA | WA | NA | NA | WA |
| 936 | WA | NA | NA | NA | NA | WA | WA | WA | WA | NA | NA | WA | NA | NA | WA | - | NA | WA | NA | NA | NA | NA | NA | NA | WA |
| 937 | WA | NA | WA | NA | NA | WA | WA | WA | NA | WA | NA | WA | NA | NA | NA | NA | WA | NA | WA | NA | WA | NA | NA | NA | NA |
| 938 | WA | WA | NA | NA | NA | NA | NA | WA | NA | NA | NA | NA | NA | NA | NA | WA | NA | WA | NA | NA | NA | WA | NA | NA | NA |
| 939 | WA | NA | NA | NA | NA | NA | NA | WA | WA | NA | NA | WA | NA | NA | NA | NA | NA | WA | NA | NA | NA | NA | NA | NA | WA |
| 940 | WA | NA | NA | NA | NA | WA | WA | WA | WA | NA | NA | NA | NA | WA | NA | NA | NA | WA | NA | NA | NA | WA | NA | NA | NA |
| 941 | WA | NA | NA | NA | NA | NA | NA | WA | NA | NA | NA | WA | NA | NA | NA | NA | WA | NA | WA | NA | NA | NA | WA | NA | NA |
| 942 | WA | NA | WA | NA | NA | WA | WA | WA | WA | NA | NA | NA | NA | NA | NA | NA | NA | NA | NA | NA | NA | WA | NA | NA | NA |
| 943 | WA | NA | NA | NA | NA | WA | WA | WA | WA | NA | NA | WA | WA | NA | NA | NA | WA | NA | WA | NA | NA | WA | NA | NA | NA |
| 944 | WA | NA | WA | NA | NA | WA | WA | WA | WA | WA | NA | NA | NA | NA | NA | NA | WA | NA | NA | NA | NA | - | NA | NA | NA |
| 945 | WA | NA | NA | NA | NA | WA | WA | WA | WA | NA | NA | NA | NA | NA | NA | NA | NA | NA | NA | NA | NA | NA | WA | NA | NA |
| 946 | WA | NA | WA | NA | NA | WA | WA | WA | WA | NA | NA | NA | NA | WA | NA | NA | WA | NA | NA | NA | NA | NA | WA | WA | WA |
| 947 | WA | NA | NA | NA | NA | NA | NA | WA | NA | WA | NA | NA | NA | WA | NA | NA | NA | NA | NA | NA | NA | WA | WA | WA | NA |
| 948 | WA | NA | NA | NA | NA | - | - | WA | - | - | NA | WA | NA | NA | - | NA | - | - | - | NA | NA | - | NA | - | - |
| 949 | WA | NA | WA | NA | NA | WA | NA | WA | NA | NA | NA | WA | NA | NA | NA | NA | NA | NA | WA | NA | NA | NA | NA | NA | WA |
| 950 | WA | NA | NA | NA | NA | NA | NA | WA | NA | NA | NA | WA | NA | NA | NA | NA | NA | NA | NA | NA | NA | NA | WA | WA | NA |
| 951 | WA | NA | WA | NA | NA | WA | WA | WA | WA | NA | NA | NA | NA | NA | NA | NA | NA | WA | WA | NA | NA | WA | NA | NA | NA |
| 952 | WA | WA | NA | NA | NA | NA | NA | WA | WA | WA | NA | WA | NA | WA | WA | WA | WA | NA | NA | NA | NA | NA | NA | NA | NA |
| 953 | WA | NA | WA | NA | NA | - | NA | WA | NA | WA | NA | NA | NA | - | NA | NA | NA | - | - | NA | - | - | NA | NA | NA |
| 954 | WA | NA | NA | NA | NA | NA | NA | WA | NA | NA | NA | NA | NA | NA | NA | NA | WA | NA | WA | NA | NA | WA | NA | NA | NA |
| 955 | WA | WA | NA | NA | NA | WA | WA | WA | WA | WA | NA | NA | NA | NA | WA | WA | NA | NA | NA | NA | NA | NA | NA | NA | NA |
| 956 | WA | NA | - | NA | NA | NA | NA | WA | NA | - | NA | - | NA | - | NA | NA | NA | NA | NA | - | NA | WA | - | - | NA |
| 957 | - | NA | - | NA | NA | WA | WA | WA | WA | NA | NA | - | - | NA | NA | NA | - | NA | - | NA | NA | - | - | - | NA |
| 958 | WA | NA | NA | NA | NA | WA | WA | WA | WA | NA | NA | NA | NA | NA | NA | NA | NA | NA | NA | NA | NA | WA | NA | NA | NA |
| 959 | WA | WA | NA | NA | NA | NA | NA | WA | NA | WA | NA | NA | NA | NA | NA | NA | NA | WA | NA | NA | NA | NA | NA | NA | NA |
| 960 | WA | - | - | NA | NA | WA | - | WA | - | NA | - | - | - | - | - | - | - | NA | NA | NA | - | NA | - | - | NA |

# Table S2. Checking of bias in samples with missed values (heat selected)

*P* -value is the probability of observing by chance the same or greater deviance if two random values are independent

| Segregant | Original database | | | Complete database | | | p-value of independence |
|---|---|---|---|---|---|---|---|
| | Total | NA | WA | Total | NA | WA | |
| chr01-0040xxx | 954 | 4 | 950 | 896 | 4 | 892 | 0.93 |
| chr01-0119xxx | 943 | 717 | 226 | 896 | 684 | 212 | 0.88 |
| chr02-0472xxx | 933 | 615 | 318 | 896 | 587 | 309 | 0.86 |
| chr02-0517xxx | 956 | 954 | 2 | 896 | 894 | 2 | 0.95 |
| chr02-0522xxx | 959 | 959 | 0 | 896 | 896 | 0 | 1.00 |
| chr04-0454xxx | 937 | 354 | 583 | 896 | 340 | 556 | 0.94 |
| chr04-0461xxx | 935 | 354 | 581 | 896 | 340 | 556 | 0.97 |
| chr04-0488xxx | 959 | 0 | 959 | 896 | 0 | 896 | 1.00 |
| chr04-0496xxx | 931 | 332 | 599 | 896 | 318 | 578 | 0.94 |
| chr04-1313xxx | 943 | 645 | 298 | 896 | 612 | 284 | 0.97 |
| chr05-0196xxx | 954 | 954 | 0 | 896 | 896 | 0 | 1.00 |
| chr07-0131xxx | 934 | 527 | 407 | 896 | 501 | 395 | 0.83 |
| chr07-0859xxx | 947 | 881 | 66 | 896 | 831 | 65 | 0.81 |
| chr09-0292xxx | 932 | 620 | 312 | 896 | 597 | 299 | 0.96 |
| chr10-0234xxx | 934 | 648 | 286 | 896 | 617 | 279 | 0.81 |
| chr10-0235xxx | 929 | 652 | 277 | 896 | 624 | 272 | 0.80 |
| chr10-0420xxx | 938 | 511 | 427 | 896 | 483 | 413 | 0.81 |
| chr12-0140xxx | 937 | 529 | 408 | 896 | 496 | 400 | 0.64 |
| chr12-0730xxx | 934 | 679 | 255 | 896 | 645 | 251 | 0.73 |
| chr13-0893xxx | 949 | 935 | 14 | 896 | 882 | 14 | 0.88 |
| chr13-0910xxx | 949 | 882 | 67 | 896 | 832 | 64 | 0.94 |
| chr14-0441xxx | 932 | 527 | 405 | 896 | 505 | 391 | 0.94 |
| chr15-0172xxx | 949 | 861 | 88 | 896 | 808 | 88 | 0.69 |
| chr15-0179xxx | 946 | 873 | 73 | 896 | 823 | 73 | 0.73 |
| chr15-1032xxx | 947 | 788 | 159 | 896 | 741 | 155 | 0.77 |

## Table S3. Non selected genotypes

| ID | chr01-0040xxx | chr01-0119xxx | chr02-0472xxx | chr02-0517xxx | chr02-0522xxx | chr04-0454xxx | chr04-0461xxx | chr04-0488xxx | chr04-0496xxx | chr04-1313xxx | chr05-0196xxx | chr07-0131xxx | chr07-0859xxx | chr09-0292xxx | chr10-0234xxx | chr10-0235xxx | chr10-0420xxx | chr12-0140xxx | chr12-0730xxx | chr13-0893xxx | chr13-0910xxx | chr14-0441xxx | chr15-0172xxx | chr15-0179xxx | chr15-1032xxx |
|---|---|---|---|---|---|---|---|---|---|---|---|---|---|---|---|---|---|---|---|---|---|---|---|---|---|
| 6738_5#1 | WA | WA | WA | NA | NA | WA | WA | WA | WA | WA | NA | NA | WA | WA | WA | WA | NA | NA | NA | WA | WA | NA | WA | WA | WA |
| 6738_5#3 | WA | WA | WA | WA | WA | WA | WA | WA | WA | NA | NA | NA | NA | NA | WA | WA | WA | NA | WA | NA | NA | NA | WA | WA | WA |
| 6738_5#4 | WA | NA | WA | NA | NA | NA | NA | NA | NA | NA | NA | NA | NA | NA | WA | WA | WA | WA | NA | WA | WA | WA | NA | NA | WA |
| 6738_5#5 | WA | NA | WA | NA | NA | WA | WA | WA | WA | WA | NA | NA | NA | WA | WA | WA | WA | WA | NA | NA | WA | WA | WA | WA | WA |
| 6738_5#6 | WA | NA | WA | NA | NA | WA | WA | WA | WA | WA | NA | NA | NA | WA | NA | NA | NA | NA | NA | WA | WA | NA | NA | NA | WA |
| 6738_5#7 | WA | NA | WA | WA | WA | WA | WA | WA | WA | WA | NA | NA | WA | NA | NA | NA | WA | NA | NA | WA | WA | WA | NA | NA | WA |
| 6738_5#8 | WA | WA | WA | WA | WA | WA | WA | WA | WA | NA | NA | NA | WA | NA | WA | WA | WA | NA | WA | NA | NA | NA | NA | NA | WA |
| 6738_5#9 | WA | NA | WA | WA | WA | WA | WA | WA | WA | NA | NA | NA | WA | NA | NA | NA | WA | WA | WA | WA | WA | NA | NA | NA | WA |
| 6738_5#10 | WA | NA | WA | NA | NA | WA | WA | WA | WA | NA | NA | WA | NA | NA | NA | NA | WA | WA | WA | NA | NA | NA | WA | WA | NA |
| 6738_5#11 | WA | WA | WA | NA | NA | WA | WA | WA | WA | WA | NA | NA | NA | WA | NA | NA | WA | WA | WA | NA | NA | WA | NA | NA | WA |
| 6738_5#12 | WA | NA | WA | NA | NA | WA | WA | WA | WA | NA | NA | WA | WA | NA | NA | NA | WA | NA | NA | WA | WA | NA | WA | WA | NA |
| 6738_5#13 | WA | WA | WA | NA | NA | WA | WA | NA | NA | WA | NA | WA | NA | NA | WA | WA | WA | NA | NA | WA | WA | WA | WA | WA | WA |
| 6738_5#14 | WA | NA | WA | NA | NA | WA | WA | WA | WA | NA | NA | NA | WA | WA | NA | NA | WA | NA | NA | WA | NA | NA | WA | WA | WA |
| 6738_5#15 | WA | WA | WA | NA | NA | WA | WA | WA | WA | NA | NA | NA | WA | WA | NA | NA | NA | NA | WA | WA | WA | WA | WA | WA | NA |
| 6738_5#16 | WA | NA | WA | NA | NA | WA | WA | WA | WA | NA | NA | WA | WA | WA | WA | WA | WA | NA | WA | NA | NA | WA | WA | WA | WA |
| 6738_5#17 | WA | NA | WA | NA | NA | NA | NA | NA | NA | NA | NA | NA | WA | WA | WA | WA | WA | WA | WA | NA | NA | NA | NA | NA | WA |
| 6738_5#18 | WA | NA | WA | WA | WA | WA | WA | WA | WA | WA | NA | NA | WA | NA | WA | WA | WA | NA | NA | WA | WA | NA | WA | WA | NA |
| 6738_5#19 | WA | WA | WA | NA | NA | NA | NA | WA | WA | WA | NA | WA | NA | NA | NA | NA | WA | NA | NA | NA | NA | WA | NA | NA | NA |
| 6738_5#20 | WA | WA | WA | NA | NA | NA | NA | NA | NA | WA | NA | WA | WA | WA | WA | WA | WA | WA | NA | NA | WA | NA | WA | WA | NA |
| 6738_5#21 | WA | NA | WA | NA | NA | WA | WA | WA | WA | NA | NA | WA | NA | NA | WA | WA | WA | WA | WA | NA | NA | WA | WA | WA | WA |
| 6738_5#22 | WA | WA | WA | NA | NA | WA | NA | WA | WA | WA | NA | WA | NA | WA | NA | NA | WA | WA | - | NA | NA | NA | NA | NA | NA |
| 6738_5#23 | WA | NA | WA | NA | NA | WA | WA | WA | WA | WA | NA | NA | WA | NA | NA | NA | NA | WA | NA | WA | WA | NA | - | WA | WA |
| 6738_5#24 | WA | NA | WA | NA | NA | NA | NA | NA | NA | NA | NA | WA | NA | WA | NA | NA | NA | NA | NA | NA | NA | NA | WA | WA | WA |
| 6738_5#25 | WA | NA | WA | NA | NA | NA | NA | NA | NA | WA | NA | NA | NA | WA | NA | NA | NA | NA | NA | NA | NA | NA | WA | WA | WA |
| 6738_5#26 | WA | NA | WA | WA | WA | WA | WA | WA | WA | NA | NA | NA | NA | WA | NA | NA | WA | NA | NA | NA | WA | WA | WA | WA | NA |
| 6738_5#27 | WA | NA | WA | NA | NA | WA | WA | WA | WA | NA | NA | NA | NA | WA | WA | WA | WA | NA | NA | WA | WA | NA | NA | NA | WA |
| 6738_5#28 | WA | WA | WA | WA | WA | NA | NA | WA | WA | WA | NA | NA | WA | WA | NA | NA | NA | NA | NA | NA | NA | WA | NA | WA | WA |
| 6738_5#29 | WA | NA | WA | NA | NA | WA | WA | WA | WA | WA | NA | NA | WA | WA | WA | WA | NA | WA | WA | NA | NA | NA | WA | WA | WA |
| 6738_5#30 | WA | WA | WA | NA | NA | WA | WA | WA | WA | NA | NA | WA | NA | NA | WA | WA | WA | NA | NA | NA | NA | NA | WA | WA | WA |
| 6738_5#31 | WA | WA | WA | NA | NA | WA | WA | WA | WA | NA | NA | WA | WA | WA | NA | NA | WA | NA | NA | NA | NA | NA | WA | WA | WA |
| 6738_5#32 | WA | WA | WA | NA | NA | WA | WA | WA | WA | NA | NA | NA | WA | NA | WA | WA | NA | NA | WA | NA | NA | NA | WA | WA | WA |
| 6738_5#33 | WA | WA | WA | NA | NA | WA | WA | WA | WA | NA | NA | NA | WA | WA | WA | WA | NA | WA | WA | NA | NA | NA | NA | WA | NA |
| 6738_5#34 | WA | NA | WA | NA | NA | WA | WA | WA | WA | WA | NA | NA | NA | WA | WA | WA | NA | WA | WA | NA | NA | NA | WA | WA | NA |
| 6738_5#35 | WA | NA | WA | NA | NA | WA | WA | WA | WA | NA | NA | NA | NA | WA | NA | NA | NA | NA | NA | WA | WA | WA | WA | WA | WA |
| 6738_5#36 | WA | WA | WA | NA | NA | WA | WA | WA | WA | NA | NA | NA | WA | WA | WA | WA | WA | NA | WA | NA | NA | NA | WA | WA | WA |
| 6738_5#37 | WA | WA | - | NA | NA | WA | WA | WA | WA | NA | NA | NA | WA | WA | WA | WA | WA | WA | WA | NA | NA | NA | WA | WA | NA |
| 6738_5#38 | WA | NA | WA | NA | NA | WA | WA | WA | WA | NA | NA | NA | NA | NA | NA | NA | WA | NA | NA | NA | NA | NA | WA | WA | WA |
| 6738_5#39 | WA | NA | WA | WA | WA | WA | WA | WA | WA | NA | NA | NA | NA | WA | WA | WA | WA | NA | WA | WA | WA | NA | WA | WA | WA |
| 6738_5#40 | WA | NA | WA | NA | NA | WA | WA | WA | WA | NA | NA | WA | NA | WA | WA | WA | NA | NA | NA | NA | NA | NA | WA | WA | WA |
| 6738_5#41 | WA | WA | WA | NA | NA | WA | WA | WA | WA | NA | NA | NA | WA | WA | NA | NA | WA | WA | WA | WA | WA | NA | WA | WA | WA |
| 6738_5#42 | WA | NA | WA | NA | NA | WA | WA | WA | WA | WA | NA | NA | NA | WA | WA | WA | WA | NA | WA | NA | NA | NA | WA | WA | NA |
| 6738_5#47 | WA | WA | WA | WA | WA | WA | NA | NA | NA | NA | NA | NA | NA | NA | NA | NA | NA | NA | NA | NA | NA | NA | WA | WA | WA |
| 6738_5#48 | WA | NA | WA | NA | NA | WA | WA | WA | WA | NA | NA | NA | NA | NA | NA | NA | WA | WA | WA | WA | WA | NA | WA | WA | WA |
| 6738_5#49 | WA | WA | WA | WA | WA | NA | NA | NA | NA | NA | NA | NA | WA | - | NA | WA | NA | WA | NA | NA | NA | WA | NA | NA | WA |
| 6738_5#51 | WA | WA | WA | NA | NA | WA | WA | WA | WA | NA | NA | WA | WA | NA | NA | NA | WA | WA | WA | NA | NA | NA | WA | WA | NA |
| 6738_5#52 | WA | NA | WA | NA | NA | WA | WA | WA | WA | WA | NA | NA | NA | WA | WA | WA | WA | NA | WA | NA | NA | NA | WA | WA | WA |
| 6738_5#53 | WA | WA | WA | NA | NA | WA | WA | WA | WA | NA | NA | NA | WA | WA | WA | WA | NA | NA | WA | NA | NA | NA | NA | NA | WA |
| 6738_5#55 | WA | NA | WA | NA | NA | WA | WA | WA | WA | NA | NA | WA | NA | WA | WA | WA | NA | NA | NA | NA | NA | NA | WA | WA | WA |
| 6738_5#56 | WA | WA | WA | WA | WA | WA | WA | WA | WA | WA | NA | WA | WA | WA | NA | NA | WA | NA | NA | NA | WA | NA | WA | WA | WA |
| 6738_5#57 | WA | WA | WA | NA | NA | NA | NA | WA | WA | NA | NA | WA | WA | WA | WA | NA | NA | NA | WA | NA | NA | NA | WA | WA | NA |
| 6738_5#58 | WA | WA | WA | WA | WA | WA | WA | WA | WA | NA | NA | NA | WA | NA | WA | WA | WA | NA | NA | NA | NA | WA | WA | WA | WA |
| 6738_5#59 | WA | WA | WA | WA | WA | WA | WA | WA | WA | WA | NA | NA | WA | NA | WA | WA | NA | WA | WA | WA | WA | WA | NA | NA | WA |
| 6738_5#60 | WA | WA | WA | WA | WA | WA | WA | WA | WA | WA | NA | NA | WA | WA | WA | WA | WA | NA | NA | NA | NA | NA | WA | WA | NA |
| 6738_5#61 | WA | WA | WA | WA | WA | WA | WA | WA | WA | NA | NA | WA | WA | WA | WA | WA | WA | NA | NA | WA | WA | NA | NA | NA | WA |
| 6738_5#62 | WA | NA | WA | NA | NA | WA | WA | WA | WA | WA | NA | NA | WA | NA | WA | WA | NA | WA | WA | NA | NA | NA | WA | WA | NA |
| 6738_5#63 | WA | NA | WA | NA | NA | WA | WA | WA | WA | NA | NA | NA | NA | - | WA | WA | WA | WA | WA | NA | WA | WA | WA | WA | NA |
| 6738_5#64 | WA | NA | WA | WA | WA | WA | WA | WA | WA | NA | NA | WA | NA | WA | WA | WA | WA | WA | WA | NA | NA | WA | NA | NA | NA |
| 6738_5#65 | WA | WA | WA | WA | WA | WA | WA | WA | WA | WA | NA | NA | WA | WA | NA | NA | WA | WA | NA | NA | WA | NA | WA | WA | WA |
| 6738_5#66 | WA | WA | WA | NA | NA | WA | WA | WA | WA | WA | NA | WA | NA | WA | NA | NA | NA | NA | NA | NA | NA | WA | WA | WA | NA |
| 6738_5#67 | WA | NA | WA | NA | NA | NA | NA | WA | WA | WA | NA | NA | WA | WA | WA | WA | WA | NA | NA | NA | NA | NA | WA | WA | NA |
| 6738_5#68 | WA | WA | WA | NA | NA | WA | WA | WA | WA | NA | NA | WA | NA | NA | NA | NA | NA | NA | WA | WA | WA | WA | WA | WA | WA |
| 6738_5#69 | WA | WA | WA | NA | NA | NA | NA | NA | WA | NA | NA | NA | NA | WA | WA | WA | WA | NA | NA | NA | NA | NA | NA | NA | NA |
| 6738_5#70 | WA | WA | WA | WA | WA | NA | NA | NA | NA | WA | NA | NA | NA | WA | WA | WA | NA | NA | WA | NA | NA | NA | NA | NA | WA |
| 6738_5#71 | WA | WA | WA | WA | WA | WA | WA | WA | WA | WA | NA | WA | WA | WA | WA | WA | WA | NA | NA | NA | NA | NA | WA | WA | NA |
| 6738_5#73 | WA | WA | WA | NA | NA | NA | NA | WA | WA | WA | NA | NA | WA | NA | WA | WA | WA | NA | WA | NA | NA | NA | WA | WA | NA |
| 6738_5#74 | WA | NA | WA | WA | WA | WA | WA | WA | WA | WA | NA | WA | NA | NA | WA | WA | WA | NA | NA | WA | WA | NA | WA | WA | NA |
| 6738_5#76 | WA | NA | WA | WA | WA | WA | WA | WA | WA | NA | NA | NA | WA | WA | WA | WA | WA | NA | WA | NA | WA | NA | WA | WA | NA |
| 6738_5#77 | WA | NA | NA | NA | NA | WA | WA | WA | WA | WA | NA | NA | WA | WA | NA | NA | WA | WA | NA | WA | WA | WA | NA | NA | WA |
| 6738_5#78 | WA | WA | WA | WA | WA | NA | NA | WA | WA | NA | NA | WA | NA | NA | WA | WA | NA | NA | WA | WA | WA | NA | WA | WA | WA |
| 6738_5#79 | WA | WA | WA | NA | NA | NA | NA | NA | NA | NA | NA | NA | NA | NA | WA | WA | WA | WA | NA | NA | NA | NA | WA | WA | WA |
| 6738_5#80 | WA | NA | WA | NA | NA | WA | WA | NA | NA | NA | NA | NA | NA | NA | WA | WA | WA | NA | NA | NA | NA | NA | WA | WA | NA |
| 6738_5#81 | WA | NA | WA | NA | NA | WA | WA | WA | WA | NA | NA | NA | WA | WA | NA | NA | NA | WA | NA | NA | NA | NA | NA | NA | NA |
| 6738_5#82 | WA | WA | WA | NA | NA | WA | WA | WA | WA | NA | NA | WA | NA | WA | WA | WA | WA | WA | NA | WA | NA | WA | WA | WA | NA |
| 6738_5#84 | WA | WA | WA | NA | NA | NA | NA | WA | WA | WA | NA | WA | WA | WA | WA | WA | WA | WA | WA | NA | NA | NA | WA | WA | NA |
| 6738_5#85 | WA | NA | WA | NA | NA | WA | WA | WA | WA | NA | NA | NA | NA | WA | NA | NA | WA | NA | WA | NA | NA | NA | WA | WA | WA |
| 6738_5#86 | WA | WA | WA | NA | NA | WA | WA | WA | WA | NA | NA | WA | WA | NA | WA | WA | WA | WA | WA | NA | NA | NA | WA | WA | WA |
| 6738_5#87 | WA | NA | WA | NA | NA | WA | WA | WA | WA | WA | NA | NA | WA | WA | WA | WA | WA | NA | WA | NA | NA | NA | WA | WA | NA |
| 6738_5#88 | WA | WA | WA | NA | NA | NA | NA | NA | NA | NA | NA | WA | WA | NA | WA | WA | WA | WA | NA | NA | NA | NA | WA | WA | NA |
| 6738_5#89 | WA | NA | WA | NA | NA | WA | WA | WA | WA | WA | NA | WA | NA | NA | NA | NA | WA | NA | NA | NA | NA | WA | WA | WA | WA |
| 6738_5#90 | WA | WA | WA | WA | WA | NA | NA | NA | NA | NA | NA | WA | WA | NA | WA | WA | WA | WA | NA | NA | NA | WA | NA | NA | WA |
| 6738_5#91 | WA | WA | WA | WA | WA | WA | WA | WA | WA | NA | NA | WA | NA | WA | NA | NA | WA | NA | WA | NA | NA | NA | WA | WA | NA |
| 6738_5#92 | WA | WA | WA | NA | NA | WA | WA | WA | WA | NA | NA | NA | WA | WA | NA | NA | WA | WA | NA | NA | NA | WA | WA | WA | NA |
| 6738_5#93 | WA | NA | WA | NA | NA | WA | WA | WA | WA | NA | NA | WA | NA | NA | WA | WA | WA | NA | NA | NA | NA | WA | NA | NA | WA |
| 6738_5#94 | WA | NA | WA | WA | WA | WA | WA | WA | WA | WA | NA | WA | WA | NA | NA | NA | NA | NA | WA | NA | NA | NA | WA | WA | NA |
| 6738_5#95 | WA | WA | WA | NA | NA | WA | WA | WA | WA | WA | NA | NA | NA | NA | WA | WA | WA | WA | WA | NA | NA | WA | NA | NA | WA |
| 6738_5#96 | WA | NA | WA | WA | WA | NA | NA | NA | NA | WA | NA | NA | NA | WA | NA | NA | WA | NA | WA | NA | NA | WA | WA | WA | WA |
| 6738_6#1 | WA | WA | NA | NA | NA | WA | WA | WA | WA | WA | NA | WA | NA | WA | NA | NA | NA | WA | NA | NA | NA | NA | WA | WA | WA |
| 6738_6#3 | WA | WA | NA | NA | NA | WA | WA | WA | WA | NA | NA | NA | WA | WA | WA | WA | WA | WA | NA | WA | NA | WA | WA | WA | NA |
| 6738_6#4 | WA | NA | NA | NA | NA | WA | WA | WA | WA | NA | NA | NA | WA | WA | NA | WA | WA | NA | NA | NA | NA | NA | NA | WA | WA |
| 6738_6#5 | WA | WA | NA | NA | NA | WA | WA | WA | WA | NA | NA | NA | NA | WA | NA | NA | WA | WA | NA | WA | WA | WA | NA | NA | NA |
| 6738_6#6 | WA | WA | NA | WA | WA | NA | NA | NA | NA | WA | NA | WA | NA | WA | WA | WA | NA | NA | NA | NA | NA | NA | WA | WA | NA |
| 6738_6#7 | WA | NA | NA | NA | NA | NA | NA | NA | NA | WA | NA | NA | WA | NA | NA | NA | WA | NA | NA | NA | NA | WA | WA | WA | WA |
| 6738_6#8 | WA | NA | NA | WA | WA | WA | WA | WA | WA | NA | NA | NA | WA | WA | NA | NA | NA | NA | NA | WA | NA | NA | WA | WA | NA |
| 6738_6#9 | WA | WA | NA | NA | NA | NA | NA | NA | NA | NA | NA | NA | WA | WA | WA | WA | NA | WA | NA | NA | NA | NA | WA | WA | WA |
| 6738_6#10 | WA | NA | NA | NA | NA | NA | NA | WA | WA | NA | NA | NA | NA | WA | WA | WA | WA | WA | WA | NA | NA | NA | NA | NA | WA |
| 6738_6#11 | WA | WA | NA | NA | NA | WA | WA | WA | WA | NA | NA | NA | NA | WA | WA | WA | WA | WA | NA | NA | NA | NA | NA | NA | WA |
| 6738_6#12 | WA | NA | NA | NA | NA | WA | WA | WA | WA | NA | NA | WA | NA | WA | WA | WA | WA | WA | - | WA | WA | NA | WA | WA | WA |
| 6738_6#13 | WA | NA | NA | NA | NA | NA | NA | NA | NA | WA | NA | NA | NA | NA | WA | WA | NA | WA | NA | NA | NA | NA | WA | WA | WA |
| 6738_6#14 | WA | WA | NA | NA | NA | WA | WA | WA | WA | NA | NA | WA | WA | NA | WA | NA | WA | WA | WA | WA | WA | NA | WA | WA | NA |
| 6738_6#15 | WA | NA | NA | WA | WA | NA | NA | NA | NA | NA | NA | NA | NA | WA | WA | WA | NA | WA | NA | NA | NA | NA | NA | NA | NA |
| 6738_6#16 | WA | NA | NA | WA | WA | WA | WA | WA | WA | NA | NA | NA | WA | WA | NA | NA | NA | WA | WA | WA | WA | NA | NA | WA | NA |
| 6738_6#17 | WA | WA | NA | WA | WA | NA | NA | NA | NA | WA | NA | WA | WA | WA | WA | WA | WA | WA | WA | NA | NA | WA | WA | WA | WA |
| 6738_6#18 | WA | NA | NA | NA | NA | WA | WA | WA | WA | NA | NA | WA | NA | WA | NA | NA | WA | NA | NA | WA | WA | WA | WA | WA | WA |
| 6738_6#19 | WA | NA | NA | WA | WA | WA | WA | WA | WA | NA | NA | WA | WA | WA | NA | NA | WA | NA | WA | WA | WA | NA | WA | WA | WA |
| 6738_6#20 | WA | WA | NA | WA | WA | WA | WA | WA | WA | NA | NA | NA | NA | NA | NA | NA | NA | WA | NA | NA | NA | NA | NA | NA | NA |
| 6738_6#21 | WA | NA | NA | NA | NA | WA | WA | WA | WA | NA | NA | NA | NA | NA | WA | WA | WA | WA | NA | WA | WA | NA | WA | WA | WA |
| 6738_6#22 | WA | NA | NA | NA | NA | WA | WA | WA | WA | WA | NA | WA | NA | WA | NA | NA | WA | NA | WA | NA | NA | NA | NA | NA | WA |
| 6738_6#23 | WA | WA | NA | NA | NA | WA | WA | WA | WA | NA | NA | NA | NA | WA | WA | WA | WA | NA | NA | WA | WA | NA | WA | WA | WA |
| 6738_6#24 | WA | WA | NA | NA | NA | NA | NA | NA | NA | WA | NA | WA | WA | WA | NA | NA | NA | NA | NA | NA | NA | WA | WA | WA | WA |
| 6738_6#25 | WA | WA | NA | NA | NA | WA | WA | WA | WA | NA | NA | WA | NA | WA | NA | NA | NA | NA | NA | NA | NA | WA | WA | WA | NA |
| 6738_6#26 | WA | WA | NA | WA | WA | WA | WA | WA | WA | WA | NA | NA | WA | WA | WA | NA | WA | WA | WA | WA | WA | WA | WA | WA | WA |
| 6738_6#27 | WA | WA | NA | WA | WA | WA | WA | WA | WA | NA | NA | WA | NA | WA | WA | WA | WA | WA | WA | NA | NA | WA | WA | WA | WA |
| 6738_6#29 | WA | NA | NA | WA | WA | WA | WA | WA | WA | WA | NA | NA | NA | NA | NA | NA | NA | WA | NA | NA | NA | WA | WA | WA | WA |
| 6738_6#30 | WA | NA | NA | NA | NA | NA | WA | WA | WA | WA | NA | WA | WA | WA | WA | WA | NA | NA | WA | NA | NA | NA | WA | WA | NA |
| 6738_6#31 | WA | NA | NA | WA | WA | WA | WA | WA | WA | NA | NA | NA | WA | NA | NA | NA | WA | WA | WA | WA | WA | WA | WA | WA | WA |
| 6738_6#33 | WA | WA | NA | NA | NA | WA | WA | WA | WA | NA | NA | WA | NA | NA | NA | NA | WA | NA | NA | NA | NA | WA | NA | NA | WA |
| 6738_6#34 | WA | WA | NA | NA | NA | WA | WA | WA | WA | NA | NA | WA | NA | NA | WA | WA | NA | NA | NA | NA | NA | NA | WA | WA | NA |
| 6738_6#35 | WA | WA | NA | NA | NA | WA | WA | WA | WA | WA | NA | WA | WA | WA | WA | WA | WA | NA | WA | WA | WA | NA | WA | WA | NA |
| 6738_6#36 | WA | WA | NA | WA | WA | WA | WA | WA | WA | NA | NA | WA | NA | WA | NA | NA | WA | NA | WA | NA | NA | WA | WA | WA | WA |
| 6738_6#37 | WA | WA | NA | NA | NA | WA | WA | WA | WA | WA | NA | NA | NA | WA | NA | NA | NA | NA | NA | NA | NA | NA | NA | NA | NA |
| 6738_6#38 | WA | NA | NA | NA | NA | WA | WA | WA | WA | NA | NA | NA | NA | WA | NA | NA | NA | WA | WA | WA | WA | WA | WA | WA | NA |
| 6738_6#39 | WA | NA | NA | NA | NA | NA | NA | NA | NA | NA | NA | NA | NA | NA | NA | NA | NA | WA | WA | NA | NA | WA | WA | WA | WA |
| 6738_6#40 | WA | WA | NA | WA | WA | WA | WA | WA | WA | WA | NA | NA | WA | WA | NA | NA | WA | NA | NA | WA | NA | NA | WA | WA | WA |
| 6738_6#41 | WA | WA | NA | WA | WA | WA | WA | WA | WA | NA | NA | NA | NA | WA | WA | WA | WA | WA | NA | NA | NA | NA | WA | WA | WA |
| 6738_6#42 | WA | WA | NA | WA | WA | WA | WA | WA | WA | NA | NA | NA | NA | NA | WA | WA | WA | WA | NA | NA | NA | NA | WA | WA | NA |
| 6738_6#43 | WA | NA | NA | NA | NA | WA | WA | WA | WA | NA | NA | WA | WA | NA | WA | WA | WA | NA | NA | NA | NA | NA | NA | NA | WA |
| 6738_6#44 | WA | NA | NA | NA | NA | WA | WA | WA | WA | WA | NA | NA | WA | NA | WA | WA | WA | NA | WA | NA | NA | NA | WA | WA | NA |
| 6738_6#47 | WA | WA | NA | NA | NA | WA | WA | WA | WA | WA | NA | NA | WA | WA | NA | NA | NA | WA | NA | WA | WA | NA | WA | WA | WA |
| 6738_6#48 | WA | WA | NA | WA | WA | WA | WA | WA | WA | NA | NA | NA | WA | WA | WA | WA | NA | WA | NA | NA | NA | WA | WA | WA | WA |
| 6738_6#49 | WA | WA | NA | WA | WA | WA | WA | WA | WA | WA | NA | WA | WA | NA | NA | NA | WA | NA | WA | NA | NA | NA | NA | NA | NA |
| 6738_6#50 | WA | WA | NA | WA | WA | WA | WA | WA | WA | NA | NA | WA | NA | WA | NA | NA | WA | WA | NA | WA | WA | NA | WA | WA | WA |
| 6738_6#51 | WA | WA | NA | NA | NA | NA | NA | NA | NA | NA | NA | NA | NA | NA | WA | WA | WA | NA | NA | NA | NA | NA | WA | WA | WA |
| 6738_6#53 | WA | WA | NA | NA | NA | WA | WA | WA | WA | WA | NA | NA | WA | WA | WA | WA | WA | WA | WA | NA | WA | NA | WA | WA | NA |
| 6738_6#54 | WA | NA | NA | NA | NA | NA | NA | NA | NA | WA | NA | WA | NA | - | NA | NA | WA | WA | NA | WA | NA | NA | WA | WA | NA |
| 6738_6#55 | WA | WA | NA | WA | WA | WA | WA | WA | WA | NA | NA | WA | WA | WA | WA | WA | WA | NA | NA | NA | NA | WA | WA | WA | NA |
| 6738_6#56 | WA | NA | NA | NA | NA | WA | WA | WA | WA | WA | NA | NA | WA | WA | WA | WA | NA | WA | WA | NA | NA | NA | NA | WA | WA |
| 6738_6#57 | WA | WA | NA | NA | NA | WA | WA | WA | WA | NA | NA | NA | WA | WA | NA | NA | WA | NA | WA | NA | NA | NA | NA | NA | NA |
| 6738_6#58 | WA | WA | NA | NA | NA | NA | NA | NA | NA | NA | NA | NA | WA | WA | NA | WA | WA | NA | WA | WA | WA | NA | WA | WA | NA |
| 6738_6#59 | WA | NA | NA | WA | WA | NA | NA | NA | NA | WA | NA | NA | WA | WA | WA | WA | NA | WA | NA | NA | NA | NA | WA | WA | NA |
| 6738_6#61 | WA | WA | NA | NA | NA | WA | WA | WA | WA | NA | NA | NA | NA | WA | NA | NA | WA | NA | NA | WA | WA | WA | WA | WA | WA |
| 6738_6#62 | WA | NA | NA | WA | WA | WA | WA | WA | WA | WA | NA | WA | WA | NA | NA | NA | WA | NA | NA | NA | NA | WA | NA | NA | NA |
| 6738_6#63 | WA | WA | NA | NA | NA | WA | WA | WA | WA | WA | NA | WA | WA | WA | WA | WA | WA | WA | WA | NA | NA | WA | WA | WA | WA |
| 6738_6#64 | WA | WA | NA | WA | WA | WA | WA | WA | WA | WA | NA | NA | WA | WA | WA | WA | NA | NA | WA | NA | NA | NA | WA | WA | NA |
| 6738_6#65 | WA | WA | NA | NA | NA | NA | NA | NA | NA | NA | NA | WA | NA | NA | WA | WA | WA | WA | NA | WA | WA | WA | NA | NA | WA |
| 6738_6#66 | WA | NA | NA | NA | NA | NA | NA | WA | WA | NA | NA | WA | NA | WA | WA | WA | WA | WA | WA | NA | NA | NA | WA | WA | NA |
| 6738_6#67 | WA | WA | NA | NA | NA | WA | WA | WA | WA | WA | NA | NA | NA | WA | WA | WA | WA | WA | NA | NA | NA | WA | WA | WA | NA |
| 6738_6#68 | WA | NA | NA | WA | WA | WA | WA | WA | WA | WA | NA | NA | NA | WA | NA | NA | WA | NA | NA | NA | NA | WA | WA | WA | WA |
| 6738_6#69 | WA | NA | NA | WA | WA | WA | WA | WA | WA | NA | NA | NA | WA | WA | NA | NA | WA | WA | NA | WA | WA | NA | WA | WA | WA |
| 6738_6#70 | WA | WA | NA | WA | WA | WA | WA | WA | WA | WA | NA | NA | NA | NA | WA | WA | WA | NA | WA | WA | WA | WA | WA | WA | WA |
| 6738_6#71 | WA | WA | NA | NA | NA | WA | WA | WA | WA | NA | NA | WA | NA | NA | NA | NA | NA | NA | NA | NA | NA | WA | NA | NA | NA |
| 6738_6#72 | WA | NA | NA | WA | WA | WA | WA | WA | WA | WA | NA | NA | WA | NA | WA | WA | WA | WA | WA | WA | WA | NA | WA | WA | NA |
| 6738_6#74 | WA | WA | NA | NA | NA | NA | NA | NA | NA | WA | NA | NA | NA | WA | NA | NA | WA | WA | WA | NA | NA | NA | WA | WA | WA |
| 6738_6#75 | WA | NA | NA | NA | NA | WA | WA | WA | WA | NA | NA | WA | NA | WA | NA | NA | WA | NA | WA | WA | WA | WA | NA | NA | WA |
| 6738_6#76 | WA | WA | NA | WA | WA | WA | WA | WA | WA | WA | NA | NA | WA | NA | NA | NA | NA | WA | NA | WA | WA | WA | WA | WA | WA |
| 6738_6#77 | WA | NA | NA | WA | WA | WA | WA | WA | WA | NA | NA | NA | WA | WA | NA | NA | WA | NA | WA | WA | WA | WA | NA | NA | NA |
| 6738_6#78 | WA | WA | NA | NA | NA | WA | WA | WA | WA | WA | NA | NA | NA | NA | NA | NA | NA | WA | NA | WA | WA | WA | WA | WA | WA |
| 6738_6#79 | WA | NA | NA | NA | NA | WA | WA | WA | WA | NA | NA | WA | NA | WA | NA | NA | WA | NA | WA | NA | NA | WA | WA | WA | WA |
| 6738_6#80 | WA | NA | NA | WA | WA | WA | WA | WA | WA | WA | NA | NA | NA | WA | WA | WA | WA | WA | WA | NA | NA | NA | WA | WA | NA |
| 6738_6#81 | WA | WA | NA | WA | WA | WA | WA | WA | WA | NA | NA | NA | WA | WA | NA | NA | WA | NA | NA | NA | NA | WA | WA | WA | NA |
| 6738_6#83 | WA | NA | NA | NA | NA | WA | WA | WA | NA | NA | NA | NA | NA | WA | WA | WA | NA | WA | WA | NA | NA | NA | WA | WA | NA |
| 6738_6#84 | WA | WA | NA | NA | NA | NA | NA | NA | NA | WA | NA | WA | NA | WA | WA | WA | WA | WA | WA | WA | NA | NA | WA | WA | NA |
| 6738_6#85 | WA | WA | NA | NA | NA | WA | WA | WA | WA | NA | NA | WA | WA | NA | WA | WA | WA | WA | NA | WA | NA | NA | WA | WA | NA |
| 6738_6#86 | WA | WA | NA | WA | WA | WA | WA | WA | WA | NA | NA | NA | WA | WA | WA | WA | NA | NA | WA | WA | WA | WA | WA | WA | WA |
| 6738_6#87 | WA | NA | NA | NA | NA | WA | WA | WA | WA | WA | NA | WA | NA | NA | WA | WA | NA | NA | NA | WA | WA | WA | WA | WA | WA |
| 6738_6#88 | WA | NA | NA | NA | NA | WA | WA | WA | WA | WA | NA | NA | WA | NA | NA | NA | WA | NA | WA | WA | WA | NA | NA | NA | WA |
| 6738_6#89 | WA | WA | NA | WA | WA | WA | WA | WA | WA | NA | NA | WA | NA | WA | WA | NA | WA | WA | WA | WA | NA | NA | WA | WA | WA |
| 6738_6#90 | WA | WA | NA | NA | NA | WA | WA | WA | WA | WA | NA | NA | NA | NA | WA | WA | WA | NA | NA | WA | NA | NA | WA | WA | WA |
| 6738_6#91 | WA | NA | NA | NA | NA | WA | WA | WA | WA | NA | NA | WA | NA | NA | NA | NA | WA | NA | NA | NA | NA | NA | WA | WA | WA |
| 6738_6#92 | WA | NA | NA | NA | NA | WA | WA | WA | WA | NA | NA | NA | NA | WA | WA | WA | WA | NA | NA | NA | NA | WA | NA | WA | WA |
| 6738_6#93 | WA | NA | NA | NA | NA | WA | WA | WA | WA | NA | NA | WA | NA | WA | NA | NA | NA | NA | WA | NA | NA | WA | NA | NA | WA |
| 6738_6#94 | WA | NA | NA | NA | NA | WA | WA | WA | WA | WA | NA | NA | WA | WA | NA | NA | NA | WA | WA | WA | NA | NA | NA | NA | WA |
| 6738_6#95 | WA | NA | NA | NA | NA | WA | WA | WA | WA | NA | NA | WA | WA | WA | NA | NA | WA | NA | NA | WA | WA | WA | WA | WA | NA |

## Table S4. Checking of bias in samples with missed values (unselected)

*P* -value is the probability of observing by chance the same or greater deviance if two random values are independent

| Segregant | Original database | | | Complete database | | | p-value of independence |
|---|---|---|---|---|---|---|---|
| | Total | NA | WA | Total | NA | WA | |
| chr01-0040xxx | 172 | 0 | 172 | 165 | 0 | 165 | 1.00 |
| chr01-0119xxx | 172 | 81 | 91 | 165 | 77 | 88 | 0.94 |
| chr02-0472xxx | 171 | 87 | 84 | 165 | 85 | 80 | 0.91 |
| chr02-0517xxx | 172 | 114 | 58 | 165 | 108 | 57 | 0.87 |
| chr02-0522xxx | 172 | 114 | 58 | 165 | 108 | 57 | 1.00 |
| chr04-0454xxx | 172 | 37 | 135 | 165 | 35 | 130 | 0.95 |
| chr04-0461xxx | 172 | 38 | 134 | 165 | 35 | 130 | 0.84 |
| chr04-0488xxx | 172 | 30 | 142 | 165 | 28 | 137 | 0.91 |
| chr04-0496xxx | 172 | 30 | 142 | 165 | 28 | 137 | 1.00 |
| chr04-1313xxx | 172 | 101 | 71 | 165 | 97 | 68 | 0.99 |
| chr05-0196xxx | 172 | 172 | 0 | 165 | 165 | 0 | 1.00 |
| chr07-0131xxx | 172 | 105 | 67 | 165 | 101 | 64 | 0.98 |
| chr07-0859xxx | 172 | 89 | 83 | 165 | 85 | 80 | 0.97 |
| chr09-0292xxx | 169 | 61 | 108 | 165 | 60 | 105 | 0.96 |
| chr10-0234xxx | 172 | 79 | 93 | 165 | 75 | 90 | 0.93 |
| chr10-0235xxx | 172 | 79 | 93 | 165 | 76 | 89 | 0.98 |
| chr10-0420xxx | 172 | 54 | 118 | 165 | 52 | 113 | 0.98 |
| chr12-0140xxx | 172 | 95 | 77 | 165 | 95 | 70 | 0.66 |
| chr12-0730xxx | 170 | 93 | 77 | 165 | 90 | 75 | 0.98 |
| chr13-0893xxx | 172 | 112 | 60 | 165 | 108 | 57 | 0.95 |
| chr13-0910xxx | 172 | 115 | 57 | 165 | 111 | 54 | 0.94 |
| chr14-0441xxx | 172 | 108 | 64 | 165 | 103 | 62 | 0.94 |
| chr15-0172xxx | 171 | 47 | 124 | 165 | 45 | 120 | 0.97 |
| chr15-0179xxx | 172 | 41 | 131 | 165 | 39 | 126 | 0.97 |
| chr15-1032xxx | 172 | 71 | 101 | 165 | 67 | 98 | 0.90 |

# Table S5. Independence test for the same loci in the heat selected and unselected pools

*P* -value is the probability of observing by chance the same or greater difference in proportion of NA and WA

(green) $0.1 < p$-value

(yellow) $0.01 < p\text{-value} < 0.1$

| Segregant | Unselected | | Heat selected | | Fisher's exact test | Chi squared |
|---|---|---|---|---|---|---|
| | NA | WA | NA | WA | | |
| chr01-0040xxx | 0 | 165 | 4 | 892 | 0.5081 | 0.3899 |
| chr01-0119xxx | 77 | 88 | 684 | 212 | <0.001 | <0.001 |
| chr02-0472xxx | 85 | 80 | 587 | 309 | <0.001 | <0.001 |
| chr02-0517xxx | 108 | 57 | 894 | 2 | <0.001 | <0.001 |
| chr02-0522xxx | 108 | 57 | 896 | 0 | <0.001 | <0.001 |
| chr04-0454xxx | 35 | 130 | 340 | 556 | <0.001 | <0.001 |
| chr04-0461xxx | 35 | 130 | 340 | 556 | <0.001 | <0.001 |
| chr04-0488xxx | 28 | 137 | 0 | 896 | <0.001 | <0.001 |
| chr04-0496xxx | 28 | 137 | 318 | 578 | <0.001 | <0.001 |
| chr04-1313xxx | 97 | 68 | 612 | 284 | 0.0116 | 0.0171 |
| chr05-0196xxx | 165 | 0 | 896 | 0 | 1.0000 | NA |
| chr07-0131xxx | 101 | 64 | 501 | 395 | 0.1194 | 0.2069 |
| chr07-0859xxx | 85 | 80 | 831 | 65 | <0.001 | <0.001 |
| chr09-0292xxx | 60 | 105 | 597 | 299 | <0.001 | <0.001 |
| chr10-0234xxx | 75 | 90 | 617 | 279 | <0.001 | <0.001 |
| chr10-0235xxx | 76 | 89 | 624 | 272 | <0.001 | <0.001 |
| chr10-0420xxx | 52 | 113 | 483 | 413 | <0.001 | <0.001 |
| chr12-0140xxx | 95 | 70 | 496 | 400 | 0.3300 | 0.5980 |
| chr12-0730xxx | 90 | 75 | 645 | 251 | <0.001 | <0.001 |
| chr13-0893xxx | 108 | 57 | 882 | 14 | <0.001 | <0.001 |
| chr13-0910xxx | 111 | 54 | 832 | 64 | <0.001 | <0.001 |
| chr14-0441xxx | 103 | 62 | 505 | 391 | 0.0863 | 0.1479 |
| chr15-0172xxx | 45 | 120 | 808 | 88 | <0.001 | <0.001 |
| chr15-0179xxx | 39 | 126 | 823 | 73 | <0.001 | <0.001 |
| chr15-1032xxx | 67 | 98 | 741 | 155 | <0.001 | <0.001 |